\documentclass[iop,twocolumn]{emulateapj}
\usepackage{amsmath, amsthm, amssymb, textcomp, verbatim, times, multirow, float, hyperref, natbib}
\newcommand\subsun{\mbox{$_{\sun}$}}

\begin{document}
\title{B- and A-Type Stars in the Taurus-Auriga Star Forming Region}
\author{Kunal Mooley\altaffilmark{1}}
\author{Lynne Hillenbrand\altaffilmark{1}}
\author{Luisa Rebull\altaffilmark{2}}
\author{Deborah Padgett \altaffilmark{2,4}}
\author{Gillian Knapp\altaffilmark{3}}
\affiliation{$^1$Department of Astronomy, California Institute of Technology, 1200 E. California Blvd., MC 249-17, Pasadena, CA 91125, USA; \href{mailto:kunal@astro.caltech.edu}{kunal@astro.caltech.edu}} 
\affiliation{$^2$Spitzer Science Center, California Institute of Technology, Pasadena, CA, USA}
\affiliation{$^3$Department of Astrophysics, Princeton University, Princeton, NJ, USA}
\affiliation{$^4$current address: Goddard Space Flight Center, Greenbelt, MD, USA}
\submitted{Draft of \today}

\begin{abstract}
We describe the results of a search for early-type 
stars associated with the Taurus-Auriga  
molecular cloud complex, a diffuse nearby star-forming region noted
as lacking young stars of intermediate and high mass.  We investigate
several sets of possible O, B and early A spectral class members. 
The first is a group of stars for which mid-infrared images show 
bright nebulae, all of which can be associated with stars of spectral type B. 
The second group consists of early-type stars compiled from 
(i) literature listings in SIMBAD; (ii) B stars with infrared 
excesses selected from the \textit{Spitzer Space Telescope} survey
of the Taurus cloud; (iii) magnitude- and color-selected 
point sources from the \textit{Two Micron All Sky Survey}; and
(iv) spectroscopically identified early-type stars from the
\textit{Sloan Digital Sky Survey} coverage of the Taurus region. 
We evaluated stars for membership in the Taurus-Auriga 
star formation region based on criteria involving:
spectroscopic and parallactic distances, proper motions and radial velocities, 
and infrared excesses or line emission indicative of stellar youth. 
For selected objects, we also model the scattered and emitted radiation 
from reflection nebulosity and compare the results with the observed 
spectral energy distributions to further test the plausibility 
of physical association of the B stars with the Taurus cloud. 
This investigation newly identifies as probable Taurus members 
three B-type stars: HR 1445 (HD 28929), $\tau$ Tau (HD 29763), 72 Tau (HD 28149), 
and two A-type stars: HD 31305 and HD 26212, thus doubling the 
number of stars A5 or earlier associated with the Taurus clouds.  
Several additional early-type sources including HD 29659 and HD 283815
meet some, but not all, of the membership criteria and therefore
are plausible, though not secure, members. 
\end{abstract}

\keywords{stars: early-type, stars: Herbig Ae/Be, stars: formation, ISM: clouds}

\section{INTRODUCTION}\label{sec.intro}

The Taurus-Auriga molecular cloud complex (hereafter ``Taurus'') 
is the quintessential region of nearby recent star formation.
It is characterized by low star-formation efficiency \citep{goldsmith2008} and the 
absence of high-mass young stars \citep{kenyon2008} and stands in contrast to more distant, massive, and dense 
star-forming environments like the Orion Molecular Clouds.
Taurus lies at a mean distance of about 140 pc with a depth of 20 pc or more \citep{kenyon1994,torres2007,torres2012} 
and spans approximately 100 square degrees on the sky, 
or about a 25 pc diameter at this distance.
The few times $10^4$ M$_\odot$ cloud currently has over 350 known members, 
mainly substellar and low-mass stellar objects with M $<0.5$ M$\subsun$, 
and only about 10 members identified with M $>1.5$ M$\subsun$.
Much effort over the past decade in Taurus has been directed 
towards completely defining the low-mass stellar and sub-stellar population.  

A comprehensive review of Taurus is given by \citet{kenyon2008}.
Major recent contributions to our knowledge include:
(i) mapping of the molecular gas \citep{goldsmith2008,narayanan2008} 
and dust \citep{lombardi2010,palmeirim2013} comprising the cloud; 
(ii) determination of the distance of individual young star members 
through Very Long Baseline Interferometry \citep[VLBI; ][]{torres2009} parallaxes; 
(iii) improvement of the young stellar object census including new stellar and
brown dwarf candidate members 
\citep{rebull2010, luhman2009,luhman2010,takita2010,rebull2011} 
as well as new companions to already known objects \citep{kraus2011};
(iv) measurement of proper motions using optical and VLBI techniques, 
\citep{torres2009,luhman2009}; 
(v) provision of evidence for mass segregation \citep{kirk2011,parker2011}; and 
(vi) searches for outflows \citep{narayanan2012, bally2012}.

\begin{figure*}[htp]
\centering
\includegraphics[width=7in,viewport=20 130 600 680,clip]{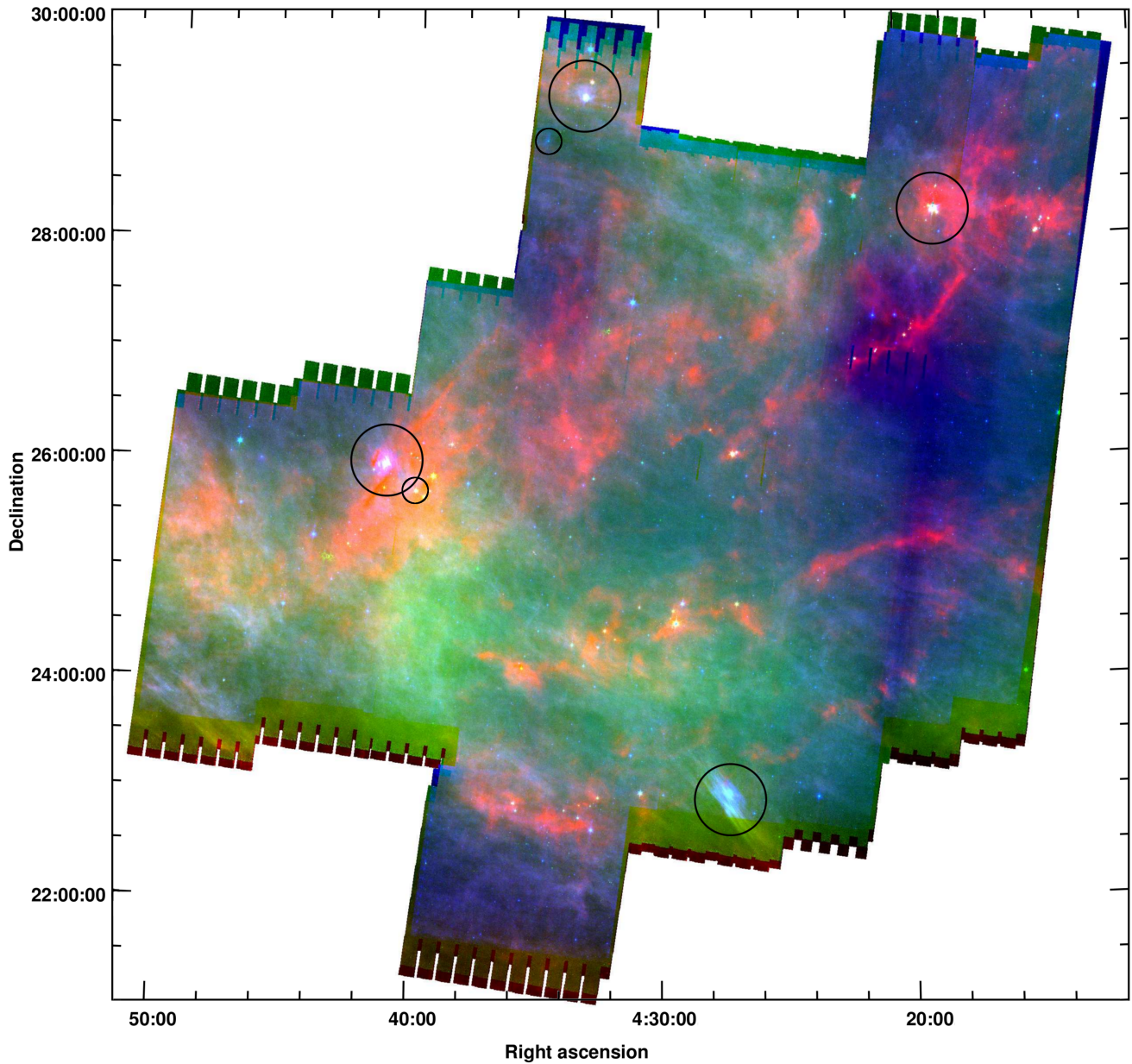}
\caption{Spitzer mosaic of IRAC and MIPS images of Taurus. 
Color coding: 8 (blue), 24 (green), and 160 (red) $\mu$m.
Four bright and large nebulous objects illuminated by B stars are evident in this mosaic. They
have been marked with large circles and are associated with: 
(i) top-middle: HD 282276, (ii) top-right: V892 Tau (Elias 1), (iii) bottom-middle: HD 28149 (72 Tau), and (iv) middle-left: HD 29647. 
Two fainter and weaker nebulae are marked with smaller circles 
and are associated with:
(v) top-middle: HD 28929 (HR 1445),
(vi) middle-left: IC~2087.}
\label{fig.mosaic}
\end{figure*}

In particular, a \textit{Spitzer} program directed at Taurus 
\citep[][PI D. Padgett]{gudel2007b} produced
large-scale multi-wavelength maps of the clouds. Photometry from this survey
has improved our understanding of both the stellar/sub-stellar membership
and the incidence of protoplanetary disks \citep{rebull2010}. 
Motivating the investigation described in this paper are four 
large and two smaller reflected and/or scattered-light nebulae 
found in these mid- and far-infrared images, shown in Figure~\ref{fig.mosaic}.  

Each of the large  infrared nebulae 
is illuminated by a point source that is a known B or A0 star.  
Two of these, HD 28149 (72 Tau) and HD 29647, have been studied 
in the literature to date \citep[e.g.][]{kenyon1994,whittet2001,whittet2004}, 
as background stars and used to derive the physical and chemical 
properties of the molecular cloud.  One source, HD 282276, was unstudied 
until noted by \cite{rebull2010,rebull2011} to have a mid-infrared excess.
Finally, V892 Tau is a well-known Herbig Ae/Be type member 
of Taurus \citep{elias1978} 
that also illuminates an optical reflection nebula --- 
an original defining characteristic of the Herbig Ae/Be population.  
The two additional smaller infrared nebulae are likewise associated with 
early-type or high luminosity stars. HD 28929 = HR 1445 has been known
as a chemically peculiar star \citep[e.g.][]{wolff_preston1978} 
but has not otherwise distinguished itself in the literature.
IC~2087 is associated with a known young stellar object \citep{elias1978}
and, like V892 Tau, illuminates an optical nebula. 
The nebular regions for all six of these sources appear brightest 
at mid-infrared wavelengths; optical nebulosity is generally absent,
but when apparent, is weaker except in the case of IC~2087.

The proximity of these early-type stars to cloud material, 
as evidenced by the mid-infrared nebulae, suggests that rather than being 
background stars as they have been traditionally considered, they may be heretofore 
unappreciated early-type members of Taurus.
The association of these stars with prominent nebulosity is not, however, 
sufficient evidence that they are genuine members of Taurus.  Instead,
they could be stars of early type that are physically unassociated 
but fortuitously located with respect to 
either the Taurus molecular cloud complex itself or smaller patches of 
locally-enhanced density in the foreground or background
interstellar medium (ISM). Reminiscent of this latter 
situation is the Pleiades star cluster, which is passing through 
and illuminating denser-than-average ISM that is physically
unassociated with the stars themselves.  In this contribution, we
explore the evidence and attempt to distinguish between these two 
scenarios.

The earliest-type Taurus members are generally considered 
\citep{kenyon2008, rebull2010} to be 
IC~2087-IR (estimated at B5 based on bolometric luminosity 
but a heavily self-embedded star with a continuum-emission optical spectrum), 
the binary system V892 Tau (Elias 1; a B8-A6 self-embedded Herbig Ae/Be star 
with a near-equal brightness companion), 
the triple system HD 28867 (B8 $+2 \times$B9.5), 
AB Aur (A0e, a Herbig Ae/Be star), and 
HD 31648 (MWC 480; A2e, another Herbig Ae/Be star).
There are no associated F stars \footnote{HD 283759 (F2--F3) and V410-Anon24 (F9--G3) have at various
times been suggested as members, but both are well underluminous 
with respect to the other stars in this list if assumed to be
at the same distance, and not particularly obscured.}
and the next earliest types are HP Tau/G2 (G0) and SU Aur (G2).
Notably, almost all of these earliest-type members of Taurus harbor 
significant amounts of circumstellar material, the HD 28867 
system \footnote{This source is to the south of the main Taurus-Auriga clouds,
in the L 1551 region, and is not otherwise discussed in this paper;
see \citet{walter2003}.} 
being the exception. 

While no systematic investigation for high-mass stars associated with
Taurus has been performed recently, there has been historical interest 
in early-type stars seen towards this canonically low mass star-forming region.
\cite{blaauw1956} identified Casseiopeia-Taurus (hereafter Cas-Tau) 
as an OB association lying between 140 and 220 pc using the 
convergent-point method. There was ensuing debate concerning the
relationship of the young Taurus-Auriga molecular cloud to the older
Cas-Tau OB association, but it has now been resolved that they are
unrelated in both space and age (see below).  
The first early type stars associated with the Taurus clouds was AB Aur, 
an original Ae/Be star associated with optical nebulosity 
\citep{herbig1960}, followed by the infrared-selected young stellar objects 
V892 Tau and IC~2087-IR \citep{elias1978}.

How many higher-mass stars should there be in Taurus?
Using the number and distribution of the known low-mass T Tauri stars and the 
assumption that weak-lined T Tauri stars (WTTS) are far more
numerous (as high as 10:1) than classical T Tauri stars (CTTS), 
\cite{walter1988} argued from a sparse X-ray survey of Taurus with 
\textit{Einstein} that there should be $\sim10^3$ low-mass members of
Taurus. 
The initial mass function (IMF) appropriate for young star 
clusters \citep{miller1979} would then predict 
the existence of tens of B-type members.
The search for these B stars culminated with \cite{walter1991} 
identifying 29 possible members of the Taurus-Auriga T association based
on spectroscopic parallax and proper motion.
However, the large number of low-mass Taurus member stars predicted by 
\cite{walter1988} were not found in the proper-motion survey 
by \cite{hartmann1991}, who suggested that many of 
the stars found in the X-ray survey by \cite{walter1988} belong to 
the older and more distant Cas-Tau group, and that the assumption of
a uniform distribution of T Tauri stars is not correct.
While \cite{walter1991} had concluded that the Tau-Aur T association
was related to the Cas-Tau OB association,
\cite{deZeeuw1999} showed that these are kinematically distinct
groups and rejected the \cite{walter1991} stars as
Cas-Tau members.

Most subsequent publications on the Taurus-Auriga stellar population
have focused on the low-mass stars.  
However, of relevance to the present study is that \cite{whittet2004}
proposed, on the basis of extinction and dust modelling, that 
HD 29647 (B9III) is located within a diffuse screen 
surrounding the dense molecular clump TMC-1 in Taurus.  
Also, \cite{rebull2010} suggested that its infrared excess
and spectroscopic parallax make HD 27659 (A0--A4V) 
a high-quality candidate member of Taurus, and that HD 282276 (B8V) 
and HD 283815 (A0V) are lower-quality, but plausible, candidate members. 
As the situation concerning associated early-type stars remains unclear, 
and considering the high level of interest in the membership and
star-formation history of Taurus, a detailed investigation 
using the wealth of new information seems warranted.

This paper is organized as follows. In \S\ref{sec.data} we describe the process of 
compiling the list of early-type stars towards Taurus (\S\ref{subsec.list}) and 
testing these stars for membership (\S\ref{subsec.membership}).
Our literature survey showed that spectral typing and age estimation of
some of the early-type stars was done decades ago 
with prism-based spectrometers, and so we considered the possibility 
that some have been assigned an incorrect spectral type.
Additionally, we considered that some early-type stars could have been 
missed due to limits on the spatial extent or photometric depth of previous investigations.
We spectroscopically followed up all probable members (\S\ref{subsec.spectroscopy}).
Given our initial motivation for this investigation,
reflection/scattered-light nebulae toward several known B stars,
we then describe the modelling procedure for the scattered/thermal 
dust emission in \S\ref{sec.dusty}.
Detailed discussion of our findings and results for individual objects 
are provided in \S\ref{sec.notes}.
We conclude with a summary and discussion in \S\ref{sec.summary}.

\section{DATA COLLECTION AND ANALYSIS}\label{sec.data}

\subsection{OVERVIEW}

Working from the evidence of early-type members provided by the 
reflection/scattered-light nebulae and the recent suggestions of 
additional B and A0 type stars as plausible members, we carried out a new 
search for stars of spectral classes O--A0 associated with Taurus.
The areal extent of our study is the region bounded by 4$^h$ and 5$^h$ 
in right ascension and 22\degr and 31\degr in declination. In galactic 
coordinates, this is roughly the region $(165,-20) \lesssim (l,b) \lesssim (180,-10)$.  
This boundary was chosen to include most of the dense cores in 
Taurus but not to be so large as to obfuscate the search 
with unassociated early-type stars at different distances.
The region south of the Taurus main cloud between 4.3$^h$ and 4.9$^h$ in 
right ascension and 16\degr and 20\degr in declination is also considered a 
part of the Taurus star-forming region, but is not included in this investigation.

We first looked at whether there is a concentration of early-type stars towards
or away from the large patch of sky under consideration.
A SIMBAD query for known O,B stars towards Taurus and eight 
neighboring regions of equal areal extent results in the distribution shown in
Table~\ref{tab.bstar_bias}. There is a higher density of known 
early-type stars in the direction of the galactic plane as expected, 
and no particular bias of early-type stars 
at the longitude of Taurus compared to adjacent longitudes.
Thus, from our study, we expect to find only a few known early-type stars, 
if any, that are genuinely associated with Taurus.

In order to compile the list of candidate early-type  stars towards Taurus, 
we gathered multiwavelength photometric and spectroscopic data and images,
and collected information from the literature.
We then passed these stars through two membership tests: appropriate
distance and appropriate kinematics.
All candidates satisfying these two criteria were labeled 
as probable members of Taurus.  These likely members along with 
other stars meeting some but not all of the criteria 
were followed up spectroscopically.


\subsection{COMPILING THE LIST OF CANDIDATE EARLY-TYPE STARS AND ANCILLARY DATA}\label{subsec.list}
\begin{figure}
\centering
\includegraphics[height=3.5in,angle=-90,viewport=10 30 570 750,clip]{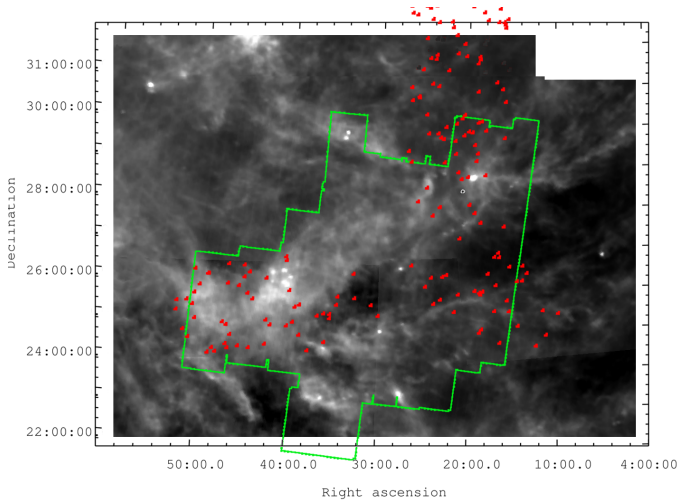}
\caption{Areal extent of the Spitzer Taurus Legacy Survey (green) and SDSS spectroscopic observations by \cite{knapp2007} (red symbols)
overlaid on a mosaic of the Taurus region at 100$\mu$m 
from the {\it IRAS} Sky Survey Atlas.}
\label{fig.coverage}
\end{figure}

Four data sets were used to assemble a list of early-type stars towards 
the Taurus region: (i) previously identified O and B-type stars listed in 
SIMBAD; (ii) proposed B and early A stars with infrared excesses selected
from the {\it Spitzer} survey of the Taurus cloud discussed in \cite{rebull2010}; 
(iii) photometrically-selected point sources from the \textit{Two Micron All Sky Survey}
\citep[2MASS-PSC;][]{skrutskie2006}; and (iv) 
spectroscopically identified early-type stars from the 
{\textit Sloan Digital Sky Survey} ({\it SDSS}) observations of the
Taurus region \citep{finkbeiner2004} presented by \cite{knapp2007}.  
As illustrated in Figure~\ref{fig.coverage}, there is only partial 
coverage of the total cloud region (see also Figure~\ref{fig.distance})
in each of {\it SDSS} and the \cite{rebull2010} {\it Spitzer} surveys, 
and the overlap between the optical and infrared photometric surveys
is also only partial. 
We now describe the collation of data from each of the four sources.

First, to select early-type stars from SIMBAD, we used the criterion 
query: $ra > 60 \ \& \ ra < 75 \ \& \ dec > 22 \ \& \ dec < 31 \ \& \ sptypes < A0$.
This query (run in early 2011) resulted in 91 stars, three of which were listed twice
with different names.
We thus obtained 88 B stars and zero O stars through the
SIMBAD database as candidates.

Second, potential Taurus members having early spectral types 
were taken from Tables 5 and 7 
in \cite{rebull2010}. One of these, JH 225, also resulted from the SIMBAD search. 
Thus, the \cite{rebull2010} paper added eight more stars with spectral 
types early A or B (O-type stars were absent).  As noted above, 
the region covered by {\it Spitzer} does not encompass the whole 
region of our search (see Figure~\ref{fig.coverage}).

\begin{figure*}[htp]
\begin{center}
\includegraphics[width=3.5in]{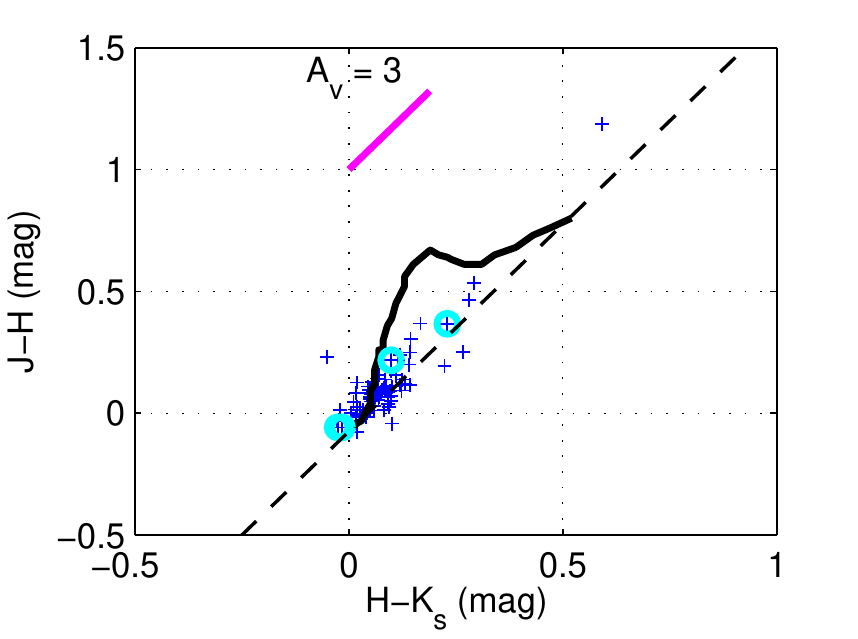}
\includegraphics[width=3.5in]{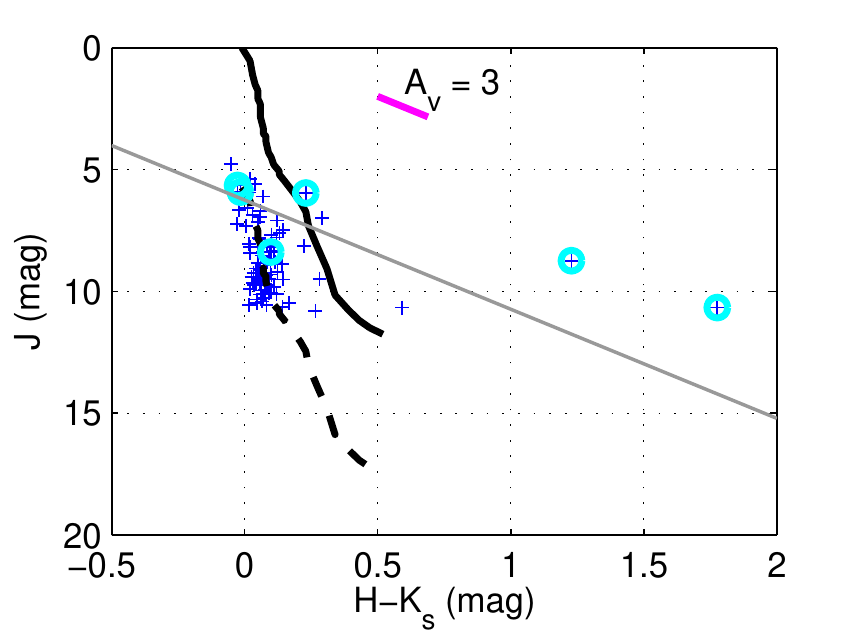}  
\includegraphics[width=3.5in]{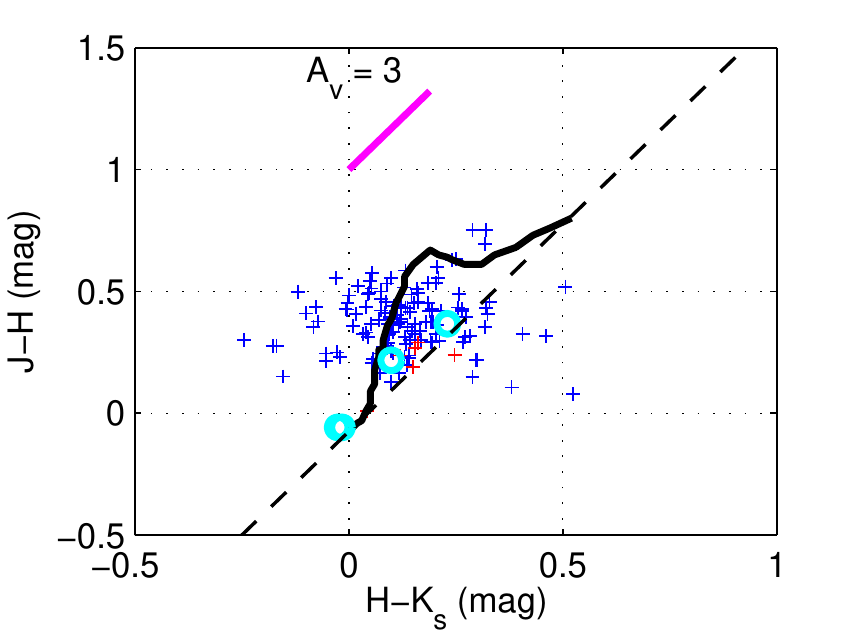}
\includegraphics[width=3.5in]{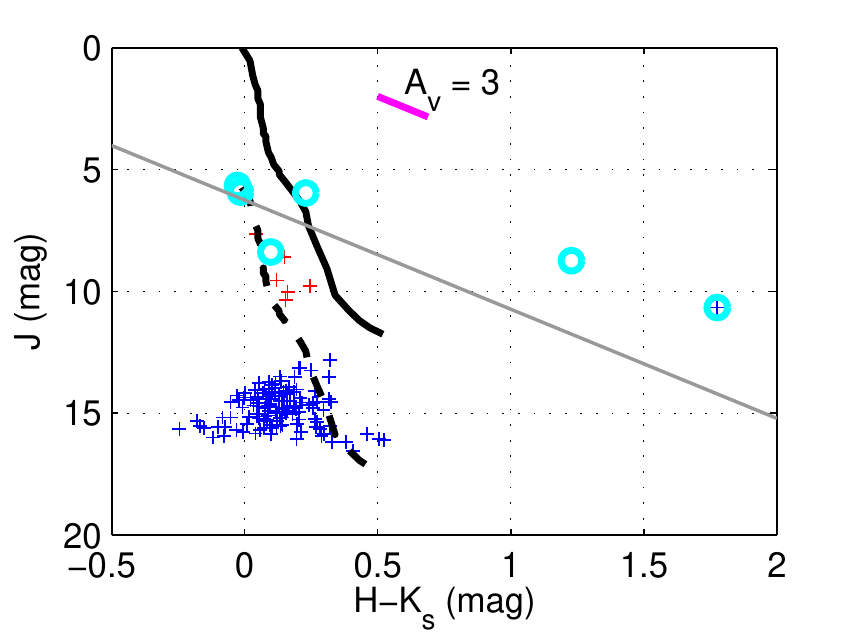}
\includegraphics[width=3.5in]{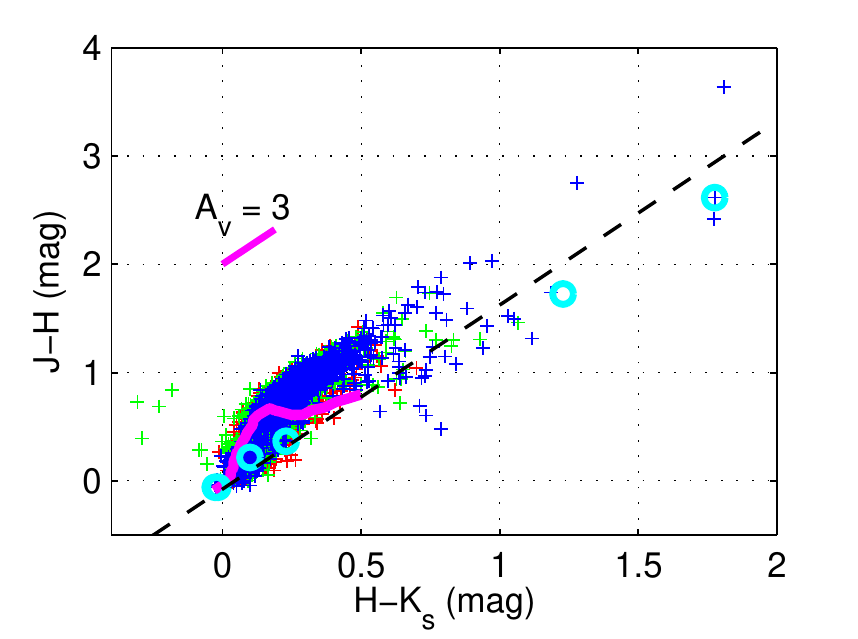}
\includegraphics[width=3.5in]{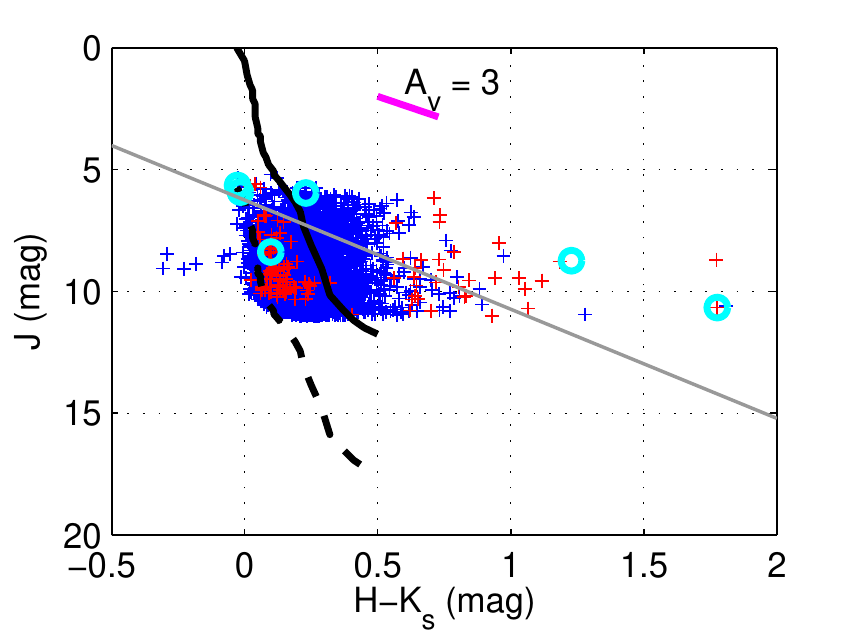}
\end{center}
\caption{2MASS color-color (left panels) and color-magnitude (right panels) diagrams for the early-type candidate stars considered in this work.
2MASS objects with contamination flags set and error in J,H,K$_s$ magnitudes greater than 0.1 were 
rejected from the plots. The mean error in J,H,K$_s$ magnitudes is about 0.02, which is smaller than the size of symbols used.
{\it Top panels}: O,B stars from SIMBAD. 
In the color-color diagram, two of these stars lie outside the range plotted: V892 Tau (Elias 1) 
and IC 2087-IR, having $(H-K_s, J-H) = (1.23,1.73), (1.78,2.62)$ respectively.
{\it Middle panels}: Stars of spectral-type A and earlier from \cite{knapp2007} are shown in blue.
The B stars proposed by \cite{rebull2010} are shown in red.
Note that the area covered by these surveys is less than that represented 
in the top and bottom panels.
{\it Bottom panels}: Left: 2MASS objects with color-coding as follows. Blue: stars with magnitude $K_s<8$, green: $8<K_s<9$, red: $9<K_s<10$. 
Right: All 2MASS objects are in blue, and those selected as possible O,B stars using 
the photometric selection criterion (described in \S\ref{subsec.list}) are shown in red.  
{\it All panels}: 
The reddening vector (magenta) from \cite{rieke_lebofsky1985} is used.
Intrinsic colors and magnitudes of main-sequence stars \citep[from Table 5 of ][]{kraus_hillenbrand2007} are shown as a thick black curve (magenta curve in the bottom panel color-color diagram).
The thick black, dashed curve in color-magnitude diagrams is the same curve, but displaced along the luminosity axis to denote the apparent magnitude of main sequence stars at 140 pc.
The thin black, dashed straight line in the color-color diagrams represents the color-selection criterion applied to the 2MASS objects (see section~\ref{subsec.list}).
The thin grey solid line in color-magnitude diagrams represents the reddening vector passing through the position of an A0V star at a distance of 140 pc.
The location of the six B stars illuminating bright IR nebulae are shown as cyan circles with two of the stars having very similar, near-zero, colors.}
\label{fig.ccd_cmd}
\end{figure*}

\begin{figure}[htp]
\centering
\includegraphics[height=3.5in,angle=-90,viewport=40 50 570 750]{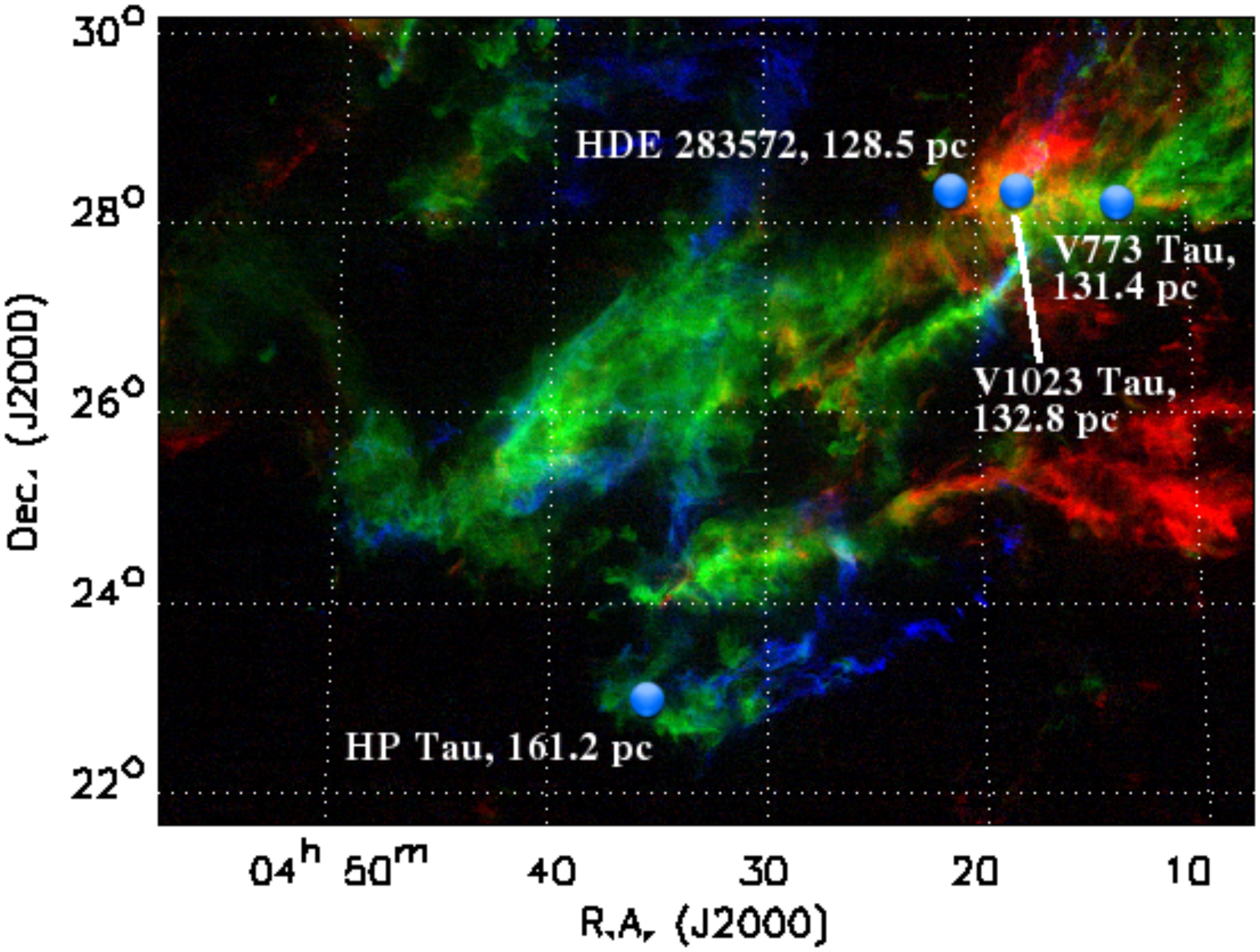}
\caption{Distances to stars in Taurus measured via VLBI, from Table~\ref{tab.distance}; 
the star T Tau is not shown since it lies south of the region of interest.
Background image is velocity-coded $^{12}CO$ map from \cite{goldsmith2008}. 
The LSR velocities are color-coded as
blue: 3--5 km s$^{-1}$, green: 5--7 km s$^{-1}$, red: 7--9 km s$^{-1}$.}
\label{fig.distance}
\end{figure}

Third, we selected from the 2MASS-PSC 
\footnote{using the multi-object search form at the Infrared Science Archive 
(\href{http://irsa.ipac.caltech.edu/}{http://irsa.ipac.caltech.edu/})} 
objects satisfying the same coordinates constraint used in the 
SIMBAD query, having $K_s < 10$ mag with $>5\sigma$ detection,
and no contamination or quality flags set.
The brightness threshold places an upper limit on the visual extinction 
for the selected stars. For example, a B8V star can have
a maximum visual extinction of $A_V \simeq 37$ to be selected, since the absolute 
K-band magnitude for such a star is $M_{K_s} = 0.11$, the distance modulus at 
140 pc is 5.73, and the reddening law for 2MASS magnitudes is $A_K = 0.112 A_V$.
Such a large value of extinction is much higher than the largest extinction 
observed for known Taurus members.
The resulting 2MASS-PSC sample appears in the lower panels of
Figure~\ref{fig.ccd_cmd}.
These objects were further filtered through the photometric color 
criterion: $J-H < 1.698 (H-K_s+0.158)$ in order to 
select stars which, when translated backwards on the 
reddening vector in the $(J-H)/(H-K_s)$ color-color diagram, 
fall on the main-sequence earlier than spectral-type A0.
For this procedure, we used intrinsic magnitudes from 
\cite{kraus_hillenbrand2007}, and the \cite{rieke_lebofsky1985} 
reddening law, which is found consistent with reddening in 
the 2MASS photometric system \citep{maheswar2010}.
There is no a priori reason to believe that this color criterion 
unconditionally, due to one or more of the following reasons:
(i) photometric errors could place stars within the reddening band 
employed in our selection; 
(ii) the reddening law (parametrized by $R_V$) is different 
for different lines of sight towards Taurus \citep[e.g.][]{dobashi2005};
(iii) stars of a given spectral type and luminosity class do not have 
unique photometric colors but tend to have a dispersion of astrophysical 
origin about the observed mean intrinsic value; 
(iv) additional emission of non-photospheric origin could change the $J-H$ and $H-K_s$ colors, 
possibly making late-type stars with infrared excesses look like earlier-type stars 
that are reddened; 
(v) stellar multiplicity is unaccounted for in our analysis.
Nevertheless, using a more relaxed criterion is subject to the risk of selecting 
a large number of unreddened K and M-type stars which lie across the 
reddening vector defined by our photometric color criterion.
Using our color criterion, we obtained 113 stars for further consideration. 
Fourteen of the stars selected in this manner were already present in the SIMBAD list 
of known early type stars
(BD+23 607, HD 25487, V1137 Tau, HD 284228, HD 282240, HD 29259, 2MASS J04395574+2545020 = IC 2087-IR, HD 283845, HD 283952, HD 31353, HD 284941, HD 284012, HD 283751, HD 283794)
and so we added 99 early-type candidates through this criterion, 
which we appended to the working list.  We also add the famous star AB Aur at this point,
which would pass our color selection criteria but is formally excluded from our analysis 
based on 2MASS-PSC flags present at K$_s$-band. 

Last, we added to our early-type candidates list the stars from 
the \cite{finkbeiner2004} survey belonging to our spatial region 
of interest which are classified as spectral class O, B, or A based on low-resolution
{\it SDSS} spectra. \citeauthor{finkbeiner2004} chose the program stars for spectroscopy as those meeting one of two criteria: 
on the basis of red colors, as part of a survey seeking M-dwarfs, 
or as stars previously known as spectral class A or F, for use as 
reddening standards.  As noted above, the region covered by 
{\it SDSS} does not encompass the whole Taurus cloud 
(see Figure~\ref{fig.coverage}).  Furthermore, there was no overlap between
these candidates and those already selected above.
We also considered a set of stars selected, similar to the 2MASS-PSC query
described above, as those having K$_s < 10$ mag and blue colors 
within {\it SDSS}, specifically $u-g < 0$.  This resulted in a short list
of a few tens of objects, nearly all of which were known to SIMBAD already
as early type stars (and thus included already among our first set of candidates),
or as late type stars (with blue colors unexpectedly blue, 
likely indicative of hot companions).

In addition to stars in the four samples considered above,
HD 31305 (A0V) is a star which we found in the vicinity of Taurus-Auriga 
due to its early spectral type and proximity to the well-known
Taurus member AB Aur, though it is not within the area of the Spitzer maps of Taurus.  

Our final list of early-type candidates for membership in the Taurus region of recent star formation thus consists 
of 329 stars. The color-color and color-magnitude diagrams for these stars, separated by the 
selection method, are shown in Figure~\ref{fig.ccd_cmd}.
We tested these objects for Taurus membership as described in the next subsection, after assembling the needed ancillary data.

For all objects in our list of candidate O, B, and A0 stars, we collected
the following astrometric and photometric information.
Proper motions were taken from the PPMXL
catalog \citep{roeser2010}, and trigonometric parallaxes 
from the Hipparcos catalog \citep{perryman1997}.
B, V, and R magnitudes listed in the NOMAD-1 catalog \citep{zacharias2005},
and J, H, K$_s$ magnitudes from 2MASS-PSC were used.
Radial velocity (RV, heliocentric) information was
extracted from \cite{gontcharov2006}
and \cite{kharchenko2007} in that order of priority. For the {\it SDSS} stars which we chose from 
\cite{finkbeiner2004}, we used an A0 template to extract their radial velocities 
using the {\it SDSS} DR7 \citep{abazajian2009}. 
In each of these catalogs, we searched for counterparts to our 
early-type candidate stars within 1 arcsec of the source position.
In cases where two counterparts were found for a particular source, 
only the closest was considered.
Finally, the spectral types were adopted from our own derivations 
for those stars which we followed-up spectroscopically
(see section ~\ref{subsec.spectroscopy}),
from \cite{rebull2010} for stars listed in that paper, from SIMBAD, 
or from the ASCC-2.5 catalog \citep{kharchenko2009}, in that order of preference. 
We also performed a thorough literature search, seeking relevant data
not available through large catalogs.

\qquad \newline
\subsection{SELECTION OF CANDIDATE MEMBERS OF TAURUS}\label{subsec.membership}
Physical association of astronomical objects can be established through 
the combination of common location and common space motion, 
with not all six dimensions available for every star.
The case at hand is that of a star forming region lying at 
a mean distance of 140 pc and having a depth and transverse extent of $\gtrsim20$ pc.
Although kinematics traditionally has been a robust mode of identification
of cluster members, uncertainties in distance may lead to discrepant space velocities.
Furthermore, the dispersion in the measured distance or kinematic quantities 
might be a significant fraction of their absolute values.
With these challenges in mind, we chose the following filters to select
(probable) members from our list of candidate early-type stars towards Taurus.

One set of criteria involved distance.
Stars with trigonometric or spectroscopic parallax between 128 and 
162 pc within 1$\sigma$ errorbar were considered.
Another set of criteria involved kinematics.
From the probability associated with a calculated $\chi^2$ statistic, 
stars having proper motion consistent with known members were selected.
Finally, radial velocity (RV) was taken into account wherever available, 
considering as members stars with $9.8 \leq RV \leq 17.5 ~\mbox{km s}^{-1}$, 
which incorporates the mean radial velocities of all Taurus groups identified 
by \cite{luhman2009} within $1\sigma$ uncertainty.
\newline 


\subsubsection{DISTANCE CRITERION AND METHODS}\label{subsubsec.parallax}
Through VLBI measurements, the distances to five Taurus members are
accurately known (see Table~\ref{tab.distance} and for context Figure~\ref{fig.distance}).
Taurus is at least as deep as it is wide \citep{torres2009}, a few tens
of parsecs in each direction.
Based on this, we assume that Taurus occupies the region between 128
to 162 pc (i.e. $6.2<\pi<7.8$ milli-arcseconds on parallax).
From our list of early-type candidate stars we chose candidate Taurus members 
such that both their Hipparcos and spectroscopic parallax distances, within $1\sigma$ error, 
were consistent with the above-stated distance criterion.
\cite{vanLeeuwen2007} has effectively demonstrated the validity of the
Hipparcos parallaxes (however, see below for an argument against 
in the case of HD 26212).
We calculated the spectroscopic parallax distances for each of the six magnitudes 
(denoted by 'X' below) --- $B,V,R,J,H$ and $K_s$ --- using the definition,
\begin{equation}
 d_X = 10^{ (X - A_X - M_X)/5 } \times 10
\label{eqn.spectDist}
\end{equation}
where $A_X = \left[B-V-(B-V)_0\right] \cdotp R_V \cdotp (A_X/A_V)$, $R_V = 3.1$, and $(A_X/A_V) = a + b/R_V$.
The parameters 'a' and 'b' are from Table 3 of \cite{cardelli1989}, and 
are the best-fit parameters to the average extinction law.
$M_V$ is from \cite{schmidt-kaler1982}, intrinsic colors, viz.
$(B-V)_0$, $(V-R)_0$ are from \cite{johnson1966}, and $(V-K)_0$,
$(J-K)_0$, $(H-K)_0$ colors are from \cite{koornneef1983}.
The \citeauthor{koornneef1983} magnitudes/colors were transformed into the 2MASS $JHK_s$
system using transformations from \cite{carpenter2001}.
This intrinsic color and magnitude information for O, B and A-type stars were compiled 
from \cite{schmidt-kaler1982}, \newline \cite{johnson1966}, \cite{koornneef1983} and \cite{corroll2006}.
$BVR$ reddening was determined using \cite{cardelli1989} with $R_V=3.1$, 
and $JHK_s$ reddening using \cite{rieke_lebofsky1985}.
Stars with missing luminosity class information were assumed to be dwarfs.
Candidates having spectral types for which the intrinsic magnitudes 
and colors are missing in our compiled tables necessitated 
interpolation between the two adjacent spectral types.

We note that the difference in spectroscopic parallax calculated 
using alternate color tables such as those of \citet{schmidt-kaler1982},
\citet{fitzgerald1970}, or \citet{johnson1966} is less than 0.5\%.
More modern empirical color references that include both dwarfs and giants
are rare and possibly non-existent.  However, considering the tables of 
Pickles (1998; based on synthetic photometry from stitched together
spectrophotometric data) or 
Bessell et al. (1998; based on synthetic photometry from model atmospheres), 
the spectroscopic parallax differences are larger, but less than 3\%. 
A more worrisome discrepancy lies in the absolute V-band magnitude, 
$M_V$, where the Pickles (1998) results differ from the Johnson (1966) values 
used in our compilation by 0.5 to 1 magnitudes, 
leading to 5--50\% disagreement in the spectroscopic parallaxes.
However, at least for B-type stars, the $M_V$ from Johnson (1966) agrees 
with observations of well-studied stars 
(e.g. the 100 brightest stars 
\footnote{\url{http://ads.harvard.edu/cgi-bin/bbrowse?book=hsaa&page=45}}) 
better than does the Pickles (1998) scale; Bessell et al. (1998)
do not quote $M_V$.

The error reported on the spectroscopic parallax distance is the 
standard deviation of the distances calculated using all of the six 
magnitudes.
As well-tested empirical estimates of intrinsic colors involving R,J,H and K$_s$ 
are not available for early-type giant stars, we calculate the distance
to the luminosity class III stars using only B,V magnitudes and color.
Other errors that could contribute but have not been
accounted for include: (i) spectral type / luminosity classification error,
(ii) error in apparent magnitudes, (iii) intrinsic colors are mean
values and do not account for astrophysical spread within the luminosity classes,
(v) error in choice of reddening model, 
(vi) presence of non-photospheric emission such as infrared excess.
For points (i) and (ii) stated above, the manner in which these criteria
impact the distance estimate can be understood quantitatively via
the discussion provided in section 3.3 of \cite{kenyon1994}.
Following that discussion, the $1\sigma$ error on spectroscopic parallax
corresponding to quantities (i) and (ii) is roughly 30 pc at a 
spectral type of A0 with $d=140$ pc.
This uncertainty would then add to our quoted error appropriately taking
into account equation~\ref{eqn.spectDist}.
Caution is thus advised in using the error bars quoted on spectroscopic
distances, especially for giant stars.


\begin{figure}[htp]
\centering
\includegraphics[height=3.0in]{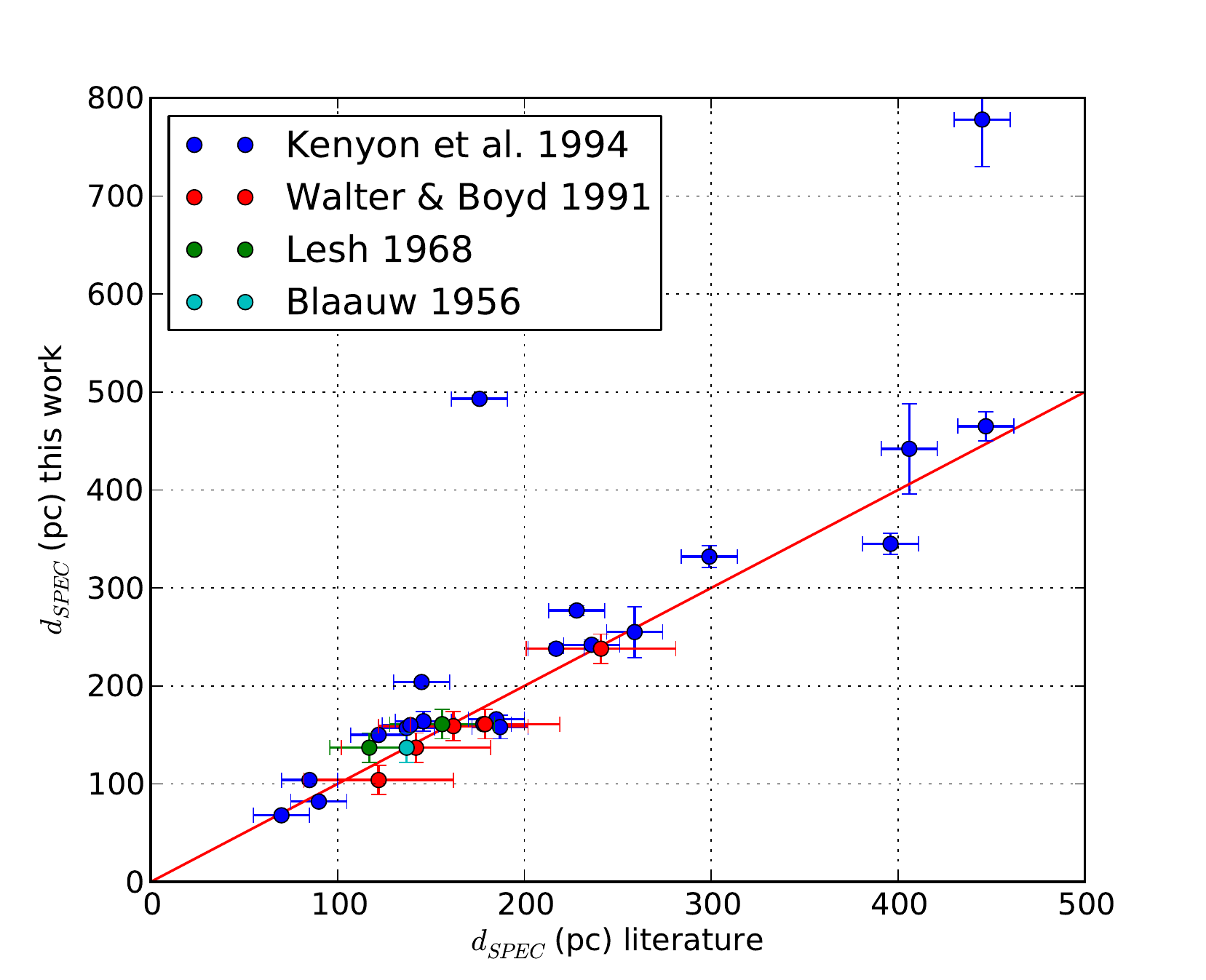}
\caption{Comparison of spectroscopic distances calculated in this work with those published in past literature.}
\label{fig.spec_parallax_comparison}
\end{figure}

Out of the 329 early-type candidates being tested for Taurus membership, 
the reddening and spectroscopic parallax distance can be calculated for 173 of them.
In general, where there is overlap, the reddening values show good 
agreement with those stated in the literature. 
We compare our averaged spectroscopic parallax distances with those determined by
\cite{kenyon1994}, \cite{walter1991}, \cite{lesh1968}, and 
\cite{blaauw1956} in Figure~\ref{fig.spec_parallax_comparison}.
For the majority of the objects, the spectroscopic parallax distances are 
in agreement within 1$\sigma$ with those determined by {\it Hipparcos}.
Notable exceptions are stars with very high reddening 
for which $R_V$ (and possibly also the reddening law
as a function of wavelength itself) would differ significantly from the value
assumed here.  Also, the method fails if the optical and/or near-infrared
photometry is dominated by circumstellar rather than photospheric emission,
or if the source is pre-main sequence rather than close to the main sequence; 
this latter condition is indeed
the case for many of the later type stars selected in the 2MASS part of the search.


In considering the appropriateness of the distance criterion we have
adopted for testing the association with Taurus, which is that all stars must lie
between 128 and 162 pc, a problem
arises in the unknown 3-dimensional shape of the molecular cloud.
The cloud may extend further along some lines of sight, or it may not have 
high enough density for current star formation along other lines of sight.
A related problem in establishing membership is that, to the west of Taurus, there is another
star-forming region along the line of sight extending behind Taurus: the Perseus 
molecular cloud at about 350 pc.
The spectroscopic parallaxes of HD 282276 and HD 283677 suggest that 
they both lie closer to the Perseus cloud, though on the sky they are aligned 
with Taurus, not the Perseus cloud; these stars also have similar proper motion (see below).
There have been suggestions of a bridge of molecular 
material connecting the Taurus-Auriga and Perseus regions \citep{ungerechts1987}.
The presence of the somewhat older Cas-Tau OB association along the 
line of sight also poses a potential contamination problem because its members span 
a range of at least 30 pc in distance \citep{deZeeuw1999} and perhaps as much as 80 pc, 
although it is securely behind the Taurus star-forming region.

A method for distinguishing chance superpositions, in addition 
to the distance criterion, is to look at the kinematics of the stars and the natal cloud. 
Consideration of proper motion and radial velocity of the stars helps in eliminating ambiguity, 
as discussed in the next sub-section.



\subsubsection{KINEMATIC CRITERIA AND METHODS}\label{subsubsec.pm_rv}
The classical studies of proper motions of stars in the vicinity 
of the Taurus clouds 
are those of \citet{jones_herbig1979} and \citet{hartmann1991}
with \citet{ducourant2005} providing the latest catalog for Taurus.
Kinematic membership probabilities are typically based on the convergent
point method, which is used for regions that cover a large part of sky
where the mean subgroup motion is changing as a function of position
(this is a purely geometric effect).
For regions less than a few degrees in size in the vicinity of 
comoving groups in Taurus, one can test the consistency of
the proper motion of one star simply with respect to the mean motion of a group.
\cite{luhman2009} computed the mean proper motions and radial velocities
of eleven distinct groups (occupying 1--10 deg$^2$ on the sky; 
see Table 8 of that paper) of Taurus members.
Seven of these groups, specifically I-V, VIII and X, lie within our region
of interest.

We checked the statistical consistency between the proper motion of the 
candidate early-type stars reported in Table \ref{tab.allBstars} and the
proper motion of the closest kinematic group from \cite{luhman2009} 
by estimating the $\chi^{2}$ probability.
The two components of proper motion, $\mu_{\alpha}$ and $\mu_{\delta}$, can 
be understood as independent Gaussian variates drawn from a normal 
distribution parametrized by the mean (which can be estimated
through the sample mean, i.e. the mean proper motion of the
presently-known Taurus members) and the dispersion (likewise estimated as
the dispersion of the sample of presently-known Taurus members).
The sum of the square of these values will then be distributed according 
to the $\chi^2$ distribution.
Strictly speaking, these components are determined through the least-squares
technique in proper-motion catalogs, and are correlated 
(the complete covariance matrix is, for example, provided by the
Hipparcos catalog).
Here, we have calculated the quantity $\chi^2$ using the definition
$\sum_{i=1}^{k}(x_i - \bar{x})/\sigma_i^2$, where $i=1,2$, 
and $x_i$ describes the components of proper motion.
The uncertainties, however, are associated with not only the proper motion
of individual stars, but also with the sample mean.
Further, we have to incorporate the internal dispersion of the 
presently-known members of Taurus.
We calculated the $\chi^2$ statistic and the associated probability using 
equations ~\ref{eq.chi2} and ~\ref{eq.prob_chi2}, after \cite{deZeeuw1999}:

\begin{eqnarray}
\chi^2_{\nu=2} = \frac{(\mu_{\alpha} - \mu_{\alpha,{\rm group}})^{2}}{(\sigma_{\mu_{\alpha}}^{2} + \sigma_{{\rm int}}^{2} + \sigma_{\mu_{\alpha,{\rm group}}}^{2})} + \qquad \qquad \qquad \nonumber \\
 \frac{(\mu_{\delta} - \mu_{\alpha,{\rm group}})^{2}}{(\sigma_{\mu_{\delta}}^{2} + \sigma_{{\rm int}}^{2} + \sigma_{\mu_{{\delta,{\rm group}}}}^{2})} 
\label{eq.chi2}
\end{eqnarray}
\begin{equation}
P(\chi|\nu)= \frac{\chi^{(\nu-2)/2} e^{-\chi/2}}{2^{\nu/2} ~\Gamma(\nu/2)} 
\label{eq.prob_chi2}
\end{equation}

\noindent
where $\mu_{\alpha}, \mu_{\delta}, \sigma_{\mu_{\alpha}}, 
\sigma_{\mu_{\delta}}$ denote the proper motion in right ascension
and declination of the star being tested and their associated 
uncertainties. The quantities $\mu_{{\rm group}}, 
\sigma_{\mu_{{\rm group}}}, \sigma_{int}$ are the proper motion
of the Taurus group closest to the star, its uncertainty, and the
intrinsic dispersion of proper motion in the group (assumed to be
2 $\mbox{mas yr}^{-1}$).
The denominator of each term is then the expected variance of
the respective numerators.
This method traditionally  has been used to find 
``proper-motion members'', but is partly biased toward stars having
a large relative uncertainty in their proper motion which reduces
the  $\chi^2$.
We are able to calculate the $\chi^2$ probability for all 
of the 329 early-type candidate stars being tested for Taurus membership. 

\begin{figure}[htp]
\centering
\includegraphics[width=3.5in]{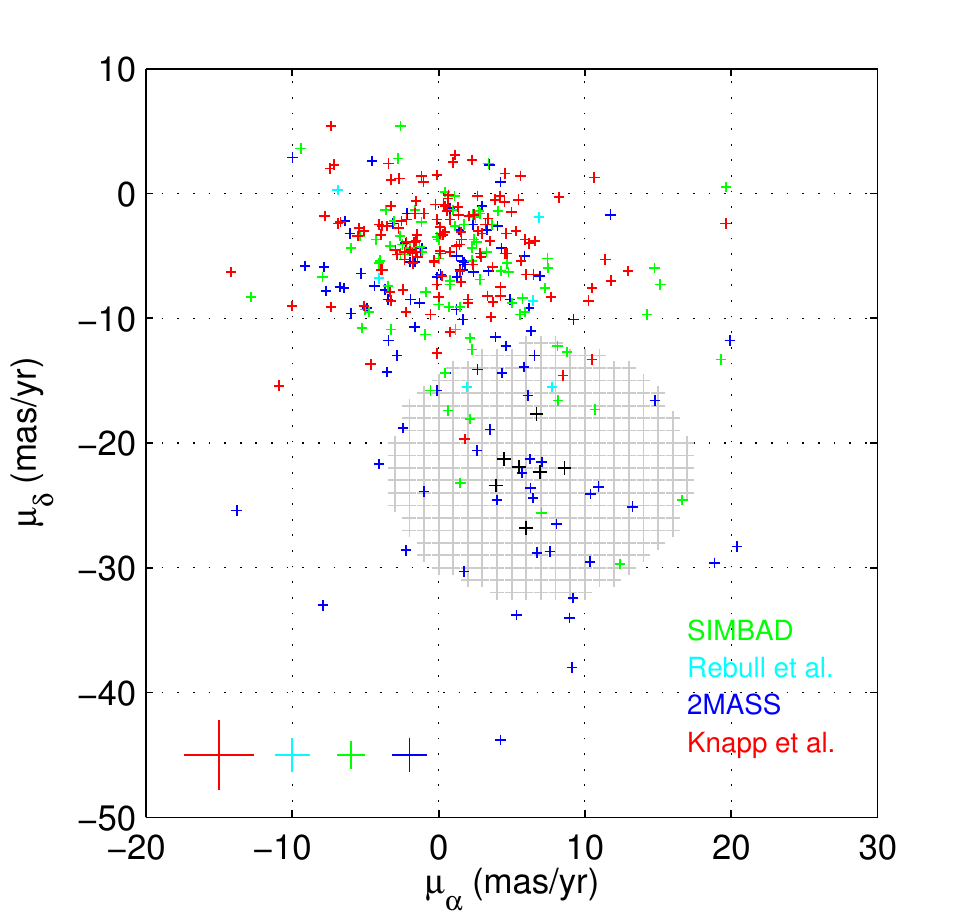}
\includegraphics[width=3.25in]{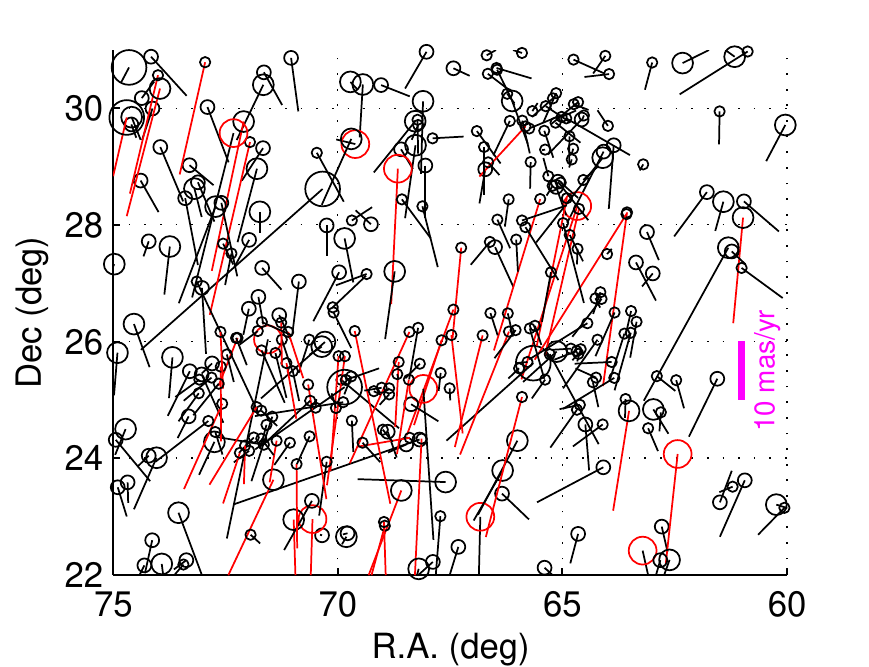}
\includegraphics[width=3in,height=0.7in,viewport=35 146 255 200,clip]{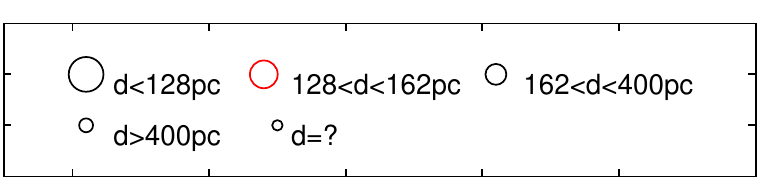}
\caption{{\it Upper panel}: Proper motions of the candidate early-type stars shown as a cloud plot 
with blue indicating objects selected from 2MASS; green, SIMBAD; cyan, the B stars proposed by \cite{rebull2010}, and red: O,B,A stars from \cite{knapp2007}.
The ``+'' symbols at the bottom-left corner denote the mean errors for each sample.
The mean proper motion of Taurus groups considered in this paper (see Section~\ref{subsubsec.pm_rv}) are shown as black symbols
The hatched reference circle indicates the area where the $\chi^2$ probability of membership is greater than 1\% with respect to the mean proper motion of Taurus.
51 stars from our list of candidate early-type stars have proper motions consistent with Taurus groups.
{\it Lower panel}: Vector diagram showing the proper motion of all the stars tested for membership.
Those which satisfy the proper motion criterion P($\chi^2 > 1 \%$) are shown in red.
Positions of the stars are indicated by the circles, whose sizes are based on the spectroscopic parallax 
distance of the respective stars (key given at the bottom).
Red circles denote stars satisfying our distance criterion for Taurus member selection (within an uncertainty of 15 pc).}
\label{fig.pm_vector_cloud}
\end{figure}

The result of this proper motion analysis is illustrated
in the upper panel of Figure~\ref{fig.pm_vector_cloud}.
The region allowed by our $\chi^2$ probability membership criterion 
(set at $>$1\%) roughly corresponds to the shaded circular region.
In the context of this figure, it is worthwhile to note that members
of the background Cas-Tau OB association as listed by \cite{deZeeuw1999} 
have $\mu_{\alpha}$ ranging from a few $\mbox{mas yr}^{-1}$ to 
50 ~$\mbox{mas yr}^{-1}$ (the mean is about 26 ~$\mbox{mas yr}^{-1}$), 
and $\mu_{\delta}$ ranging from a negative few $\mbox{mas yr}^{-1}$ to
$-40~ \mbox{mas yr}^{-1}$ (mean is about $-19 ~\mbox{mas yr}^{-1}$; 
from the PPMXL catalog). 
Thus, a few of the stars studied herein are probably Cas-Tau members.
A combined diagram showing the spectroscopic parallax distance and the 
proper motion of the early-type candidate stars is shown in the lower 
panel of Figure~\ref{fig.pm_vector_cloud}.

\begin{figure}[htp]
\centering
\includegraphics[width=2.8in,angle=-90,viewport=170 220 430 650,clip]{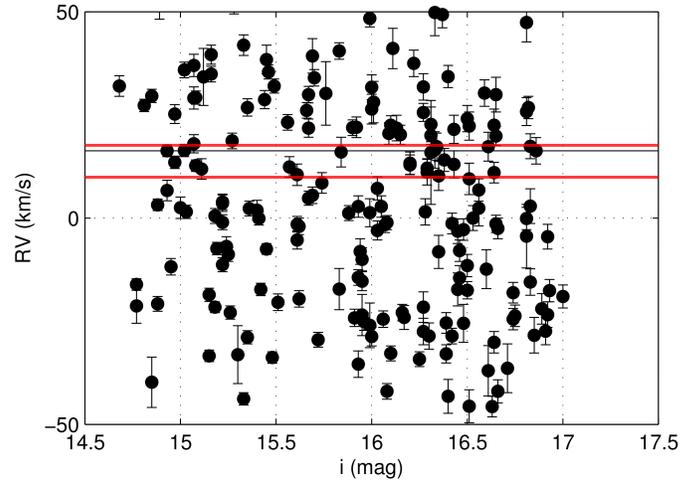}
\caption{
Radial velocity versus {\it SDSS} i-band 
magnitude for the early-type stars selected from \cite{knapp2007}. 
The y-axis has been rescaled to show only the stars with RV 
between $\pm50$ km s$^{-1}$.
The mean RV of accepted members of Taurus \citep{luhman2009}, 
15.8 km s$^{-1}$, is shown with a black horizontal line.
The neighboring red lines denote the region $9.8 \leq RV \leq 17.5$, our RV-member selection criterion.
No concentration at the Taurus velocity is seen, and
among those stars satisfying this RV criterion, most are 
too faint to be probable members of Taurus.
}
\label{fig.rv_jillstars}
\end{figure}

The RV dimension was not included in the $\chi^2$ analysis because this 
quantity is unknown for most stars. In cases where it is known, 
the uncertainties are generally quite large. 
Thus, as the second component of our kinematics investigation, 
we compared the RV of each of the early-type candidate member stars 
(where available and as reported in Table \ref{tab.allBstars}) 
with that of the nearest Taurus group listed in \cite{luhman2009}.
We chose as likely members the stars which, 
within 1$\sigma$, satisfied the criterion $9.8 \leq RV \leq 17.5$, corresponding 
to the range in the mean radial velocities of the Taurus groups. 
For the {\it SDSS}-selected early-type stars we show the radial velocities
in Figure~\ref{fig.rv_jillstars}. Most of the {\it SDSS}-selected stars 
satisfying the nominal RV-selection criterion are too faint 
in the near-infrared to be probable members of Taurus.

\subsubsection{RESULTS OF CANDIDATE SELECTION}\label{subsubsec.results}

The information in Table~\ref{tab.allBstars} was used along with 
the procedures outlined above to arrive at the list of likely early-type 
members of Taurus.
In all, 52 stars independently satisfy the proper motion membership 
criteria and 28 the distance membership criteria, with 18 satisfying both.
However, not all sources have known values for all or any of 
the quantities we consider. As the samples of early-type stars under
consideration were derived from four different sources, we discuss the 
details of our analysis as appropriate for each sample in what follows.

For the SIMBAD sample of known early-type stars and for the 
\citeauthor{rebull2010} infrared excess sample, spectral types exist 
in the literature.
Also, we have followed up some of these stars spectroscopically ourselves
in order to verify or revise their spectral classification.
Fairly accurate spectroscopic parallax distances have thus been 
used to test the membership of these two sets of stars with Taurus.
Radial velocity measurements also exist for some of these stars.
We found that two stars, HD 28929 (B8V; also known as HR 1445) 
and HD 29763 (B3V; also known as $\tau$ Tau), 
satisfy all the tested criteria for membership of Taurus: 
parallax distance, proper motion, and radial velocity.
V892 Tau (A0; also known as Elias 1) 
satisfies the first two criteria, leaving the third unknown due to 
insufficient information, but this star is already an accepted member 
of Taurus on the basis of its circumstellar disk attributes.

In the 2MASS photometric sample, accurate spectral classification 
is absent for many of the stars, and for these calculation 
of accurate spectroscopic parallax has not been possible.
In this set of stars, twelve satisfy two criteria leaving the third 
criterion indeterminate due to insufficient information,  
while two stars satisfies all the criteria for membership of Taurus.
However, many of these stars are late type T Tauri stars 
that are already known members of Taurus but selected by our methods
because they have large enough near-infrared excesses to push them into the
region of the 2MASS color-color diagram occupied by reddened earlier type stars. 
We note that the 2MASS search does select the mid-A stars HD 26212 and HD 31648  
(along with AB Aur, if we ignore the 2MASS-PSC flag at $K_s$-band)
as candidate early-type members, but finds no new B stars.

For the {\it SDSS} spectroscopic sample, the magnitude range precludes the 
availability of any {\it Hipparcos} parallax values, but since spectral types
are available for many stars, spectroscopic parallaxes can be calculated.
RV is also measured by the {\it SDSS} analysis pipeline.
Some of the {\it SDSS}-selected stars satisfy the kinematic criteria 
(see Figures~\ref{fig.pm_vector_cloud} and \ref{fig.rv_jillstars});
however, they are under-luminous with respect to the expectations
for reddened early type stars at the Taurus distance 
(see middle panels of Figure~\ref{fig.ccd_cmd}) and
indeed have much larger spectroscopic distance estimates 
(Table~\ref{tab.allBstars}).
Most of these early-type stars are likely in the Galactic halo.

%

Finally, the spectroscopic parallax distance suggests that HD 31305 
(located near AB Aur and discussed by \cite{cody2013}) 
lies just beyond the distance range defined by our member-selection criterion  
($<$10\% in excess of the standard deviation among calculated 
$d_{SPEC}$ values).  Nevertheless,
its proper motion conforms with that of known Taurus members.
Hence, we consider in what follows that this star could well be a member of Taurus.

In summary, we newly advocate the membership of two B-type stars
(HD 28929 and HD 29763) and one A-type star (HD 26212) in Taurus 
using distance and kinematic arguments, 
and in addition find one A0-type star (HD 31305) 
to be a probable member (though it is not yet confirmed due to
the lack of radial velocity information).
Below, in section ~\ref{sec.notes}, we discuss these stars in more detail 
and also revisit some of those which were rejected in the above procedures. 
After consideration of other evidence which might point towards 
their association with Taurus, we find several additional B and A0-type 
to be plausible Taurus members. In order to inform our further 
assessments regarding the likelihood of cluster membership, 
additional data on many of these stars were collected.

\subsection{FOLLOW-UP SPECTROSCOPY}\label{subsec.spectroscopy}
We performed follow-up spectroscopy of selected stars
with an aim of (i) confirming or revising their spectral types 
based on temperature and surface gravity diagnostics, (ii) 
measuring radial velocities, and (iii) 
determining more precise stellar parameters so as to estimate ages. 
We obtained optical spectra for a subset of the Taurus early-type candidates 
which were found to satisfy several of our membership criteria, 
or which illuminated a nebula in the Spitzer image.
Some of these sources appeared to be better candidates at the time
we obtained the spectra than later re-analysis revealed. 
We also observed for comparison a grid of dwarf B stars from \cite{abt2002}.
and, for calibration, RV standard stars 
and spectrophotometric standards.

The optical spectra were obtained at the Palomar 200-inch Hale telescope
on 4 December 2010 using the Double Spectrograph (DBSP). The data have
medium spectral resolution ($R \simeq 7800$ and 10419 in the blue and
red channels respectively).
We used a dichroic at 5500\AA\ to split the optical light into blue and
red channels with a 1200 lines mm$^{-1}$ grating blazed at 4700\AA, at a grating angle 
$34.92\degr$ on the blue side and 1200 lines mm$^{-1}$, 7100\AA\ blaze, 
and $42.73\degr$ on the red side.  The spectral range covered 
was $\sim3480-5020$ \AA\ at 0.55 \AA\ pixel$^{-1}$ (blue) and $\sim 6440-7110$ \AA\ 
at 1.4 \AA\ pixel$^{-1}$ (red).  For wavelength calibration we used an Fe-Ar lamp 
in the blue and a He-Ne-Ar lamp in the red.
Spectra for two stars (HD 27659 and HD 26212) were taken on 2
September 2011 using a different configuration resulting in a much lower
resolution and a larger wavelength coverage.

We reduced the data using the Image Reduction and Analysis Facility (IRAF)
 \emph{ccdred} and \emph{onedspec} packages.
Spectra were extracted with the \emph{apall} task after trimming, 
bias-subtraction, and flat-fielding of the images.
The wavelength solution was then applied using \emph{dispcor}.
In the case of stars for which we had multiple short-exposure observations, the spectra were 
coadded using \textit{scombine} to get a higher signal-to-noise ratio. 
We normalized all the spectra with \textit{splot}.

\begin{figure}[htp]
\begin{center}
\includegraphics[width=2.8in,viewport=4 23 240 180,clip]{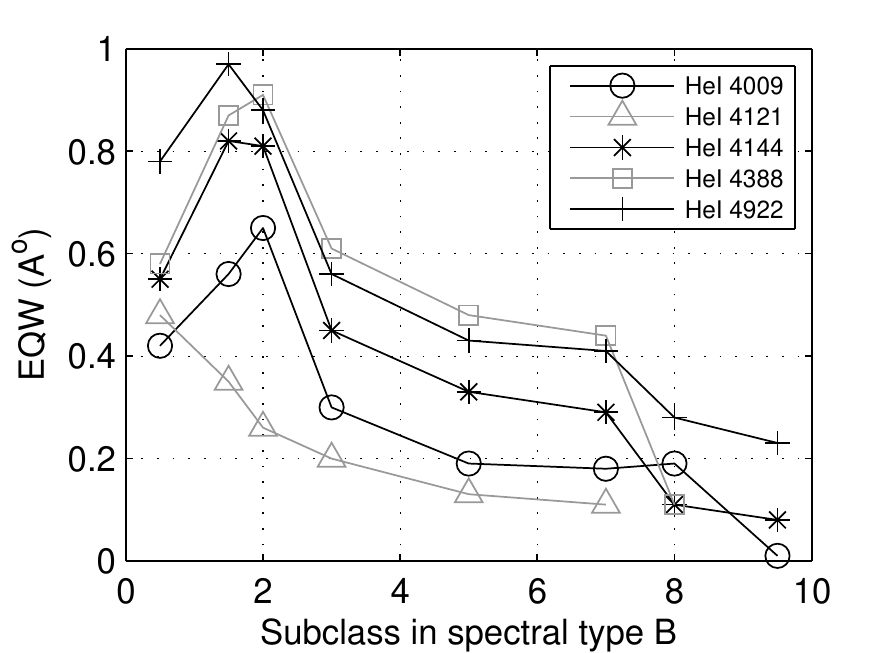}
\includegraphics[width=2.8in,viewport=4 23 240 175,clip]{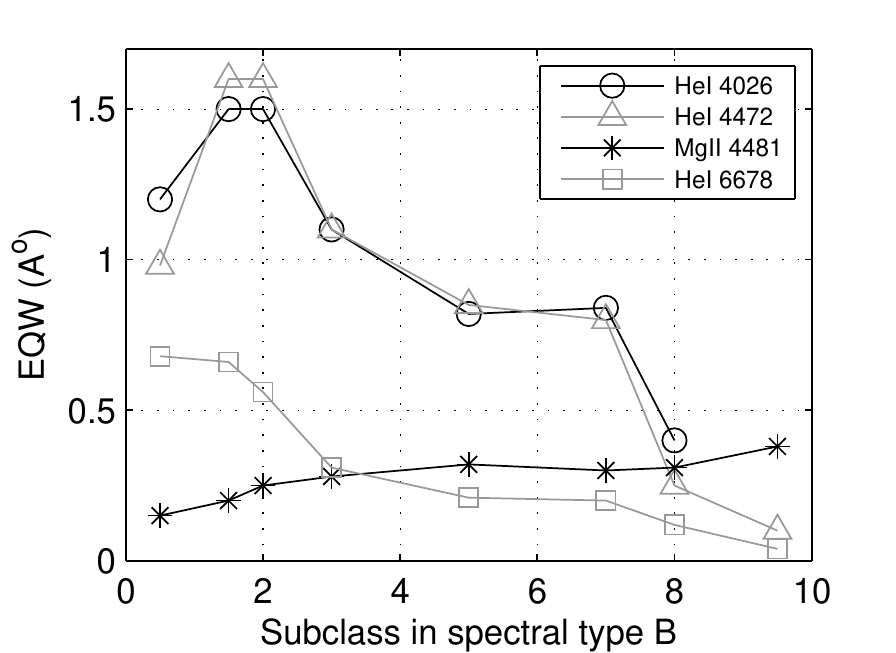}
\includegraphics[width=2.8in,viewport=2 0 240 180,clip]{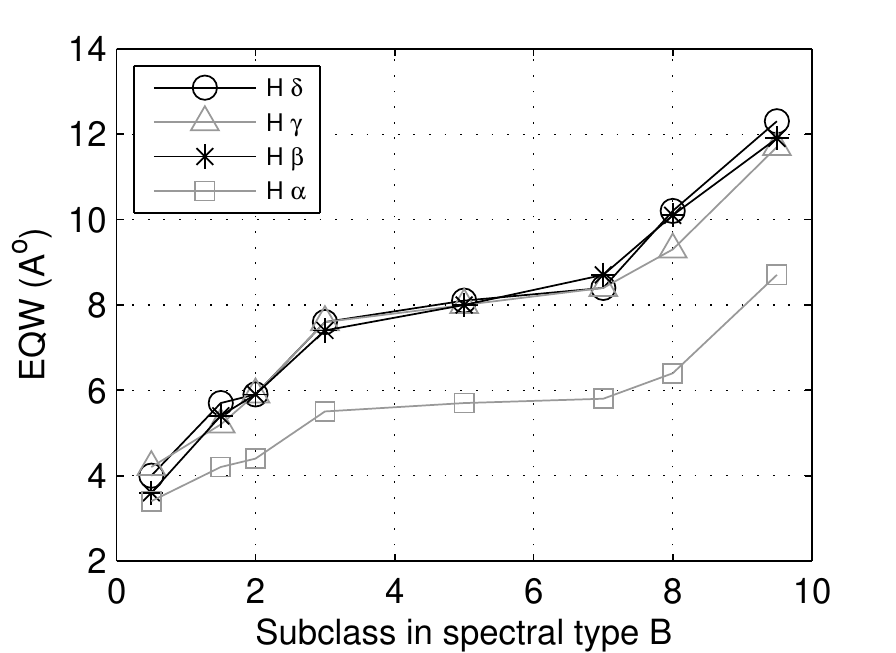}
\end{center}
\begin{center}
\caption{Equivalent widths of various absorption lines measured in the grid of B-type
spectral standard stars (luminosity class V only) that were observed 
for comparison with the Taurus candidate early-type stars.}
\label{fig.eqw_grid}
\end{center}
\end{figure}

\begin{figure*}[htp]
\centering
\includegraphics[height=8in,viewport=55 55 595 810,clip]{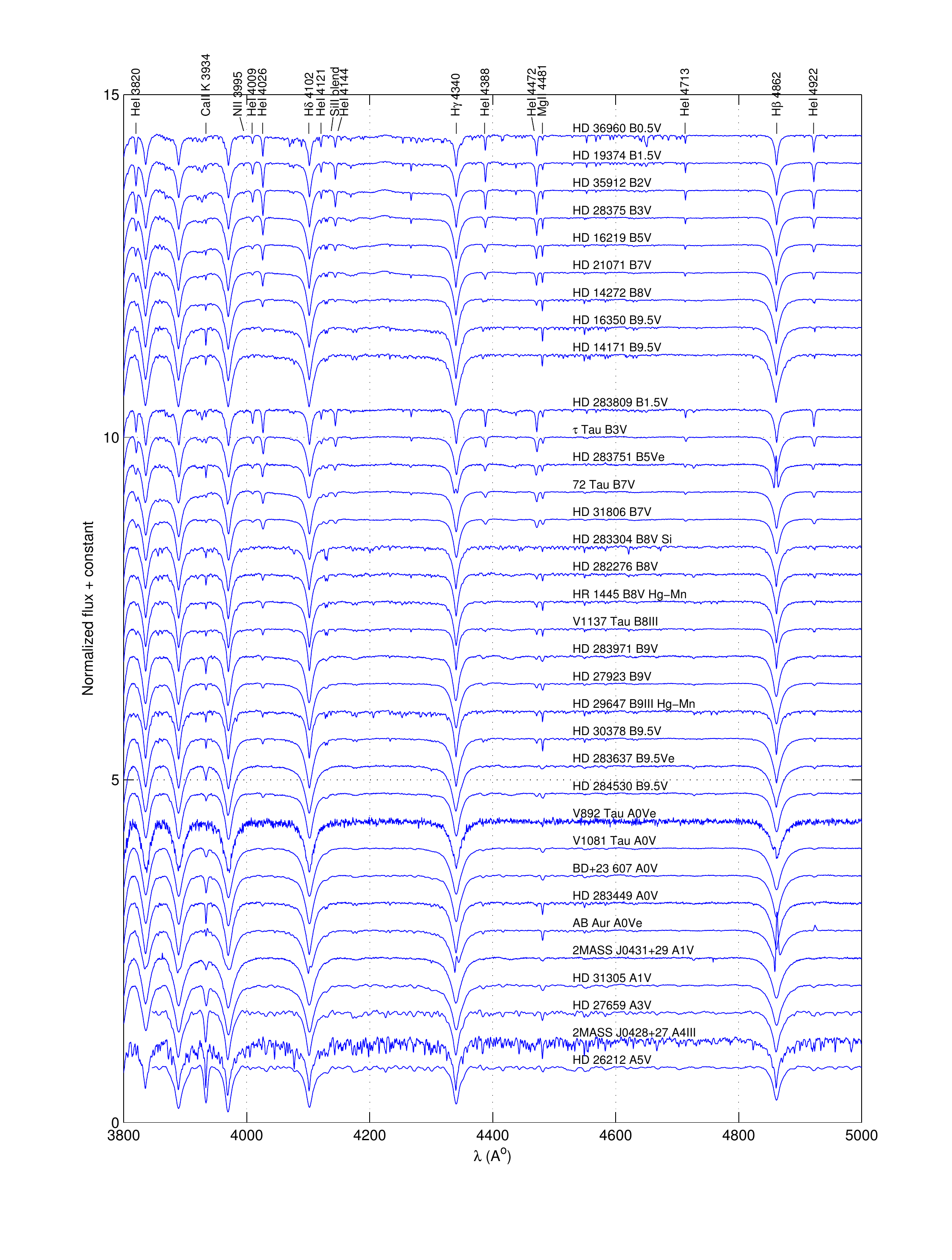}
\includegraphics[height=8in,viewport=5 55 130 810,clip]{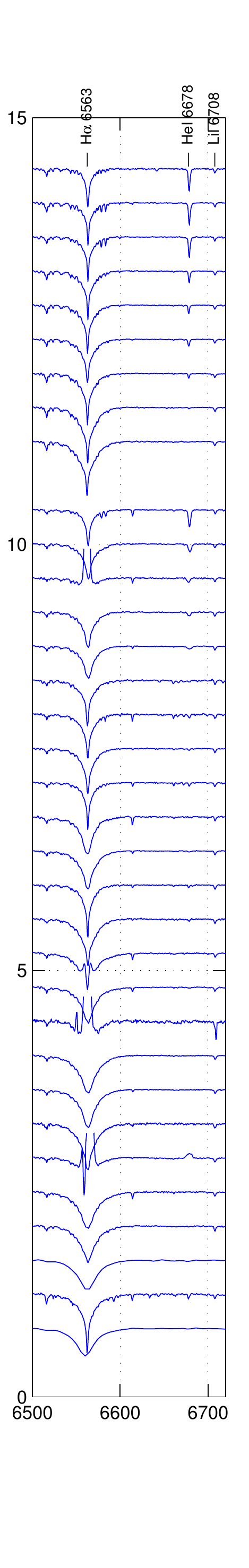}
\caption{Normalized optical spectra obtained at the Palomar 200-inch Hale telescope.
The upper set of stars comprise the grid of B-type spectral standard stars 
used for assigning spectral types to the program stars (lower set of spectra).
All spectra have an arbitrary offset along the ordinate.
Note the hydrogen emission lines or line cores in 
HD 283751, HD 283637, V892 Tau, and AB Aur.  Approximately half of the program stars
have diffuse interstellar band signatures: a narrow absorption at 6614 \AA\
and a broader shallow feature at 4428 \AA.  The feature labeled as 
\ion{Li}{1} 6708\AA\ in the right panel is probably anomalous as this line is
not expected to be present in these early-type young stars, and especially
not in the spectral standards (including our white dwarf flux standard 
which is not shown); we suspect a possible terrestrial atmosphere source, perhaps 
associated with the Geminid meteor shower; this hypothesis is supported
by lack of absorption at this wavelength in the two spectra towards the bottom
of the sequence that were taken on a different night from all others.}
\label{fig.spec1}
\end{figure*}

For spectral typing the program stars, we measured the equivalent 
widths of several diagnostic absorption lines using \textit{splot}, 
and then compared them with those of reference-grid stars 
(Figure~\ref{fig.eqw_grid}), guided by the graphics in \cite{didelon1982}.
The normalized spectra of the reference-grid stars and the program stars 
are shown in Figure~\ref{fig.spec1} with the reference types adopted from
the literature and the program star types derived by us. 
We also compared our spectra with templates by 
Gray\footnote{\href{http://ned.ipac.caltech.edu/level5/Gray/Gray_contents.html}{http://ned.ipac.caltech.edu/level5/Gray/Gray\_contents.html}} 
and \cite{morgan1943}\footnote{\href{http://ned.ipac.caltech.edu/level5/ASS_Atlas/MK_contents.html}{http://ned.ipac.caltech.edu/level5/ASS\_Atlas/MK\_contents.html}}.
The results of the spectroscopic analysis are given in Table~\ref{tab.spec}.
Estimates for the effective temperature (T$_{\rm eff}$), projected rotational velocity ($v \cdotp$ sin $i$ ) and the 
surface gravity (log g) and were made by fitting the spectra with templates from \cite{munari2005}.
For the template spectra, the comparison grid resolution 
was 500--1000 K in T$_{\rm eff}$ and 0.5 in log g, 
while the grid in $v \cdotp$ sin $i$ was: 
0, 10, 20, 30, 40, 50, 75, 100, 150, 200, 250, and 300 km s$^{-1}$;
hence, our derived values are no more accurate than this.
Some stars have equally good fits between a higher temperature and gravity point, 
versus a lower temperature and gravity point one grid spacing away; in these cases
we generally preferred the dwarf to the giant solution.
The physical parameters derived from this fitting are given in Table~\ref{tab.kurucz_param}.
In combination with the set of intrinsic stellar parameters discussed above, 
we thus derived a second set of spectral types for the stars 
that were spectroscopically followed up.
These spectral types generally agree with those derived using equivalent widths in Table~\ref{tab.spec}.
Due to the coarse spectral grid of templates and degeneracies involved in the fitting process,
the spectral types derived from our equivalent width analysis usually take
precedence over those derived from spectral fitting.

Unfortunately, radial velocity information could not be derived from our spectra
at the expected performance of the instrument
(given our care in taking source-by-source comparison lamp calibration frames),
perhaps due to poorly understood flexure effects.  We note
that an error as small as 1 \AA\ in the wavelength calibration leads to an error of about 66 km s$^{-1}$ at 4500 \AA.
Shifts of this order have been experienced between contiguous exposures while working with DBSP data.
Due to our short exposure times, there are no sky lines 
in the blue part of the spectrum that could aid in more accurate wavelength calibration.
While the red channel spectra have ample telluric absorption, too few 
photospheric absorption lines are available to provide a good fit.
Hence, we defer the estimation of RV to a later time with another data set.

Notably, hydrogen emission lines or line cores are seen in 
HD 283751, HD 283637, V892 Tau, and AB Aur (see Figure~\ref{fig.spec1}). 
While the emission properties of the last two stars in this list are well known,
they have not been reported previously for the first two objects.
Emission lines are often taken as a signature of activity
associated with stellar youth, although evolved early-B type stars
may exhibit a ``Be phenomenon''.  We note that the derived spectral types 
of these emission-line objects are B5e and B9.5e, later than typical
of evolved Be stars, but the infrared excesses detected by \cite{rebull2010}
are more typical of evolved Be stars than of young accretion disk systems.  
Neither star can be associated with Taurus by kinematic or distance arguments, 
and thus they appear to be interesting background interlopers.

A further note concerning the spectra is the appearance in Figure~\ref{fig.spec1} 
of what is likely diffuse interstellar band absorption at 6614 \AA\ in
about half of the program stars.  Corresponding broad absorption at 4428 \AA\ 
is also seen.  There is excellent correlation between the presence of this
feature and the spectroscopic parallax distance estimates reported in
Table \ref{tab.allBstars}.  Distant stars have the absorption while closer
stars do not.  Notably, none of the stars we eventually conclude in this work
to be associated with Taurus have these interstellar absorption features.

Finally, we call attention to ubiquitous absorption at 6708 \AA, 
coincident with the \ion{Li}{1} line seen in young low mass stars, 
that is seen in all of our spectra taken on 2010, December 4 
(Figure~\ref{fig.spec1}).  The feature is not likely to be astrophysical 
and we suspect a (currently unexplained) terrestrial origin, 
perhaps related to a meteor shower.

\begin{figure}[htp]
\centering
\includegraphics[width=3.1in,height=1.7in,viewport=35 120 720 560,clip]{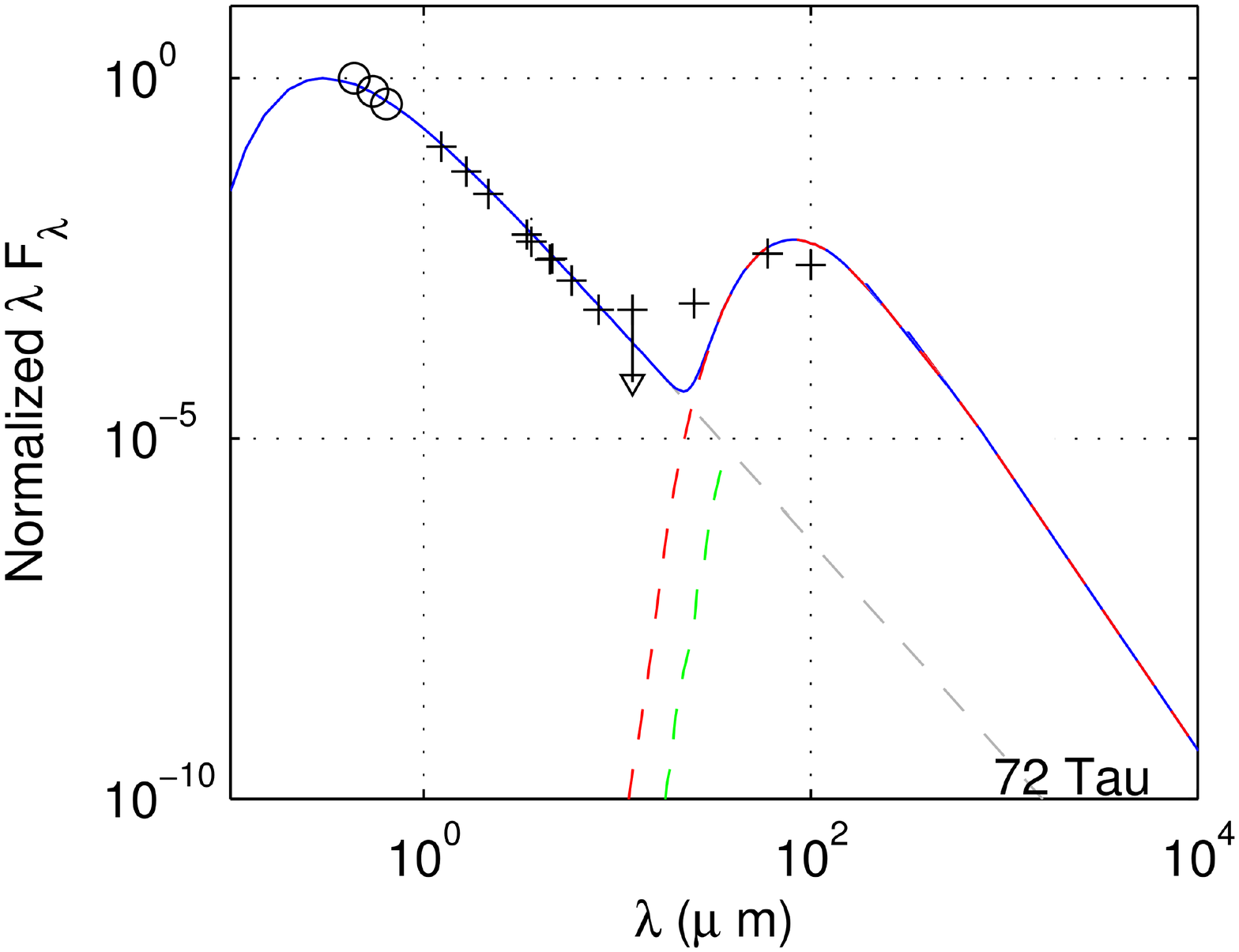}
\includegraphics[width=3.1in,height=1.7in,viewport=15 110 720 555,clip]{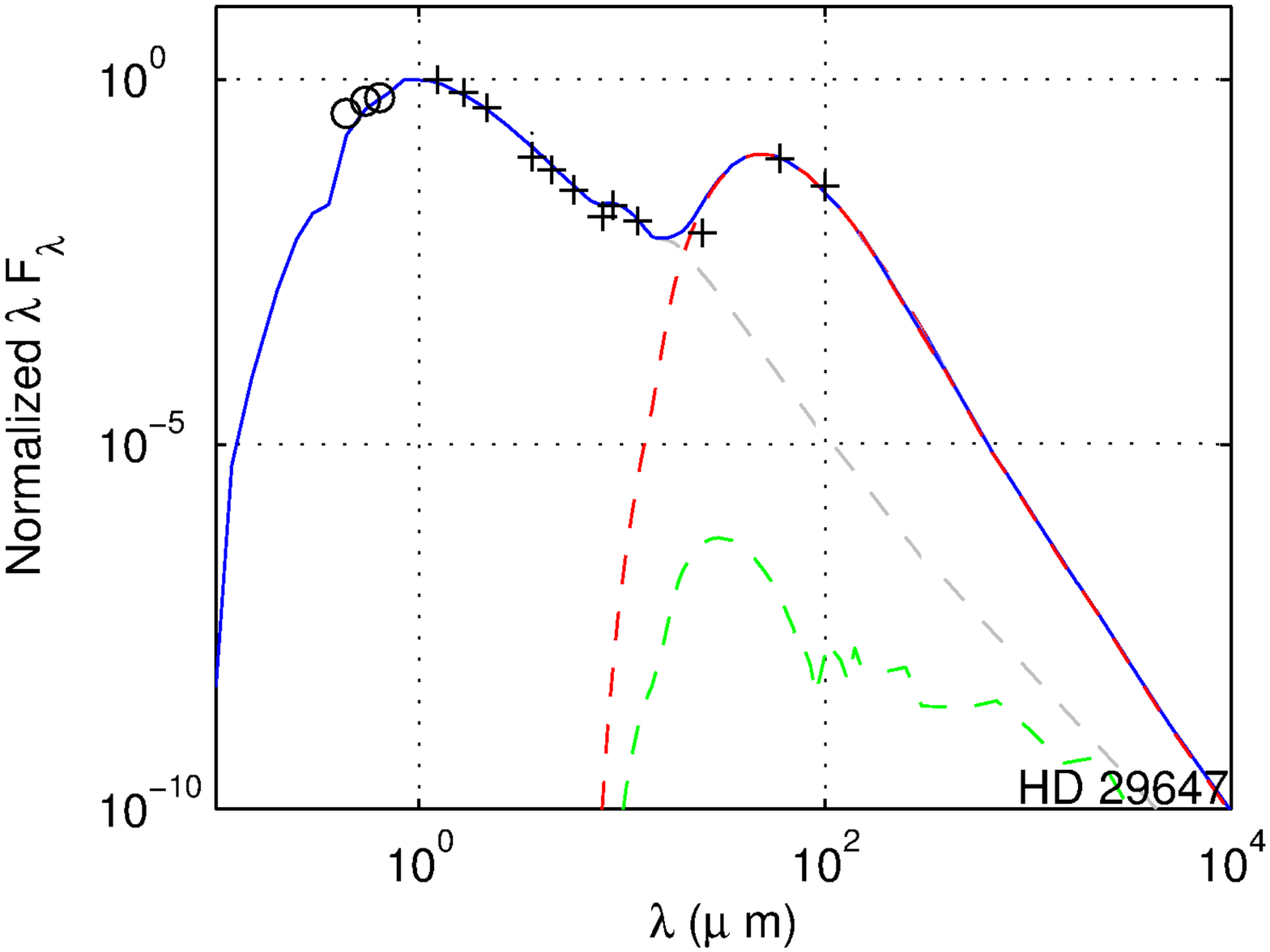}
\includegraphics[width=3.1in,height=1.7in,viewport= 5 110 720 550,clip]{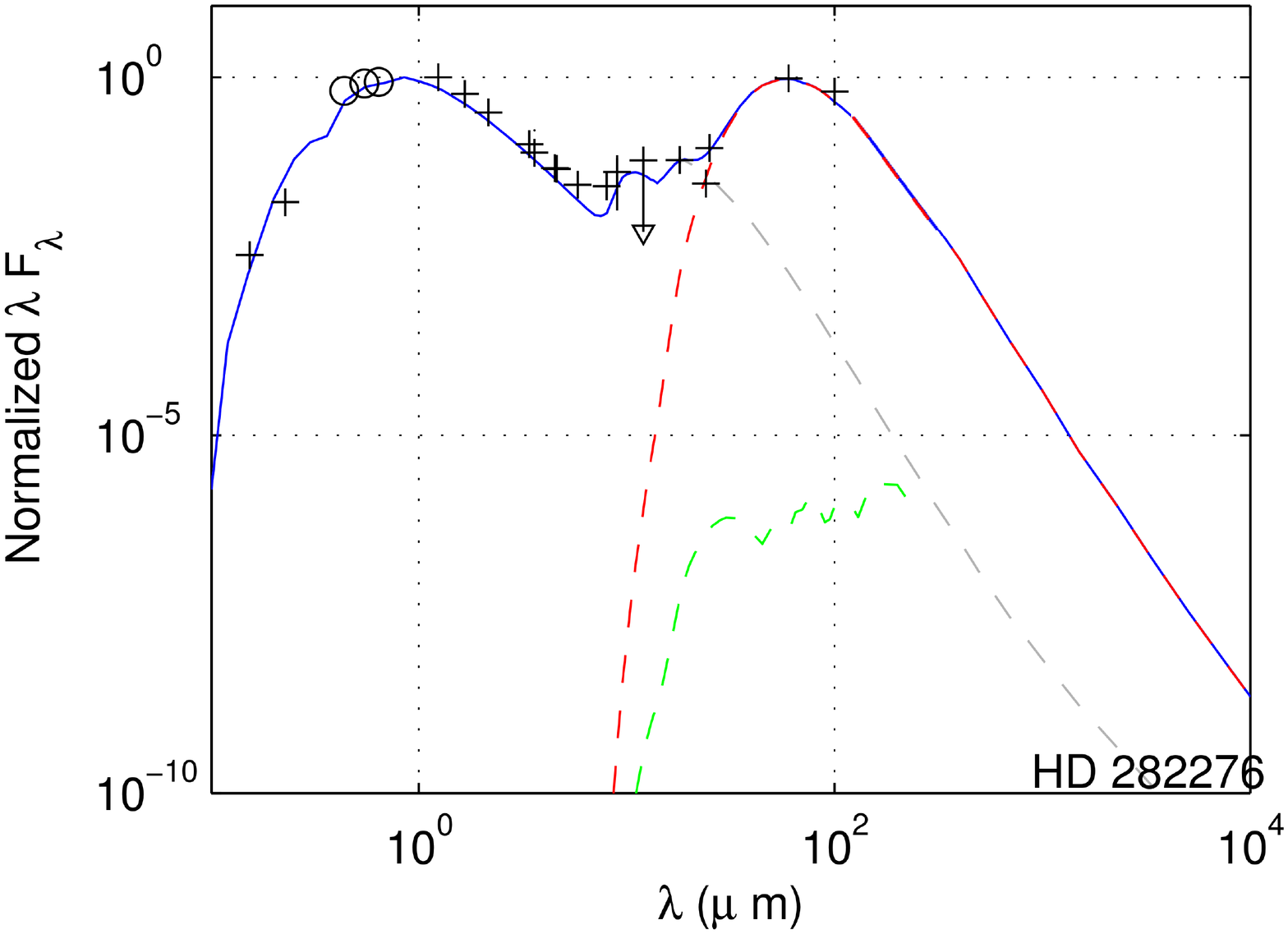}
\includegraphics[width=3.1in,height=2.0in,viewport=55  40 670 555,clip]{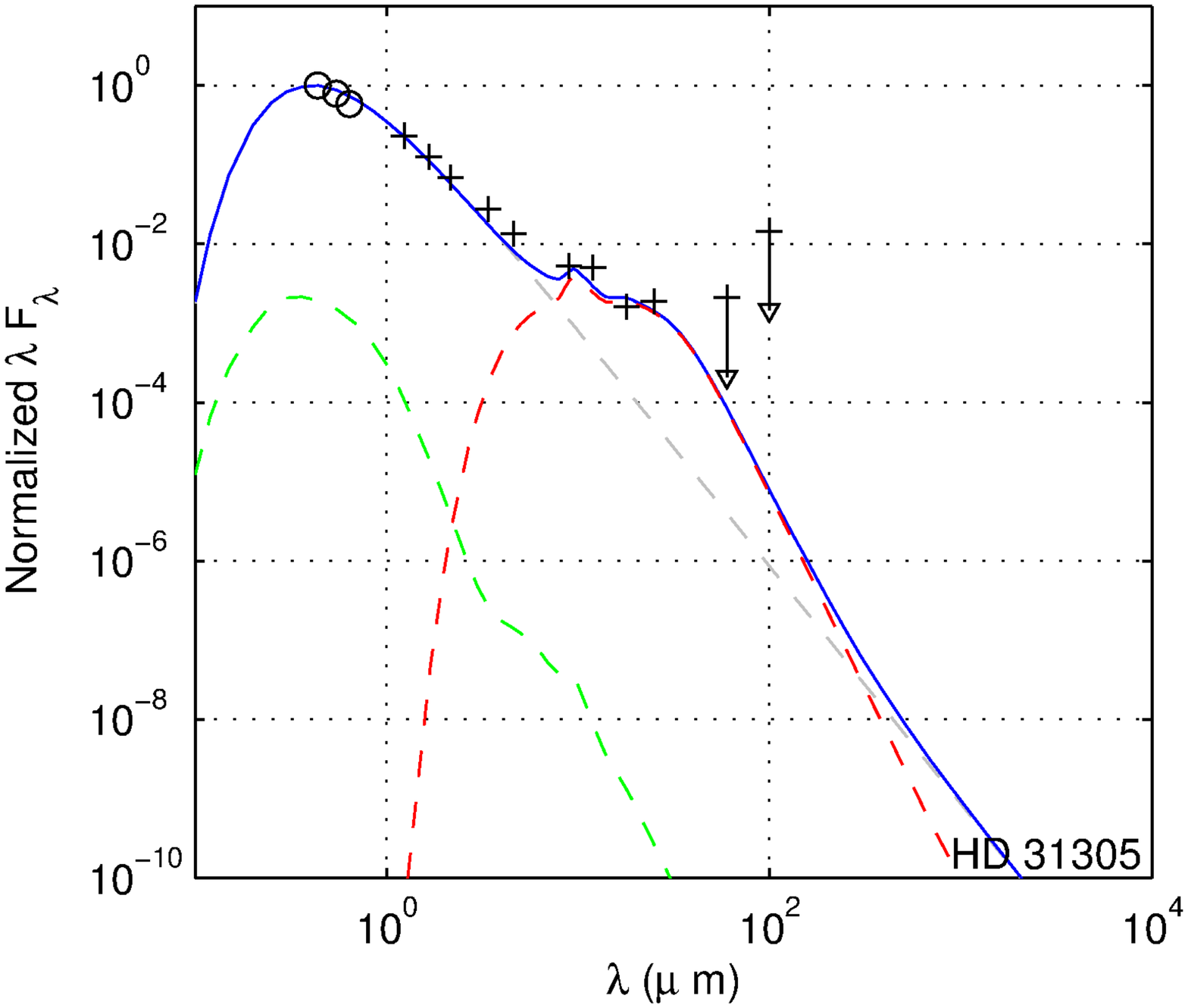}
\caption{Modelling the SEDs (black data points) of some sources having far-infrared excess 
using \textit{DUSTY}. For each of the sources, 
the attenuated blackbody representing the star is the grey dashed line, 
the contribution from the thermal and scattered emission components 
from the associated dust slab are shown as red and green dashed lines, respectively,
and the blue solid line represents the total model output.
Photometric points with error bars are shown; the circles represent BVR photometry that is reported without errors.
{\it 72 Tau}: A dust slab having 40 K temperature at the boundary closest to 72 Tau roughly reproduces the SED.
{\it HD 29647 Tau}: Two slab components are required to reproduce the SED, one chosen to 
have a dust temperature of 500 K to reproduce the $10\mu$m silicate feature, 
and another one at about 45 K.
{\it HD 282276}: Two dust slabs, having inner-edge temperatures of 200 K and 35 K are required to 
faithfully reproduce the SED. 
{\it HD 31305}: The inner-edge dust temperature was found to be 350 K. 
}
\label{fig.dusty}
\end{figure}

\subsection{MODELLING THE DUST EMISSION WITH \textit{DUSTY}}\label{sec.dusty}
In this section we consider the nebular structures 
associated with several of the early type sources
and prepare to model in the next section 
their multiwavelength image morphology.  
Reflection nebulae appear bluish when illuminated by light from a nearby star 
on account of the scattering properties of dust. 
The star and the dust may be physically related, or the encounter between the star 
and a cloud of overdense interstellar medium may be by chance.  
Infrared emission is also associated with the illuminated clouds, due to the warm dust, 
but compared to, e.g., \ion{H}{2} regions, 
the infrared luminosity is lower and there is a lack of radio emission.

DUSTY\footnote{\url{http://www.pa.uky.edu/~moshe/dusty/}} solves the radiation transfer problem for a light source embedded 
in dust through an integral equation for the spectral energy density \citep{ivezic1997}.
The code takes the following input parameters: type of external radiation source, dust composition, 
grain size distribution, dust temperature at the edge nearest to the external source, geometry of the cloud 
(spherical shell / planar slab), density profile, and the optical depth.
The dust temperature and optical depth together define the amount of radiation present at the edge of the photodissociation region.
Ideally, given the dust properties and parameters of the illuminating star,
the inner temperature of dust can be calculated giving us the separation from the star.
Following \cite{tielens2005}, we can calculate the dust temperature in a slab geometry as,
\begin{align}
(T_d/Kelvin)^5 = &2.7 \times 10^5 \, G_0 \, e^{-1.8 \, A_V} \nonumber \\
        &+ 4.1 \times 10^{-4} \, \big[ 0.42 - \ln(4.3 \times 10^{-4} \, G_0)\, G_0^{6/5} \big] \nonumber \\
        &+ 2.7^5 
\end{align}
where $A_V$ is the reddening caused by the slab.
The assumption here is a simplistic model in which the absorption efficiency of the 
dust is directly proportional to the wavelength for $\lambda < \lambda_0 = 1000 ~\AA{}$ and is unity elsewhere; the dust size is $a=1~\mu$m.
$G_0$, the far-ultraviolet (FUV, $h\nu>13.6$ eV) radiation field in terms of the average  interstellar radiation field 
($1.6 \times 10^{-3} ~\mbox{erg cm}^{-2}~\mbox{s}^{-1}$), is given by,
\begin{equation}
G_0 = 1.8 \times \left(\frac{L_*}{100 \, L\sun}\right) \left(\frac{\chi}{3.6 \times 10^{-4}}\right) \left(\frac{d}{0.02~\mbox{pc}}\right)^{-2}
\end{equation}
\noindent where $\chi$ is the fraction of the star's luminosity ($L_*$) above 6 eV and 
$d$ is the distance from the star with the normalization constant appropriate for a B8V star.


In the cases of interest here, the brightest part of the illuminated nebula is a 
few arcminutes wide, which, at a fiducial distance of 140 pc corresponds to $\sim$0.1 pc.
For comparison, using the \citep{tielens2005} formulation, 
the \ion{H}{2} region expected for a constant density pure hydrogen 
region with electron number density $n_e = 10^3 ~\mbox{cm}^{-3}$ surrounding a 
B0V star is 0.4 pc (and would produce detectable radio emission) 
while for a B5V or B8V star the Str\"omgren radius, $\mathcal{R}_s$, is about 0.03 pc 
or 0.02 pc respectively (and usually would not be detectable in the radio).
Note that $\mathcal{R}_s \propto n_e$, so local overdensities can allow 
the photodissociation regions (PDRs) to exist closer to the star thus heating them 
to higher temperatures.
Therefore, if a dust slab were located at a few hundredths of a parsec from a B8 dwarf, 
the FUV radiation field it would experience is $G_0 \simeq 2$. 
Using $A_V \simeq 1$ for the slab, we then get $T_d \sim 10$ K. However, this over-simplified 
picture overlooks important processes within PDRs such as cooling through trace species such as [OI] and [CII], 
and the inhomogeneities in the dust cloud as indicated by the changing morphology of the 
nebulae towards the B stars with wavelength.


Our calculations with \textit{DUSTY} were conducted assuming a 
power-law grain size distribution according to the 
Mathis-Rumpl-Nordsieck model \citep{mathis1977}. 
As expected, when the optical depth or the physical thickness of the 
simulated slab is reduced, 
the scattered and thermal components of the output flux both decrease. 
If the dust temperature is lowered, the thermal emission peak 
increases relative to the primary (scattered stellar flux) peak.
Notably, the shape of the thermal bump flattens and broadens if the relative abundance 
of amorphous carbon to graphite increases, whereas 
increasing the relative abundance of silicates produces the broad 10 and 18 $\mu$m features.

We model the dust surrounding 72 Tau, HD 29647, HD 282276 and HD 31305 as slabs, 
and make an attempt to reproduce their SEDs using the \textit{DUSTY} code.
Our results appear in Figure~\ref{fig.dusty} and are discussed for
the individual stars in the next section.

\clearpage
\begin{figure*}[htp]
\fbox{\includegraphics[width=6in]{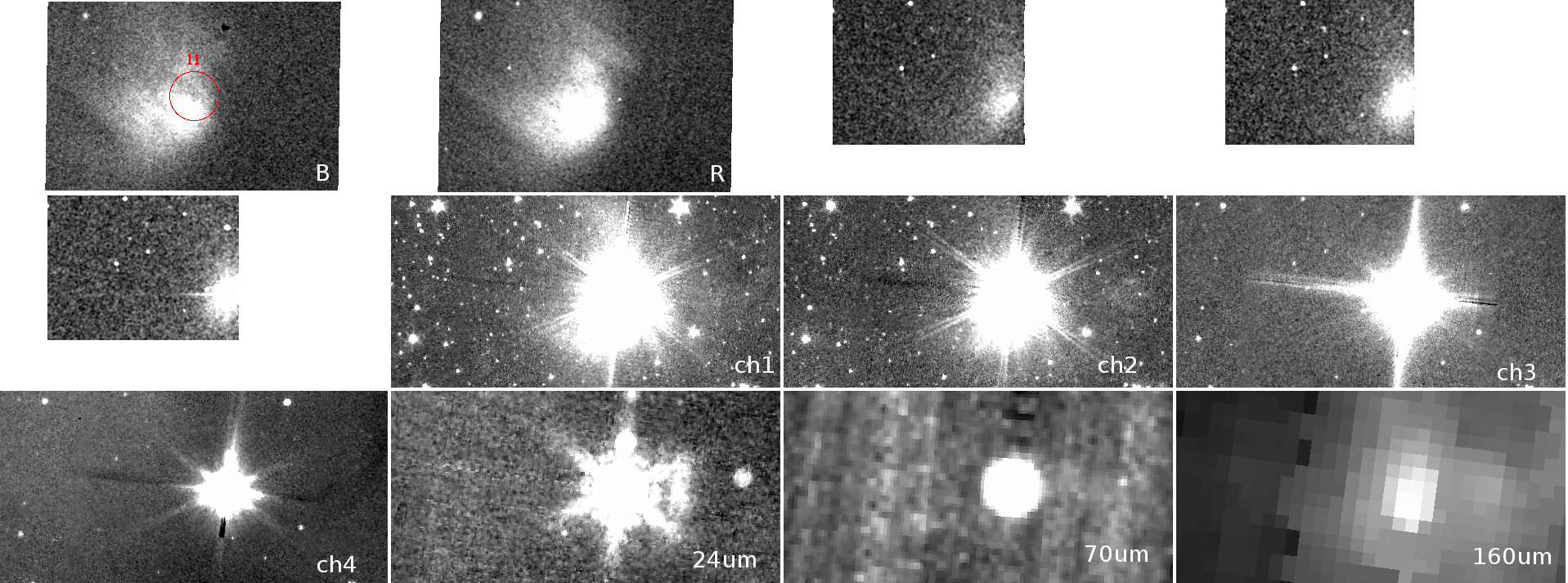}}
\caption{Cutouts of the IC 2087 region. 
Left-to-right are: B, R images from the Palomar Observatory Sky Survey (POSS-I) and J, H images from 2MASS (Row 1), 
Ks band from 2MASS, and IRAC channels 3.6$\mu$m, 4.5$\mu$m, 5.8$\mu$m (Row 2), 
IRAC 8$\mu$m, and MIPS images 24$\mu$m, 70$\mu$m, and 160$\mu$m (Row 3).
The red circle in the POSS-I B band image has a diameter of 1 arcmin, for scale.
}
\label{fig.ic2087}
\end{figure*}


\begin{figure*}[htp]
\fbox{\includegraphics[width=6in]{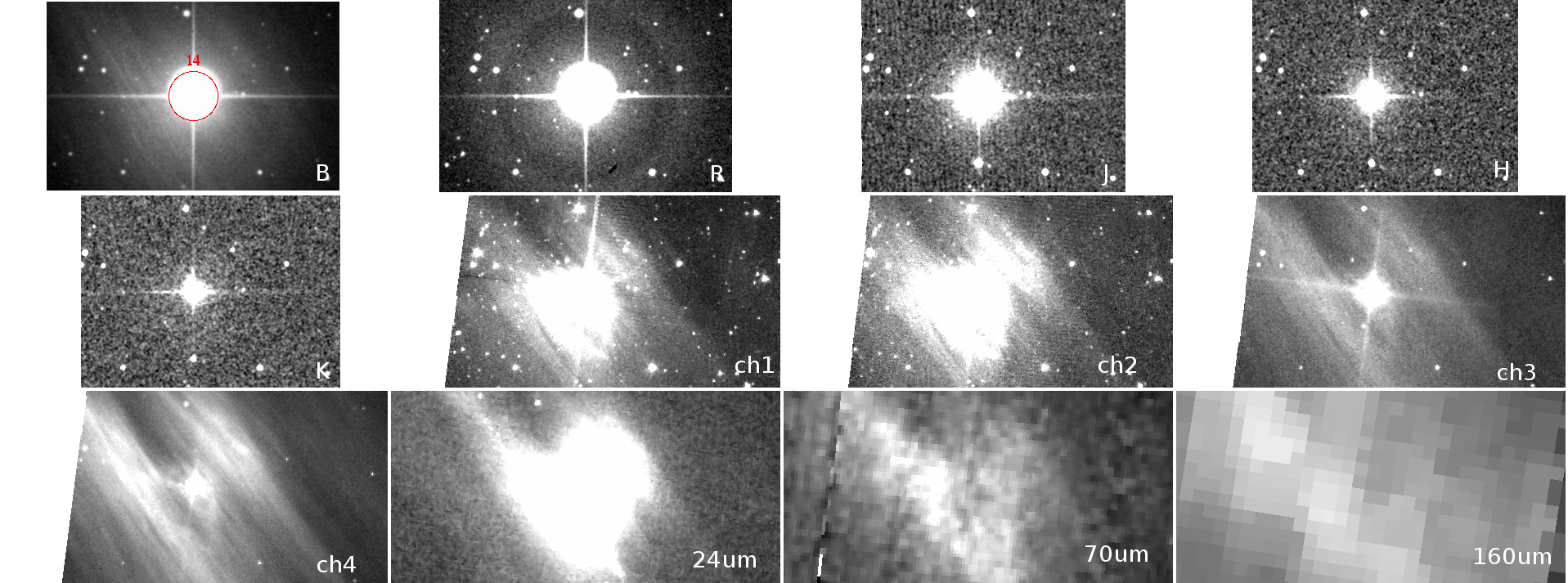}}
\caption{Same as Figure~\ref{fig.ic2087} but for 72 Tau.}
\label{fig.72tau}
\end{figure*}


\begin{figure*}[htp]
\fbox{\includegraphics[width=6in]{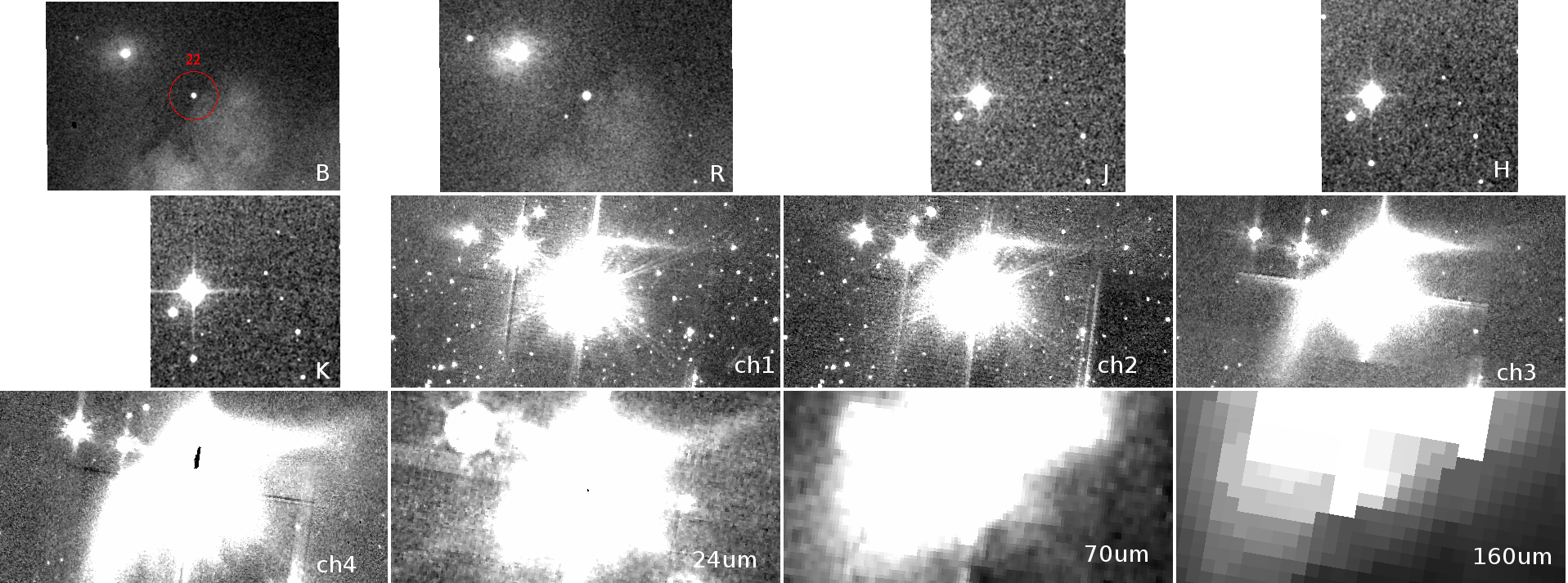}}
\caption{Same as Figure~\ref{fig.ic2087} but for V892 Tau / Elias 1.}
\label{fig.elias1}
\end{figure*}

\begin{figure*}[htp]
\fbox{\includegraphics[width=6in]{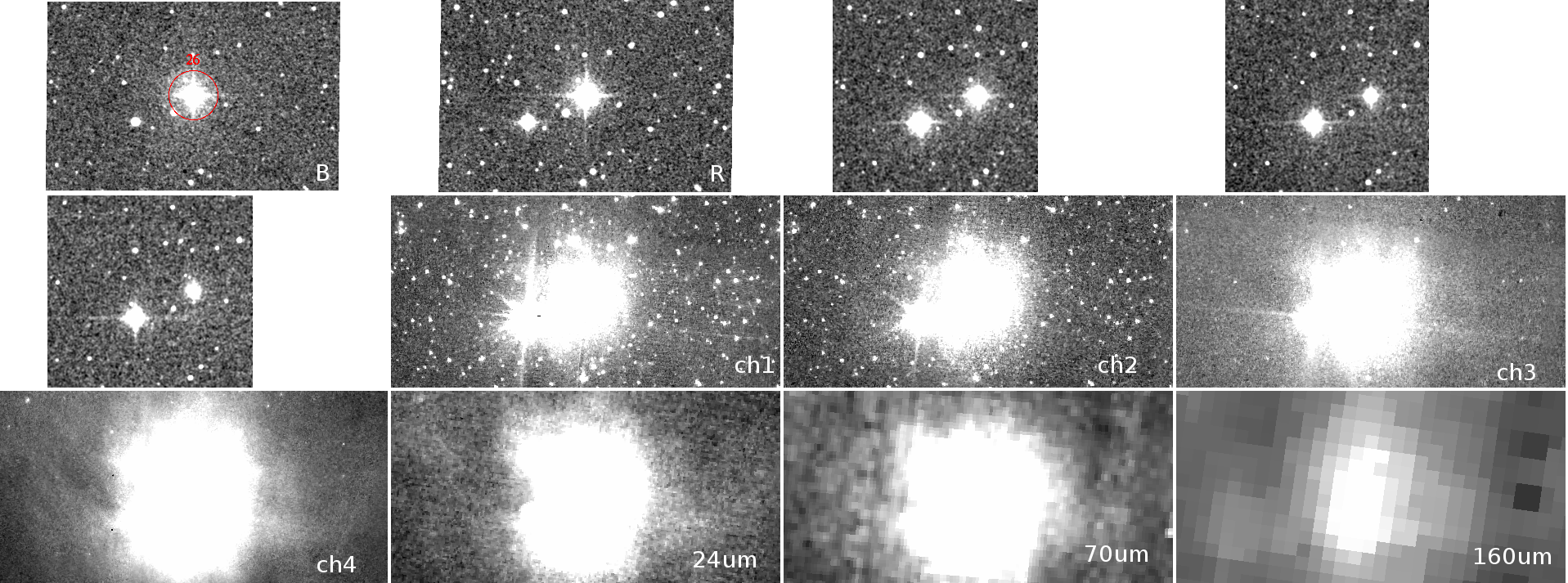}}
\caption{Same as Figure~\ref{fig.ic2087} but for HD 282276.}
\label{fig.hd282276}
\end{figure*}

\begin{figure*}[htp]
\fbox{\includegraphics[width=6in]{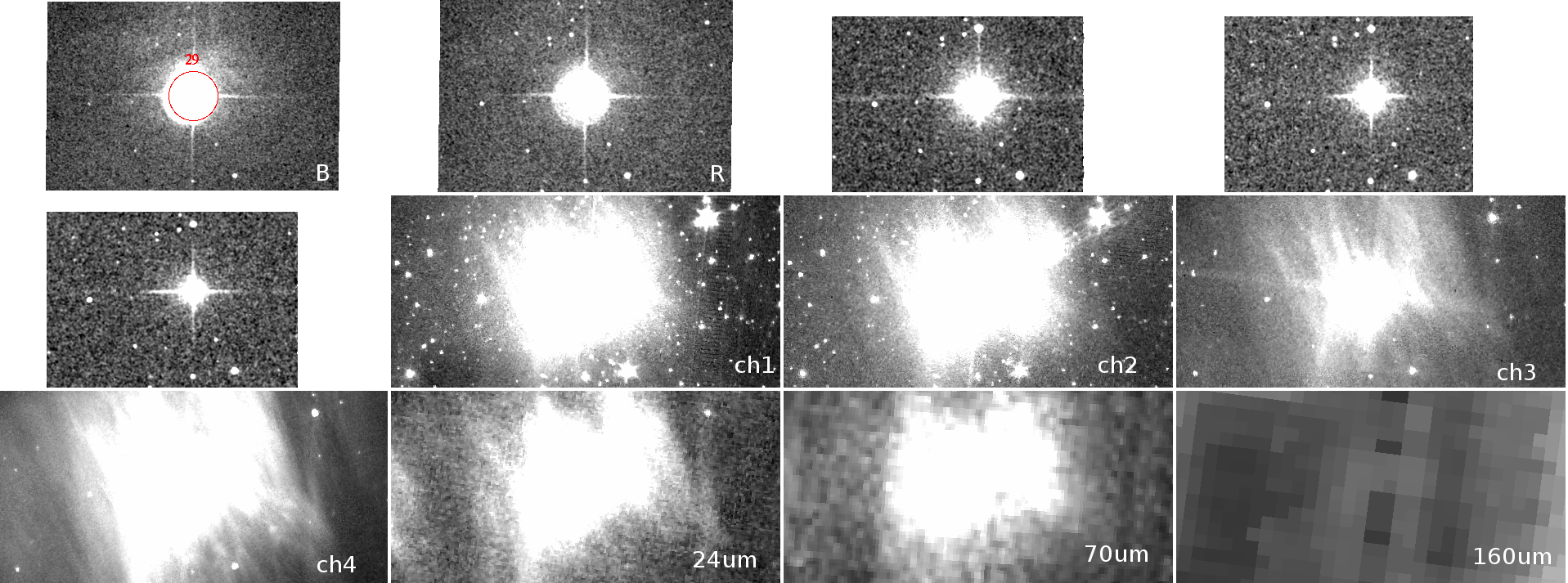}}
\caption{Same as Figure~\ref{fig.ic2087} but for HD 29647.}
\label{fig.hd29647}
\end{figure*}

\begin{figure*}[htp]
\fbox{\includegraphics[width=6in]{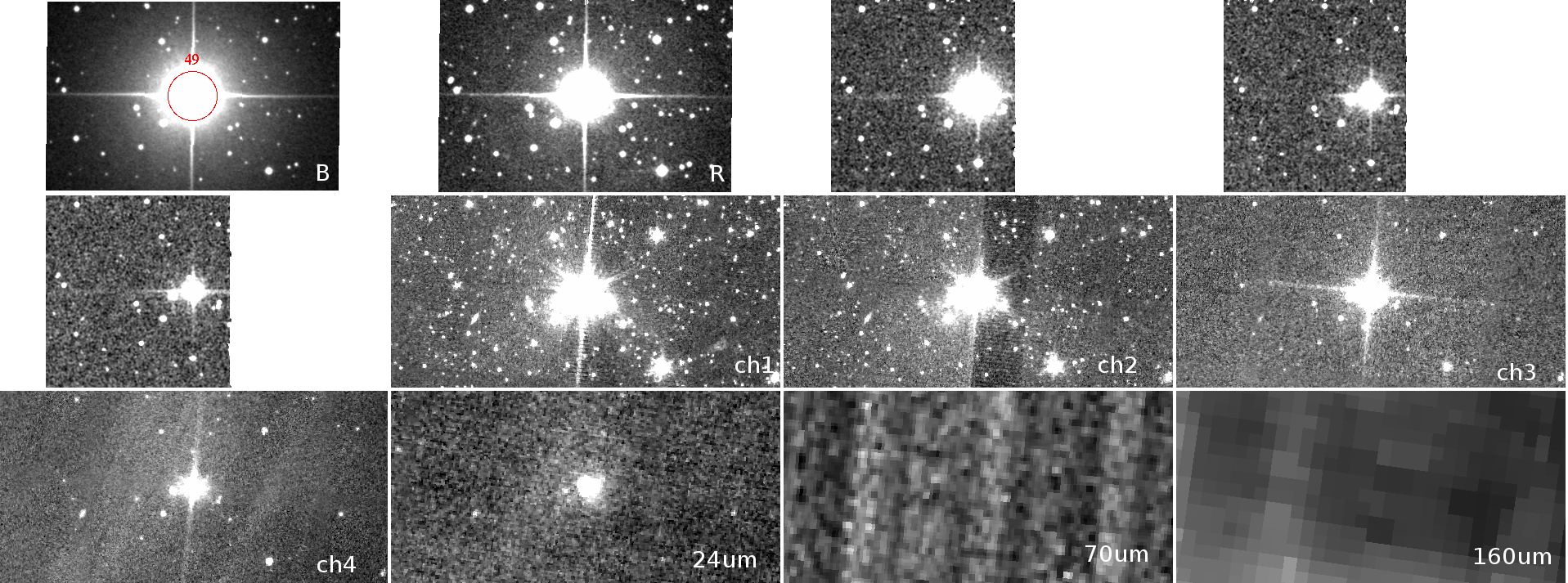}}
\caption{Same as Figure~\ref{fig.ic2087} but for HR 1445.}
\label{fig.hr1445}
\end{figure*}

\section{INDIVIDUAL EARLY TYPE OBJECTS PLAUSIBLY ASSOCIATED WITH TAURUS}\label{sec.notes}

In this section we consider the collective set of plausible early-type members 
of Taurus based on the various lines of evidence 
for their physical association with the clouds.

Above we discussed the kinematic and distance evidence for association of 
HD 28929 (HR 1445), HD 29763 ($\tau$ Tau), HD 26212, and HD 31305  with Taurus.
Additionally, the early type sources HD 31648, AB Aur, HD 27659, and HD 283815
have several lines of evidence that favor their association with Taurus
but do not meet all of our stated criteria, mostly due to missing data. 

Further, within the footprint observed by Spitzer, 
six B-type stars (IC 2087-IR, 72 Tau, V892 Tau, HD 282276, HD 29647 and HR 1445) 
are seen in the multiband Spitzer images to illuminate mid-infrared
reflection/scattered-light nebulae, not all of which can be kinematically associated with Taurus.
As mentioned above, HR 1445 was also picked out by our distance and 
kinematic membership selection criteria. 
The nebular structure is illustrated in Figures~\ref{fig.ic2087} -- \ref{fig.hr1445}.
With the exception of IC 2087 which has a bright optical nebula, optical scattered light is weak or absent among our sample. 
Furthermore, there is relatively little extended emission in the near-infrared ($J,H,K_s$ bands), 
with the wavelength of peak emission in the nebular regions typically being 8 or 24 $\mu$m.
The morphologies of the nebulae are quite varied. 
They extend up to a few arcminutes and can appear circular or squarish, some of them being asymmetric and highly striated.

\begin{figure*}[htp]
\centering
\includegraphics[width=7.5in]{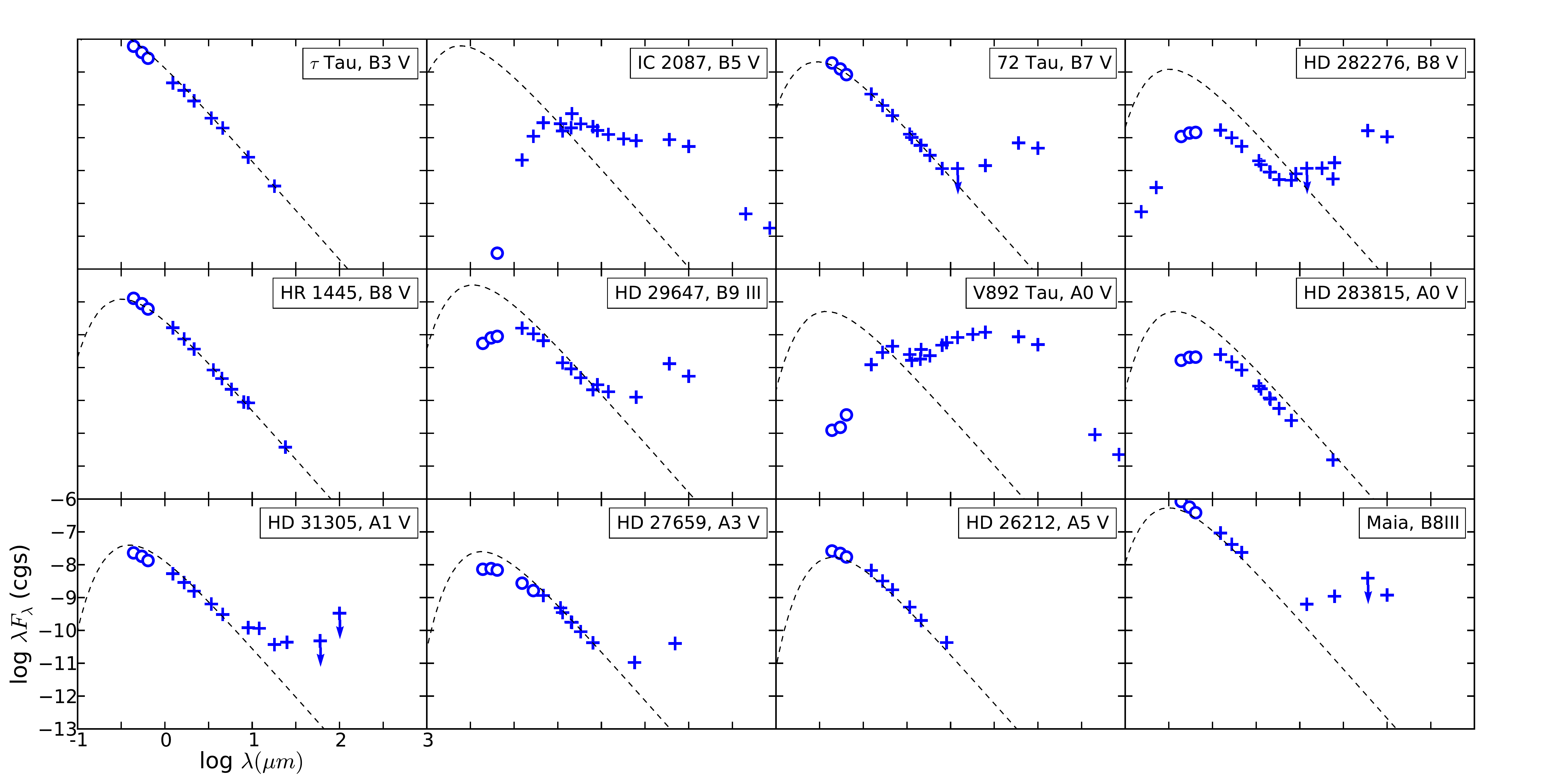}
\caption{Spectral energy distributions (SEDs) of those sources exhibiting 
infrared nebulae in Figure~\ref{fig.mosaic} plus the candidate early-type stars 
we conclude are probable members of Taurus based on our assessment of distance 
and kinematics.  Infrared excess is apparent in many objects. 
This may be due to the presence of a circumstellar disk associated with
a pre-main sequence stars, to a debris disk in a somewhat older main sequence star,
to a dusty atmosphere in the case of an evolved giant star, 
or to a chance superposition of a hot star with a nearby diffuse cloud.  
For comparison, an example of the chance-superposition case is also shown in the lower right panel: 
the Pleiades member Maia, whose SED exhibits an apparent infrared excess. 
The data sources include GALEX (ultraviolet), NOMAD (optical, $BVR$), 2MASS-PSC
(near-infrared, $JHK_s$), {\it Spitzer}, AKARI and IRAS (mid-far infrared), and SCUBA (sub-mm).
Photometric error bars are generally smaller than the symbol size; circles denote photometry 
lacking uncertainty (usually values from NOMAD).
The dashed line in each panel represents a blackbody at 140 pc characterized by the effective temperature and radius of the star whose SED is represented in that panel.  No correction for reddening has been applied
though the existence of reddening can be inferred from the location of short wavelength photometry
well below the nominal blackbody.}
\label{fig.seds}
\end{figure*}

For both kinematically selected and nebular-selected objects, 
we constructed spectral energy distributions (SEDs) as shown in
Figure~\ref{fig.seds}. We used the following data in making the SEDs: 
(i) sub-mm: SCUBA/\cite{andrews2005}; (ii) infrared: {\it Spitzer}, 2MASS, AKARI, IRAS; (iii) optical/UV: NOMAD, GALEX.
Counterparts within one arcsecond of the source were chosen, with the exception of IRAS counterparts.


As reviewed by \citet{williams_cieza},
for young pre-main sequence stars 
red {\it Spitzer}/IRAC colors indicate excess emission from circumstellar 
disks and envelopes, whereas excess emission at 24 $\mu$m but not in the 
shorter wavelength IRAC bands is indicative of a disk with an inner hole.
At slightly later stages the dust is attributed to second generation `debris' rather than primordial material. 
Infrared excess also could be attributed to dust shells around evolved stars, or illumination of nearby 
interstellar material, irrespective of any physical association of it with the star.

\begin{figure}[htp]
\centering
\includegraphics[width=3in,angle=-90,viewport=190 250 430 590,clip]{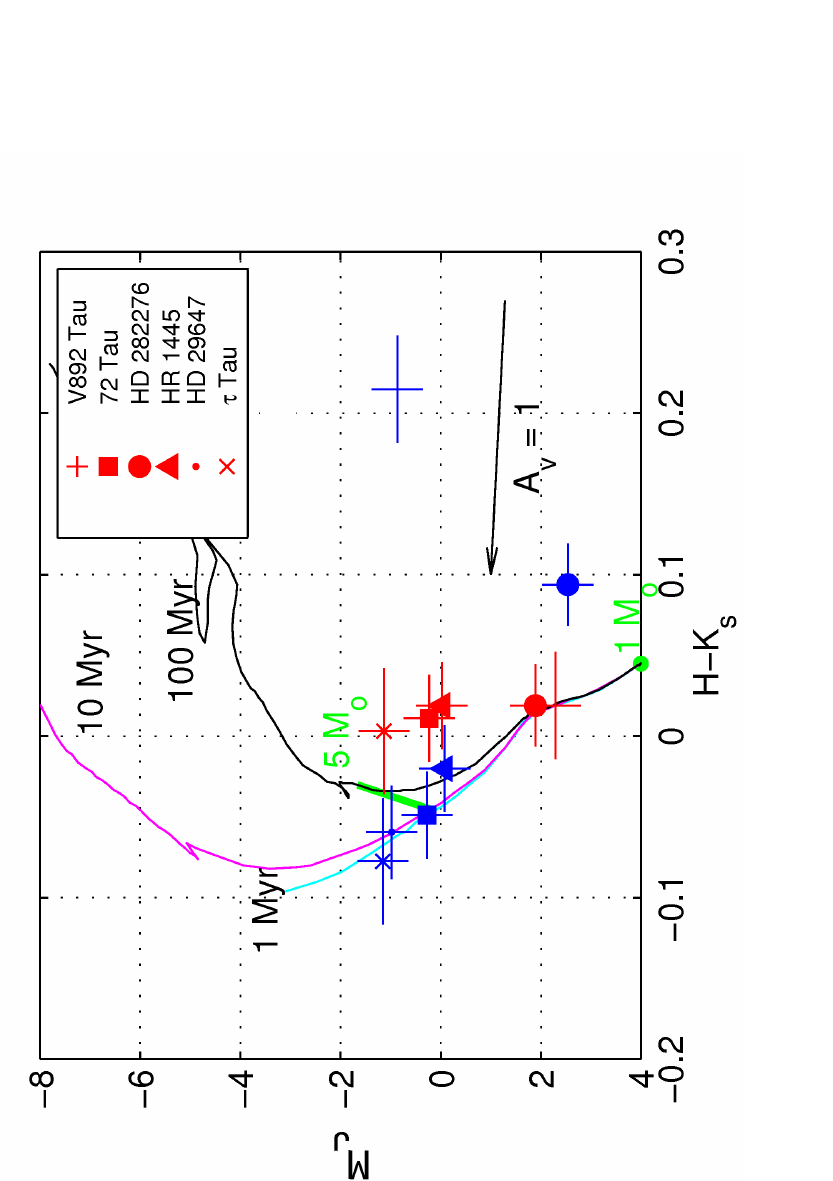}
\caption{Estimation of the isochronal age of some of B stars showing infrared reflection nebulae and $\tau$ Tau. 
Isochrones are from \cite{girardi2002}.
Horizontal errorbars represent the error in the 2MASS magnitudes and vertical errorbars represent the uncertainty in the distance ($128<d<162$pc).
Blue markers are plotted using the reddening parameters from literature, while red markers represent values derived using $R_V = 3.1$ and 
the \cite{cardelli1989} reddening.}
\label{fig.age}
\end{figure}

We also constructed various color-magnitude diagrams and overplotted isochrones 
(see Figure~\ref{fig.age}) to assist in the assessment of stellar age, assuming 
that the distance of Taurus is appropriate for each source. 
The unknown stellar multiplicity and photometric 
error add substantial uncertainty to the stellar age estimate. 
Also, because of the rapid evolution of high mass stars, the age derived 
via isochrones is very sensitive to the reddening correction, which  
is not insignificant, and which we have derived assuming a plausibly
invalid constant $R_V$ of 3.1.
All these factors together preclude accurate determination 
of the stellar ages of our sample but we present the resulting color-magnitude
diagram for completeness.

We now discuss our findings for individual sources, beginning with those
illuminating nebulae in the \textit{Spitzer} and then moving on to other
candidates that we have assessed.

\subsection{IC 2087}
The IC 2087 nebula (Figure~\ref{fig.ic2087}) is brightest at optical wavelengths 
and is less prominent at {\it Spitzer} wavelengths, a unique illumination 
pattern among our sample.  The SED (Figure~\ref{fig.seds}) 
of the associated point source (IC 2087-IR = IRAS 04369+2539 = Kim 1-41) 
is consistent with an early-type Class I -- Class II young stellar object
seen through $\sim$15 magnitudes of visual extinction.
The foreground extinction is claimed to be $A_V < 1$ \citep{frerking1982}, 
and hence the majority of the extinction toward IC 2087-IR should be due to circumstellar material \citep{shuping2001}.

This source is a known YSO member of Taurus. 
High veiling has precluded accurate determination of its spectral type but \cite{elias1978} 
classified it as B5 based on the bolometric luminosity.
We were unable to derive a spectral type from either the {\it SDSS} spectrum 
or our own follow-up optical spectroscopy of this source.
No spectroscopic parallax is possible given the vague spectral type.
However it appears that a proper motion measurement exists for IC 2087-IR.
\cite{white2004} quoted a radial velocity of $22 \pm 8$ km s$^{-1}$ but with the
large error bar it is hard to tell whether the measurement is consistent 
with the RV membership criterion used herein; furthermore, the lines
used may have been dominated by outflow kinematics rather than photosphere.
Nonetheless, the source is currently an accepted member of Taurus, and based on multiwavelength information, we propose that an early 
spectral type of B5--B8 is most appropriate.

\cite{rebull2010} confirmed that the source has a flat spectrum in the near-to-far-infrared, and report 
$L_{IR}/L_{total}=0.41$.
Weak molecular outflows, the presence of Herbig-Haro knots, mid-IR absorption features and 
associated reflection nebulosity provide additional evidence 
of the pre-main sequence nature of this object \citep[e.g.][]{furlan2008}.
\cite{hillenbrand2012} provide an extensive discussion of this source.


The nebula is apparently the result of complex radiative transfer
of an emitting source seen through a three-dimensional, non-uniform distribution
of circumstellar dust that both obscures the central source 
and produces significant amounts of scattered light asymmetrically distributed
with respect to the infrared point source. 
We do not attempt to model the emission here, but see \cite{hillenbrand2012} 
for an accretion disk and weak envelope fit to the spectral energy distribution.

\subsection{72 Tau}
The nebula associated with 72 Tau (HD 28149) is brightest at about $\sim 60~\mu$m.
It is prominent in all IRAS and {\it Spitzer} bands and is also discernible 
in blue optical bands.
From the image cutouts (Figure~\ref{fig.72tau})  
it is evident that the optical depth of the associated nebula 
is quite different at different wavelengths in the infrared.
The blue optical nebula appears to have been known by \citet{cederblad}
and thus this feature should be referred to as Ced 34.
The striated nebular morphology is similar to that observed 
for the Maia nebula, illuminated by the Pleiades stars.
The SED (Figure~\ref{fig.seds}) of 72 Tau is consistent
shortward of about 10 $\mu$m with an ideal blackbody having very low reddening,
but there is a longer wavelength excess.  The SED morphology is similar to that
of Maia as well.
\cite{kalas2002} noted that 72 Tau is a Vega-like source associated with a gas overdensity, but no rigorous analysis was performed.

We arrive at a spectral classification of B7V for 72 Tau based on our follow-up spectroscopy. 
 The Hipparcos distance to this star is $127\pm12$ pc while our estimate of spectroscopic parallax is $161\pm3$ pc.
\cite{walter1991} quoted a spectroscopic parallax distance of 179 pc and proposed that this star is a member of Taurus and the Cas-Tau OB association.
The study by \cite{deZeeuw1999} however does not acknowledge this star as a member of Cas-Tau association.
\citet{kenyon1994} found a similar distance of 178 pc based on a B5V spectral type.
If the Hipparcos distance is accurate, 72 Tau would lie close to the nearer edge of the Taurus cloud whereas if the further distance estimates are correct
the star is at the back edge of (our distance) or behind (literature distances) the cloud. 
Both the proper motion and the radial velocity of this star are consistent with the Taurus group \citep[as defined by][]{luhman2009}.
We conclude that it is very likely that 72 Tau belongs to Taurus, although
in the past \citep{whittet2001}
the star has been considered a background source in
studies of the Taurus clouds. 

Previous age analysis in the literature
\citep{westin1985} suggested an age of 20 Myr based on ubvy$\beta$ photometry and theoretical stellar evolution models.  
From our spectral analysis, 72 Tau is clearly a main-sequence star which, 
after consideration of its spectral type, tells us that it is at least several Myr old and at most 170 Myr old;
we are unable to place any further constraints on its age.

We were able to reproduce the SED with \textit{DUSTY} using the parameters:
(i) $T_{\rm eff}=12,500$ K blackbody as the external radiation source; 
(ii) dust composition with silicates from \cite{draine1984} and amorphous carbon 
from \cite{hanner1988} in a ratio of 1:4 (however, output not very sensitive to this 
ratio so long as silicate fraction is less than 1); 
(iii) a 40 K dust temperature at the edge nearest to the external source;
(iv) a 0.55 $\mu$m optical depth of 0.1 (corresponding to $A_V \simeq 0.1$).
The resultant ratio of the dust contribution 
to the observed flux of the central star, a B7V-type blackbody, is about 1:200.
This final model of the SED is shown in Figure~\ref{fig.dusty}.
Since the fraction of the observed flux contributed by dust emission is much less 
compared to that contributed by the central star itself, 
we conclude that the nebula is not a photodissociation region, and we do not 
expect any free-free emission (indicative of an ionizing front) from it.

\subsection{V892 Tau}
In the optical bands, V892 Tau (also known as Elias 1 or more properly Elias 3-1) 
appears to be a point source with an associated faint reflection nebula
(Figure~\ref{fig.elias1}).
The nebular structure is brightest at about 24$\mu$m. 
Like IC 2087-IR, it is heavily reddened with a significant envelope component to its circumstellar environment.
The SED (Figure~\ref{fig.seds}) is that of a Class I -- Class II source.
\cite{rebull2010} report it as a flat-spectrum source in the near-to-far-infrared, having 
$L_{IR}/L_{total}=0.089$.
V892 Tau is a double-star system \citep{smith2005} with a circumbinary disk \citep{monnier2008}. 

This Herbig Ae/Be system is a well-accepted member of Taurus with reported spectral types ranging from B8 to A6.
Our spectral type is B8.5V to $\sim$A0Ve.
The optical spectrum also shows Balmer emission lines, further evidence of its youth and likely membership.
Our derivation of the spectroscopic parallax distance for this star has a large uncertainty,
possibly enhanced by reddening law uncertainties.
The probability associated with the $\chi^2$ analysis confirms that this source 
is a proper-motion member of Taurus.
As for IC 2087-IR, the circumstellar environment is complex
and we do not attempt to model the emission.

\subsection{HD 282276}

The nebula associated with HD 282276 (Figure~\ref{fig.hd282276}) is evident only in 
the {\it Spitzer}/IRAC and MIPS bands, and not in the 2MASS or shorter wavelength bands.
The nebula appears to be circular and optically thick at {\it Spitzer} wavelengths.
This source shows considerable infrared excess in its SED (Figure~\ref{fig.seds}) 
beyond 10 $\mu$m.
The two peaks in the infrared excess suggest two different dust components 
from either two different radii or two different compositions
contributing to the infrared emission. 

\cite{rebull2010} have classified HD 282276 as a pending member of Taurus 
which needs additional follow-up; they tentatively categorized it as a Class II YSO.
The spectrum of this star is B8V in our analysis but it does not show any emission lines. 
Both the spectroscopic parallax distance and the proper-motion are inconsistent with Taurus.  
We estimated the reddening to be $A_V=2.8$, and a spectroscopic parallax 
distance of $422 \pm 52$ pc, which means that it lies closer to
the distance of Perseus than Taurus.  We speculate that this star 
could be associated with the Perseus star forming region, 
which has a mean proper motion of about 
$\mu_{\alpha}=5$ mas yr$^{-1}$, $\mu_{\delta}=-12$ mas yr$^{-1}$ 
(\cite{deZeeuw1999} and PPMXL catalog), quite consistent with 
the measured values for HD 282276.
However, the derived distance is somewhat large, even for Perseus,
and the star is offset to the east of the main Perseus cloud.
There is no radial velocity measurement.

The fact that the star illuminates a reflection nebula in the infrared
(see Figure~\ref{fig.mosaic}), suggests the presence of cloud material, 
perhaps associated with a low column density extension of the classical Perseus region.
It is also possible that HD 282276 is similar to several Pleiades 
member stars, with the star randomly encountering a local overdensity in 
the interstellar medium.

The $\sim 3$ arcmin-wide size of the nebula corresponds to a roughly 0.4 pc diameter
structure at 420 pc. The \ion{H}{2} region for a B8V star surrounded by a pure-hydrogen nebula
would be about 0.02 pc in diameter and as the two
values are inconsistent (even if our distance estimate is wrong by a factor of 3), 
we believe that the nebula is not a PDR immediately
surrounding the \ion{H}{2} region. The SED shows two bumps in the infrared,
indicative of contribution from two different dust 
components. We can reproduce this SED with {\it DUSTY} using 
the parameters for slab A: (i) a blackbody with T$_{\rm eff} =$ 11,400 K
as the external radiation source;
(ii) silicate grains from \cite{draine1984};
(iii) a 200 K dust temperature at the edge nearest to the 
external source; (iv) a 0.55 $\mu$m optical depth of 1.4.
The output of slab A summed with a 6:1 contribution from 
the star (blackbody) was fed into slab B, which
was modeled to have the same dust composition and
grain-size distribution of slab A, and
a dust temperature as 35 K at the edge, with visual
optical depth of 1.6. Together, the optical depth adds up
to three, corresponding to $A_V \simeq 2.8$, approximately
the average extinction through the Taurus dark cloud.
The dust composition is only suggestive; the bump at 10$\mu$m is much too
broad to guess the silicon abundance. However, noting 
that the increase of amorphous carbon abundance 
relative to silicon results in a broad spectral 
feature at $\sim 0.12\mu$m and flattening of the 
thermal bump, we expect the relative amorphous 
carbon content to be small. Also, this broad bump 
makes it difficult to interpret the temperature of 
slab A, and we note that the edge temperature for
this slab can lie anywhere between 100 and 300 K.

\subsection{HD 29647}

Multi-wavelength images of HD 29647 (Figure~\ref{fig.hd29647}) 
clearly show striated nebulosity similar to that associated with Pleiades stars.
Even the SED is similar to that of Maia, a Pleiades member;
both are shown in Figure~\ref{fig.seds}. 
The source is frequently used in interstellar column density 
and extinction law studies, and has a rich literature.

HD 29647 is known as a heavily reddened Hg-Mn B6-7 IV or B8 III  star.
From our spectrum, we classify it as a B9 III spectral type 
and can confirm the presence of Hg absorption. 
Our analysis results in a spectroscopic parallax distance of 160 $\pm$ 1 pc, 
consistent with Taurus membership. However,
as we do not have reliable estimates for the intrinsic colors of B-type giant stars,
this distance estimate might have systematic error much greater than the quoted uncertainty.
The Hipparcos parallax distance to HD 29647 is $177 \pm 35$ pc.
Based on proper motion analysis, the star is probably a non-member. 
There is no radial velocity measurement; however, 
\citet{2001MNRAS.328.1144A}
make a statement that the radial velocity of the star ``closely matches that
of the foreground cloud" with unfortunately no quantitative detail given.

The fact that HD 29647 illuminates a bright infrared nebula is, however,
a compelling reason to associate this star with the far side of Taurus. 
\cite{whittet2004} argue that the visual extinction of 3.6 magnitudes for HD 29647 
and 5.3 magnitudes toward an adjacent sightline suggests that this star lies 
within a diffuse cloud that is slightly beyond the molecular gas constituting TMC-1.  These authors also mention the existence of infrared excess based on
IRAS data as an argument for the proximity of HD 29647 to small grains
associated with the cloud.
A scenario in which HD 29647 was born from the Taurus cloud and ended up having different motion
can be envisaged (for example, through ejection from a binary/multiple-star system).
Given the relatively low space velocity as estimated from the small proper motion, 
however, this scenario seems unlikely. 
Perhaps it was born from a cloud lying between the present-day Taurus and Perseus clouds.
Being a giant star, an age between 90 Myr and 120 Myr as deduced 
from stellar evolution models, may be appropriate for this $\sim$5 M$_\sun$ star.
We conclude that the nebulosity in this case is a situation similar to that of Pleiades stars.


Although we have argued that this source is probably not associated 
with Taurus, 
it does illuminate nebulosity and has a substantial infrared excess. 
We see on inspection of the SED that there is a broad silicate 
feature evident at $10\mu$m, and another thermal bump at about $60\mu$m.
On modelling this emission with {\it DUSTY}, we find that two dust components 
are required to explain this.
One produces the broad silicate feature, requiring dust temperature at 
the inner boundary between 400 and 2000 K.
The cooler limit comes from a fit to the thermal bump of the dust continuum emission. 
The hotter temperature limit is defined by the fit of the SED to optical data points.
We note that as dust evaporates somewhere between 1000 K and 2000 K 
depending on the exact composition \citep[e.g.][]{duschl1996,speck2000}, 
dust temperatures at the upper end of the range become unphysical.
We assume somewhat arbitrarily a 500 K temperature for the first component 
but the use of different temperatures within the range specified above does 
not affect the overall fit substantially.  
The second dust component is characterized by a temperature of 
about 45 K, which is defined by the peak of the thermal bump seen at tens 
of microns.  To reproduce the SED as shown in Figure~\ref{fig.dusty} we used
(i) a blackbody with $T_{\rm eff}=10,900$ K as the external radiation source; 
(ii) dust composition of silicates from \cite{draine1984} 
(with the output not very sensitive to the composition); 
(iii) dust temperature at the edge nearest to the external source: 500 K for slab A and 46.5 K for slab B; 
(iv) optical depth of 0.1 and 4 at 0.55$\mu$m for slabs A and B respectively 
(we need $\tau_A + \tau_B \simeq 4$, so that the total $A_V$ for this star equals about 3.6).
The output from slab A was passed as input to slab B and thus the overall output was calculated.
To the output of slab B, another contribution from the output of slab A had to be added (in a 1:10 ratio) 
to the result of \textit{DUSTY} in order to reproduce the observed flux. 

\subsection{HR 1445}

The slight nebulosity associated with HR 1445 = HD 28929  is discernible only in the 
8 and 24 $\mu$m bands, with a striated morphology evident at 8$\mu$m
(Figure~\ref{fig.hr1445}).
The SED (Figure~\ref{fig.seds}) does not reveal any infrared excess out to 24 $\mu$m
and neither the star nor any nebular emission is seen at 70 and 160 $\mu$m.

HR 1445 is known as a peculiar star, an Hg-Mn main-sequence star of spectral type B8.
Our spectral analysis also suggests a B8 (dwarf) star and we do see a weak Hg signature.
The spectroscopic parallax distance is $136 \pm 15$  pc, and in good agreement with
its Hipparcos parallax distance of $143 \pm 17$ pc. 
\cite{walter1991} derived a spectroscopic parallax distance of 158 pc
while \cite{kenyon1994} found 137 pc.
From our $\chi^2$ probability test, we deduce secure proper motion membership. 
HR 1445 also has a radial velocity consistent with that of Taurus.
The good agreement of the distance, proper motion, and
radial velocity of this star with that expected for Taurus members provides 
strong support to the idea of HR 1445 being a Taurus member.

The age of this star, because it is main sequence, can be constrained only to 
less than a few hundred Myr, within $1\sigma$ uncertainty.
\cite{westin1985} quoted the age of this star as 60 Myr, 
in which case it would be an unlikely member of Taurus.
HR 1445 is located toward a region of Taurus which is devoid of 
dense material along the line of sight.  One might argue, to explain the nebulosity, 
that HR 1445 is located within a diffuse dust screen behind Taurus, just like HD 29647.

While HR 1445 shows hints of nebulosity, there is no infrared excess 
and because of the wide range of unconstrained parameter space allowed 
in the models we do not attempt a \textit{DUSTY} model.

\subsection{$\tau$ Tau}

The $\tau$ Tau = HD 29763 region was not covered by any of the {\it Spitzer} surveys.
However, examination of {\it WISE} image data reveals a compact, 
nebulous feature at 12 and 22 $\mu$m, similar to that observed for HR 1445.
Also similar to the situation for HR 1445,
the SED (Figure~\ref{fig.seds}) of $\tau$ Tau 
does not reveal any infrared excess out to 22 $\mu$m and
the spectrum does not show the presence of any emission lines.

This source is a well-studied binary system 
composed of B and early A dwarfs \citep[e.g.][]{cvetkovic2010}.
From our B3V spectral type we derive a spectroscopic parallax distance of $137 \pm 9$ pc.
The Hipparcos parallax distance is $122 \pm 13$ pc.  
The value from \cite{walter1991} is 142 pc.
Both the proper motion and system radial velocity of $\tau$ Tau are consistent 
with Taurus group V \citep[as defined in ][]{luhman2009}, towards which it lies.
If it is a main-sequence star, as the spectrum indicates, 
then we can constrain the age as being $\lesssim 40$ Myr.
\cite{westin1985} derived 20 Myr. 
Since the spectral type is so early $\tau$ Tau is plausibly coeval 
with the low mass T Tauri members of Taurus.

$\tau$ Tau also lies toward that region of Taurus which is devoid of dense material.
A visual extinction of about 0.4 mag is derived.  Only 1.6 degrees 
away from $\tau$ Tau is HP Tau/G2 which has had its distance 
accurately measured via VLBA at 161 pc, whereas we find 130 pc as the distance to the binary system;
this would suggest a mainly line-of-sight separation of about 30 pc between the two stars.
Indeed, HP Tau/G2 has higher reddening with $A_V=2.6$ quoted by \cite{rebull2010}.
We thus conclude the $\tau$ Tau is associated with the near side of the Taurus clouds.

As for HR 1445, given that 
$\tau$ Tau shows hints of nebulosity but no infrared excess, 
we do not attempt a \textit{DUSTY} model.

\subsection{AB Aur and HD 31305}

AB Aur = HD 31293 is the prototypical Herbig Ae star with an assigned a spectral type of B9--A0e, consistent with our spectrum.  
It is a known member of Taurus with a Class II SED.
We do not produce a model for this well-studied source.

HD 31305, though not part of the area mapped under the guise of the {\it Spitzer} Taurus Legacy Survey, 
was covered in the ``C2D" (PI N. Evans) maps. 
The source is not associated with nebulosity in any mid-infrared, near-infrared, or optical wavelength. 
However, it came to our attention through its proximity to known early-type member AB Aur.
In the literature, HD 31305 mainly has been used as a reference star for variability studies
of AB Aur and other nearby young stars, though itself turns out to be a variable \citep{cody2013}.

We derive a spectral type of A1V 
with very little reddening ($A_V \simeq 0.1$)
for HD 31305, and a spectroscopic parallax distance of $174\pm 11$ pc.  
Although this star lies at 1$\sigma$ just outside the formal distance range that we consider 
for Taurus membership, (up to 162 pc), the star is seemingly a good candidate for membership.
A secure proper-motion membership probability is found, but no radial velocity measurements exist.
Keeping these points in mind, we conclude that HD 31305 is likely
a newly appreciated member of Taurus.

There is no nebulosity apparent but the SED shows infrared excess in the near-to-far infrared. 
Investigation of its SED relative to a range of {\it DUSTY} models
tells us that the ~10 $\mu$m silicate feature is probably strong, 
above the blackbody line, with the second thermal bump at a longer wavelengths 
broad and flat. This SED morphology indicates a similar abundance of silicon and
amorphous carbon. The silicate feature remains significant if the edge
temperature of the dust slab is set to lie between 100 and 500 K, but
this is also dependent on the optical depth.
We model this SED using the following parameters for a single slab.
(i) blackbody with T$_{\rm eff} =$ 9,400 K (=A1V) as the external radiation source;
(ii) silicate and graphite grains from \cite{draine1984} in a 10:1 ratio; 
(iii) a 350 K dust temperature at the edge nearest to the external source; 
(iv) 0.55~$\mu$m optical depth of 0.05.
The contribution of the central star 
relative to the thermal and scattered dust emission is 4:1.
The dust emission could equivalently be modeled using a 2-slab model 
to get a better fit, but this is not justified by the low extinction
and the appearance of the nebula on 2MASS image cutouts.

\subsection{HD 26212}
HD 26212 also lies outside the region of the {\it Spitzer} Taurus Legacy Survey, but was not 
covered in C2D.
Inspection of DSS, 2MASS, WISE and IRAS images does not reveal any nebulosity towards this star.
It does not have any infrared excess out to 24 $\mu$m as evidenced by its SED (Figure~\ref{fig.seds}).

We arrived at a spectral classification of A5V for HD 26212. The corresponding 
spectroscopic parallax is $115 \pm 4$ pc which is in agreement (within the $2\sigma$ error)
with both its larger Hipparcos parallax as well as the new Hipparcos reduction \citep{vanLeeuwen2007}, $d_{_{HIP}} = 100^{+7.8}_{-6.7}$. 
The proper motion of this star is consistent with Taurus, which means that it is tangentially comoving with the members of this cloud.
The mean radial velocity quoted in the literature is about 20 $~\mbox{km s}^{-1}$, and within 1$\sigma$, agrees with that observed for Taurus members.
The derived spectroscopic parallax yielded a visual reddening of $A_V = 0.16$.
The low reddening and near distance suggests that HD 26212 lies in the outskirts of the L1498 region in Taurus.
Since its spectrum suggests a main-sequence star and the spectral type is late, 
we can not meaningfully constrain the age.

HD 26212 shows neither significant infrared excess 
nor scattered light, and therefore cannot be modelled with \textit{DUSTY}.

\subsection{HD 27659, HD 283815}
Both of these stars are listed as having spectral type A0 and as probable members of Taurus based on infrared excess by \cite{rebull2010}.

The multiwavelength image cutouts of HD 27659 show extended emission at 8$\mu$m.
The object is seen to be fuzzy but somewhat compact beyond this wavelength, a 
feature which is similar to multi-wavelength image morphology of HR 1445 and $\tau$ Tau.
HD 27659 also shows considerable infrared excess beyond 25 $\mu$m (Figure~\ref{fig.seds}).
\cite{rebull2010} find a Class II SED with $L_{IR}/L_{total} = 6 \times 10^{-4}$.
\cite{kenyon1994} classified HD 27659 as an A4V star and derived a spectroscopic parallax distance of $d=146$ pc.
\cite{belikov2002} list HD 27659 as a member of the Perseus OB2 star forming complex, 
but its proper motion renders this improbable.

For HD 283815, there is no associated infrared nebulosity but
\cite{rebull2010} found the 8 $\mu$m to 24 $\mu$m flux ratio to indicate a Class III
SED exhibiting a weak infrared excess, though this is not readily apparent 
from examination of the SED (Figure~\ref{fig.seds}).
There is sparse literature for HD 283815. 

In terms of spectroscopic parallax, HD 27659 has an implied 
distance based on our derivation of an A3 spectral type of 164 pc
which matches our criteria of membership with Taurus. 
For HD 283815 we did not obtain a spectrum, but using the literature spectral type
of A0 shows that it does not meet our distance criteria, as can be discerned 
from the underluminous SED of Figure~\ref{fig.seds} when a 140 pc distance is assumed.
In proper motion, however, the inverse is true with
HD 27659 showing proper motion inconsistent with Taurus, but HD 283815 a high probability proper motion member.  
Radial velocity measurements exist for neither star.  

Given the evidence, at present no strong statements can be made 
about the membership of either star.  They could be early type members 
with peculiar motions, or background stars reddened due to the Taurus cloud.
In both cases, there is some evidence for associated dust.
HD 27659 has just two SED points within the region showing infrared excess, 
and we hesitate to attempt fitting of a unique model to it.
HD 283815 shows neither significant infrared excess 
nor scattered light, and therefore cannot be modelled.

Of possible relevance is that lying only a quarter of a degree away from HD 27659 is HDE 283572, 
a Taurus member whose distance has been measured precisely by VLBA techniques. 
The implied distance between the two stars would be about 35 pc.
From Figure ~\ref{fig.distance}, the proximity with a high surface density part of the cloud is evident
and there is significant reddening implied for the further star HD 27659.

\section{SUMMARY \& CONCLUSION}\label{sec.summary}

Early-type stars illuminating infrared nebulae found in {\it Spitzer} IRAC/MIPS maps of the Taurus-Auriga Molecular Cloud complex
led us to carry out a more comprehensive search for early-type stars in this star-forming region.
We compiled a list of 329 candidate early-type stars (see Table~\ref{tab.allBstars}) towards Taurus from 
(i) previously-known O, B stars listed by SIMBAD, 
(ii) proposed B stars from \citeauthor{rebull2010} selected to have infrared excess and followed up spectroscopically,
(iii) stars from the \textit{Two Micron All Sky Survey} Point Source Catalog selected on the basis of photometric color, and 
(iv) early-type stars spectroscopically identified in the \textit{Sloan Digital Sky Survey}.
This set of stars was then tested against various membership criteria including distance, kinematics,
and age criteria.

First, we provided accurate spectral type information for about 20 stars which were spectroscopically followed 
up from our initial sample at the 200-inch Hale telescope at the Palomar Observatory.
This, along with the magnitudes available in the literature has helped in deriving 
spectroscopic parallax distances to the stars, accounting for extinction.
Notably, the presence of several diffuse interstellar bands is well-correlated with the
estimated distances; none of the stars we believe associated based on distance or kinematic
arguments with the Taurus star forming region exhibit these features, 
which are seen in only more distant stars.
Several of the spectra show emission lines which in two cases are associated with known young stellar
objects and in the other two cases appear associated with background Be stars also exhibiting
infrared excesses. 

Hipparcos parallaxes and spectroscopic parallaxes were used to select stars between 128 and 162 pc, allowing for error.
Proper motion membership was tested by calculating the $\chi^2$ probability using the proper motion of various comoving groups in Taurus \citep[defined by ][]{luhman2009}.
Radial velocity, wherever available, has been compared with that of the previously-known members of Taurus.
We have also specially considered all early-type sources illuminating nebulae, regardless
of whether they meet the distance and kinematic criteria.

Our final assessment of membership is shown in Table~\ref{tab.members}.
Through this work, we have found three stars of spectral type B, 
and two of spectral type A to be newly-appreciated members of Taurus.
Specifically, HR 1445 (HD 28929), $\tau$ Tau (HD 29763), 72 Tau (HD 28149), HD 31305, HD 26212 meet 
the kinematic and distance criteria while HD 27659, and HD 283815 show ambiguous 
kinematic and/or distance properties that make their membership plausible but not secure.
Additional or improved space velocity information for these and several other stars could confirm their membership.
These sources should be considered along with the currently accepted early-type members:
IC2087-IR (B5-like luminosity but heavily self-embedded source with a continuum-emission optical and infrared spectrum), 
the binary system V892 Tau (Elias 1; a B8-A6 self-embedded Herbig Ae/Be star 
with a near-equal brightness companion), the triple system HD 28867
(B8 $+2 \times$B9.5), AB Aur (A0e, a Herbig Ae/Be star), and 
HD 31648 (MWC 480; A2e, another Herbig Ae/Be star).
While HD 28867 is located to the south of the main Taurus-Auriga clouds and 
therefore not recovered in our search, the other known early type members were recovered,
although to varying degrees of security.  Notably, more than half -- but not all -- 
of the stars listed above distinguish themselves through illumination of optical or infrared nebulae. 
Furthermore, two-thirds -- but not all -- of the stars have mid-infrared excesses which may be due
to an extended nebula or to a more compact protoplanetary or later stage debris disk.

Among the stars with infrared reflection/scattered-light nebulae, 
two sources (HD 28149 = 72 Tau, and HD 29647) have been used 
as intrinsically bright background ``candles'' in studies of 
the molecular cloud's physical and chemical properties.
In the case of 72 Tau, the kinematics and distance are both consistent 
with those of previously-known members of Taurus.  While chance superposition 
with the Taurus cloud is possible, this main-sequence star was probably 
born near its present environment and is truly associated with the cloud. 
Our assessment of HD 29647 suggests an object at similar distance to 
though just behind the Taurus cloud 
(the same conclusion was reached by \citet{whittet2004})
and with different kinematics.  Both HD 29647 and 72 Tau 
have nebulae which appear to be morphologically similar 
to the striated nebulosity toward the Pleiades stars Merope and Maia.
The difference is that the Taurus nebulae under study here are much fainter 
and much less distinct at visible wavelengths than the Pleiades cases, 
probably due to the amount of foreground extinction 
or to the spectrum of incident radiation 
(and less likely due to the dust properties).

Another bright nebulous source, HD 282276 
(northernmost circle in Figure~\ref{fig.mosaic})
is likely a B-type star lying background to Taurus
and close to the distance of the Perseus star-forming region, 
although well east of it. 
The molecular material may also be at that further distance.

We modelled the nebular dust emission of a subset of interesting stars also showing infrared excess.
The dust temperatures found via this study, which account for the thermal dust emission peaks in the SEDs, are usually about 45 K.
In some cases (viz. HD 282276 and HD 29647) there is evidence of two dust components contributing to the infrared emission, where the 
higher-temperature component is between 100 K and 500 K.
The distance between the nebulae and the early-type stars can be estimated by balancing the heating and cooling rates.
However, the unknown composition, complicated morphology and the self-similar nature of the radiation transfer makes the problem very challenging.
Nevertheless, the infrared excess tells us that the star is sufficiently close to the nebula to heat it.
Kinematic information of the stars supplement this proximity information to help distinguish
between association with the cloud and chance superposition.

We call attention to the fact that we have doubled the number
of stars with spectral type A5 or earlier that can be associated with
the Taurus Molecular Cloud.  This includes the earliest
type star claimed yet as a member of Taurus: the B3V star $\tau$ Tau.
Nevertheless, the cloud still seems to be a factor of two short 
in early type members, considering a standard log-normal form 
for the initial mass function and given its low mass and brown dwarf population 
that numbers at least 350.

\acknowledgments{\emph{
We acknowledge the contributions of Caer McCabe, Alberto Noreiga-Crespo,
Sean Carey, Karl Stapelfeldt, Tim Brooke, Tracy Huard, and Misato Fukagawa in the 
production of the Spitzer maps that inspired this analysis.
We thank John Carpenter, Varun Bhalerao, Eric Mamajek 
for their various suggestions and helpful advice. 
This research has made use of: 
the SIMBAD and VizieR online database services, IRAF  
distributed by the National Optical Astronomy Observatories, 
which are operated by the AURA under cooperative agreement with NSF,
and the DUSTY code developed by Gary Ferland. 
}

\facility{FACILITIES: Spitzer, 2MASS, SDSS, WISE, Palomar 200-inch}

\bibliography{taurus_bstars}

\begin{table}
\centering
\tabletypesize{\footnotesize}
\caption{Number of O,B stars known to SIMBAD towards Taurus (central rectangle)
and neighboring regions of equivalent area in coordinates of
Galactic latitude and Galactic longitude.}
\label{tab.bstar_bias}
\begin{tabular}{cc|cc|cc|cc|c} 
\multirow{2}{*}{0$\degr$} & & & & & & & & \\
\cline{2-9}
&  & \multicolumn{2}{c|}{\multirow{2}{*}{304}} & \multicolumn{2}{c|}{\multirow{2}{*}{513}} & \multicolumn{2}{c|}{\multirow{2}{*}{342}} & \\
\multirow{2}{*}{-10$\degr$} & & & & & & & \\
\cline{2-8} 
&  & \multicolumn{2}{c|}{\multirow{2}{*}{100}} & \multicolumn{2}{c|}{\multirow{2}{*}{117}} & \multicolumn{2}{c|}{\multirow{2}{*}{122}} & \\
\multirow{2}{*}{-20$\degr$} & & & & & & & \\
\cline{2-8}
&  & \multicolumn{2}{c|}{\multirow{2}{*}{32}} & \multicolumn{2}{c|}{\multirow{2}{*}{46}} & \multicolumn{2}{c|}{\multirow{2}{*}{33}} & \\
\multirow{2}{*}{-30$\degr$} & & & & & & & \\
\cline{2-8}
& & & & & & & & \\
& \multicolumn{2}{c}{195$\degr$} & \multicolumn{2}{c}{180$\degr$} & \multicolumn{2}{c}{165$\degr$} & \multicolumn{2}{c}{150$\degr$}\\
\end{tabular}
\end{table}

\begin{table}[htp]
\centering
\tabletypesize{\footnotesize}
\caption{Distances to known Taurus members measured through VLBI techniques.}
\label{tab.distance}
\begin{tabular}{lllll}
\hline \hline
Star       & $\alpha_{J2000}$ & $\delta_{J2000}$        & Distance & Ref.\\
           & (h,m,s)          & (\degr,\arcmin,\arcsec) & (pc)     &    \\
\hline 
HDE 283572 & 04 21 58.847     & 28 18 06.51             & $128.5\pm0.6$ & 1 \\
V773 Tau   & 04 14 12.922     & 28 12 12.30             & $131.4\pm2.4$ & 2 \\ 
V1023 Tau (Hubble 4) & 04 18 47.037     &  28 20 07.32  & $132.8\pm0.5$ & 1 \\
T Tau      & 04 21 59.434     & 19 32 06.42             & $146.7\pm0.6$ & 3 \\
HP Tau/G2  & 04 35 54.152     & 22 54 13.46             & $161.4\pm0.9$ & 4 \\
\hline
\multicolumn{5}{p{3.3in}}{Refs: (1) \cite{torres2007}, (2) \cite{torres2012}, (3) \cite{loinard2007}, (4) \cite{torres2009}}
\end{tabular}
\end{table}
\clearpage
\LongTables
\begin{deluxetable*}{p{0.1cm} p{2.2cm} p{1cm}p{1cm} p{0.8cm}r rrr l ll}
\tabletypesize{\tiny}
\tablecaption{List of candidate early-type stars with
sections separating various selection methods. 
}
\tablehead{\colhead{B\#} & \colhead{Star} & \colhead{$\alpha$} & \colhead{$\delta$} & \colhead{$d_{_{\rm HIP}}$} & \colhead{$d_{_{\rm SPEC}}$} & \colhead{$\mu_{\alpha} cos\delta$} & \colhead{$\mu_{\delta}$} &\colhead{RV} & \colhead{SpT} & \multicolumn{2}{c}{Candidate?} \\
\cline{11-12}
\colhead{} & \colhead{} & \multicolumn{2}{c}{(ICRS 2000, degrees)} & \multicolumn{2}{c}{(pc)} &  \multicolumn{2}{c}{(mas/yr)} & \colhead{(km s$^{-1}$)} & \colhead{} & \colhead{$d$} & \colhead{$\mu/RV$}
\label{tab.allBstars}
}
\startdata
\\
\multicolumn{12}{c}{O,B, and A0 stars from SIMBAD}\\
\\\hline
1 & HD 25063 & 60.036389 & 29.710836 &\nodata & $195 \pm 5$ & $3.7 \pm 0.9 $ & $-6.2 \pm 0.8 $ &\nodata & B9 & N & N/? \\
2 & HD 25201 & 60.236744 & 23.201484 & $332_{-103}^{+274}$ & $189 \pm 5$ & $-34.6 \pm 15.2$ & $-9.2 \pm 15.2$ & $13.8 \pm 1.1$ & B9V & N/N & N/Y \\
3 & HD 281490  & 60.871667 & 30.970278 &\nodata & $792 \pm 123$ & $13.0 \pm 1.6 $ & $-7.3 \pm 1.7$ &\nodata & B9/A3 & N & N/? \\
4 & BD+23 607 & 60.936736 & 23.620942 &\nodata & $643 \pm 94$ & $5.1 \pm 1.3 $ & $-9.7 \pm 1.3 $ &\nodata & A0V & N & N/? \\
5 & HD 25487 & 60.976316 & 28.125973 & $301_{-76}^{+153}$ & $342 \pm 32$ & $1.9 \pm 1.7 $ & $-18.1 \pm 1.5 $ &\nodata & B8Ve+K0IV & N/N & Y/? \\
6 & HD 283304 & 61.226290 & 27.545890 &\nodata & $739 \pm 69$ & $-2.7 \pm 1.3 $ & $-2.2 \pm 1.3 $ &\nodata & B8V Si & N & N/? \\
7 & HD 284130 & 61.487807 & 23.241767 &\nodata & $943 \pm 20$ & $-2.4 \pm 1.0 $ & $5.4 \pm 1.0 $ &\nodata & B8 & N & N/? \\
8 & HD 283363 & 61.780708 & 28.560528 &\nodata & $528 \pm 25$ & $6.4 \pm 1.5 $ & $-7.6 \pm 1.6 $ &\nodata & B9 & N & N/? \\
9 & HD 281679 & 62.320792 & 30.775956 & $185_{-44}^{+85}$ & $373 \pm 11$ & $-8.1 \pm 1.5 $ & $3.6 \pm 1.2 $ &\nodata & B8 & Y/N & N/? \\
10 & HD 284179 & 62.785610 & 22.825372 &\nodata & $745 \pm 41$ & $-1.2 \pm 1.2 $ & $-4.3 \pm 1.2 $ &\nodata & B8 & N & N/? \\
11 & HD 283507 & 62.969714 & 24.837432 &\nodata & $405 \pm 18$ & $-3.0 \pm 1.2 $ & $-4.2 \pm 1.3 $ &\nodata & B9 & N & N/? \\
12 & HD 283449 & 63.112000 & 27.876861 &\nodata & $1613 \pm 652$ & $-2.3 \pm 1.8 $ & $-4.8 \pm 1.8$ &\nodata & A0V & N & N/? \\
13 & HD 26571 & 63.213520 & 22.413458 & $316_{-66}^{+113}$ & $159 \pm 1$ & $-3.0 \pm 0.5 $ & $-10.9 \pm 0.5 $ & $9.3 \pm 3$ & B8III & N/Y & N/Y \\
14 & HD 281815 & 63.852452 & 29.365851 &\nodata & $548 \pm 57$ & $1.0 \pm 1.6 $ & $-10.9 \pm 1.6 $ &\nodata & B8 & N & N/? \\
15 & HD 281818 & 64.079170 & 29.256597 &\nodata & $749 \pm 67$ & $7.1 \pm 1.3 $ & $-12.2 \pm 1.3 $ &\nodata & B8 & N & N/? \\
16 & HD 284228 & 64.085078 & 23.841866 &\nodata & $839 \pm 32$ & $13.5 \pm 1.2 $ & $-6.0 \pm 1.2 $ &\nodata & B5 & N & N/? \\
17 & HD 283553 & 64.488019 & 24.577938 & $248_{-62}^{+122}$ & $465 \pm 15$ & $-0.8 \pm 0.7 $ & $-7.9 \pm 0.7 $ &\nodata & B8 & N/N & N/? \\
18 & HD 281920 & 64.569998 & 29.807023 &\nodata & $777 \pm 69$ & $6.5 \pm 1.2 $ & $-6.0 \pm 1.0 $ &\nodata & B5 & N & N/? \\
19 & V892 Tau & 64.669250 & 28.320972 &\nodata & $1697 \pm 1548$ & $10.9 \pm 5.1 $ & $-29.7 \pm 5.1 $ &\nodata & $\sim$A0V & Y: & Y/? \\
20 & HD 27405 & 65.126585 & 25.827451 & $229_{-44}^{+71}$ & $242 \pm 5$ & $6.7 \pm 1.4 $ & $-5.2 \pm 1.5 $ &\nodata & B9 & N/N & N/? \\
21 & HD 283567 & 65.137783 & 28.652376 &\nodata & $655 \pm 44$ & $-2.0 \pm 1.5 $ & $-4.9 \pm 1.6 $ &\nodata & B9 & N & N/? \\
22 & HD 27638 & 65.645598 & 25.629314 & $82_{-6}^{+7}$ & $104 \pm 2$ & $17.4 \pm 2.0 $ & $-13.3 \pm 1.9 $ & $15.3 \pm 3.4$ & B9V & N/N & N/Y \\
23 & HD 27778 & 65.999008 & 24.300992 & $223_{-44}^{+73}$ & $238 \pm 5$ & $8.0 \pm 1.5 $ & $-12.7 \pm 1.5 $ &\nodata & B3V & N/N & N/? \\
24 & HD 284427 & 66.341599 & 23.389317 &\nodata & $577 \pm 56$ & $-5.5 \pm 1.2 $ & $-4.4 \pm 1.2 $ &\nodata & B9V & N & N/? \\
25 & HD 28149 & 66.822698 & 22.996337 & $127_{-11}^{+13}$ & $161 \pm 3$ & $0.4 \pm 0.5 $ & $-14.4 \pm 0.5 $ & $7.3 \pm 2.6$ & B7V & Y/Y & Y/Y \\
26 & HD 284479 & 67.311975 & 22.478510 &\nodata & $1153 \pm 158$ & $2.6 \pm 1.5 $ & $-6.9 \pm 1.5$ &\nodata & B8V & N & N/? \\
27 & HD 282151 & 67.424917 & 30.684083 &\nodata & $771 \pm 15$ & $-3.1 \pm 1.6 $ & $-1.3 \pm 1.6 $ &\nodata & B9V & N & N/? \\
28 & HD 28482 & 67.593513 & 23.588859 & $301_{-64}^{+110}$ & $204 \pm 1$ & $18.0 \pm 0.6 $ & $0.5 \pm 0.6 $ &\nodata & B8III & N/N & N/? \\
29 & HD 284487 & 67.879371 & 22.220857 &\nodata & $1102 \pm 177$ & $3.4 \pm 1.4 $ & $-2.5 \pm 1.4$ &\nodata & B9V & N & N/? \\
30 & HD 282240 & 68.022275 & 30.967710 &\nodata & $824 \pm 31$ & $4.1 \pm 1.3 $ & $-6.3 \pm 1.4$ &\nodata & B9V & N & N/? \\
31 & HD 283677 & 68.045575 & 29.015574 & $344_{-102}^{+248}$ & $541 \pm 15$ & $0.0 \pm 0.8 $ & $-8.9 \pm 0.7 $ &\nodata & B5V & N/N & N/? \\
32 & CoKu HK Tau G1 & 68.173333 & 24.317778 &\nodata & $5976 \pm 4466$ & $37.7 \pm 4.5 $ & $-11.1 \pm 4.5$ & $20 \pm 4.5$ & B2 & N & N/Y \\
33 & HD 282276 & 68.267637 & 29.363863 &\nodata & $422 \pm 52$ & $2.0 \pm 1.6 $ & $-12.5 \pm 1.7 $ &\nodata & B8V & N & N/? \\
34 & 2MASS J04335319+2414080 & 68.471667 & 24.235583 &\nodata & $1244 \pm 177$ & $3.7 \pm 1.6 $ & $-1.4 \pm 1.6$ &\nodata & B8 & N & N/? \\
35 & HD 282278 & 68.585583 & 29.297556 &\nodata & $439 \pm 30$ & $-2.3 \pm 1.3 $ & $-3.4 \pm 1.4 $ &\nodata & B9 & N & N/? \\
36 & HD 28929 & 68.658299 & 28.961151 & $143_{-16}^{+20}$ & $157 \pm 3$ & $1.3 \pm 2.2 $ & $-23.2 \pm 2.0 $ & $12.6 \pm 2.2$ & B8V Hg-Mn & Y/Y & Y/Y \\
37 & HD 283701 & 68.729070 & 27.203107 &\nodata & $307 \pm 3$ & $1.9 \pm 1.3 $ & $-11.6 \pm 1.3 $ &\nodata & B8III & N & N/? \\
38 & 2MASS J04353218+2427069 & 68.884125 & 24.451944 &\nodata & $1104 \pm 201$ & $-1.8 \pm 4.6 $ & $-4.4 \pm 4.6$ &\nodata & B8 & N & N/? \\
39 & HD 283740 & 69.260698 & 28.011207 &\nodata & $778 \pm 48$ & $3.0 \pm 1.5 $ & $2.4 \pm 1.5 $ &\nodata & B8V & N & N/? \\
40 & HD 29259 & 69.438853 & 30.407638 & $667_{-298}^{+2782}$ & $312 \pm 10$ & $0.6 \pm 1.0 $ & $-9.1 \pm 0.8 $ &\nodata & B9 & N/N & N/? \\
41 & HD 282380 & 69.715330 & 30.449740 &\nodata & $287 \pm 50$ & $-1.4 \pm 1.5 $ & $-1.3 \pm 1.5 $ &\nodata & B8 & N & N/? \\
42 & HD 284583 & 69.777811 & 22.712064 &\nodata & $1017 \pm 25$ & $1.0 \pm 1.3 $ & $-0.2 \pm 1.3 $ &\nodata & B5 & N & N/? \\
43 & HD 29450 & 69.806437 & 22.652255 & $429_{-132}^{+340}$ & $345 \pm 11$ & $0.9 \pm 0.8 $ & $-1.3 \pm 0.7 $ & $-10.9 \pm 2.6$ & B9 & N/N & N/N \\
44 & JH 225 & 69.916538 & 25.342850 &\nodata & $540 \pm 125$ & $-11.6 \pm 4.6 $ & $-8.3 \pm 4.6 $ &\nodata & B9 & N & N/? \\
45 & HD 283794 & 69.964024 & 27.188951 &\nodata & $428 \pm 12$ & $5.1 \pm 1.2 $ & $-8.4 \pm 1.3 $ &\nodata & B9V & N & N/? \\
46 & IC 2087-IR & 69.982292 & 25.750556 &\nodata &\nodata & $1.6 \pm 5.9 $ & $-19.7 \pm 5.9$ &\nodata & B5 & ? & Y/? \\
47 & HD 283772 & 70.246863 & 27.990404 &\nodata & $519 \pm 11$ & $0.0 \pm 1.3 $ & $-5.2 \pm 1.3 $ &\nodata & B9V & N & N/? \\
48 & HD 29647 & 70.283524 & 25.992765 & $177_{-29}^{+43}$ & $160 \pm 1$ & $12.8 \pm 0.9 $ & $-9.7 \pm 0.7 $ &\nodata & B9III Hg-Mn & Y/Y & N/? \\
49 & HD 29681 & 70.349520 & 22.676800 &\nodata & $699 \pm 18$ & $0.4 \pm 1.1 $ & $0.1 \pm 1.1 $ &\nodata & B8V & N & N/? \\
50 & HD 283809 & 70.353006 & 25.913469 &\nodata & $695 \pm 230$ & $2.5 \pm 1.6 $ & $-1.4 \pm 1.7 $ &\nodata & B1.5V & N & N/? \\
51 & HD 29763 & 70.561257 & 22.956926 & $123_{-11}^{+13}$ & $137 \pm 9$ & $0.6 \pm 0.2 $ & $-17.4 \pm 0.2 $ & $12.3 \pm 4.1$ & B3V & Y/Y & Y/Y \\
52 & HD 283800 & 70.863805 & 27.026948 &\nodata & $564 \pm 42$ & $1.3 \pm 1.5 $ & $-9.1 \pm 1.5 $ &\nodata & B8 & N & N/? \\
53 & HD 29935 & 70.975003 & 22.944410 & $90_{-9}^{+11}$ & $175 \pm 2$ & $-0.5 \pm 0.7 $ & $-15.8 \pm 0.6 $ & $32 \pm 4.8$ & A0V & N/N & Y/N \\
54 & HD 283805 & 71.301213 & 26.453282 &\nodata & $846 \pm 17$ & $-2.2 \pm 1.4 $ & $-4.1 \pm 1.4 $ &\nodata & B8V/A3 & N & N/? \\
55 & HD 30122 & 71.426948 & 23.627996 & $216_{-29}^{+40}$ & $290 \pm 1$ & $9.8 \pm 0.6 $ & $-17.3 \pm 0.5 $ & $23.2 \pm 2$ & B5III & N/N & Y/N \\
56 & HD 282430 & 71.644099 & 30.615554 &\nodata & $918 \pm 97$ & $-3.5 \pm 1.7 $ & $-5.6 \pm 1.7 $ &\nodata & B5 & N & N/? \\
57 & HD 282431 & 71.654007 & 30.404109 &\nodata & $400 \pm 16$ & $5.1 \pm 1.6 $ & $-9.5 \pm 1.5 $ &\nodata & B9/A0 & N & N/? \\
58 & HD 282485 & 71.658884 & 29.317718 &\nodata & $483 \pm 16$ & $4.1 \pm 1.5 $ & $-5.6 \pm 1.5 $ &\nodata & B9V & N & N/? \\
59 & HD 283851 & 71.678804 & 27.261715 &\nodata & $637 \pm 46$ & $-3.8 \pm 1.2 $ & $-3.7 \pm 1.2 $ &\nodata & B9V & N & N/? \\
60 & HD 283836 & 71.729035 & 28.225371 &\nodata & $336 \pm 29$ & $0.0 \pm 1.3 $ & $-3.7 \pm 1.3 $ &\nodata & B9 & N & N/? \\
61 & HD 283854 & 71.765792 & 26.765833 &\nodata & $809 \pm 29$ & $-1.0 \pm 1.4 $ & $-4.7 \pm 1.4$ &\nodata & B9V/A7 & N & N/? \\
62 & HD 283845 & 71.967880 & 27.744442 &\nodata & $750 \pm 23$ & $2.9 \pm 1.5 $ & $-4.7 \pm 1.5 $ &\nodata & B3V & N & N/? \\
63 & HD 30378 & 72.094763 & 29.773003 & $196_{-33}^{+49}$ & $239 \pm 5$ & $6.1 \pm 0.6 $ & $-25.6 \pm 0.6 $ & $21.6 \pm 3.5$ & B9.5V & N/N & Y/N \\
64 & HD 30675 & 72.716687 & 28.314066 & $368_{-109}^{+269}$ & $332 \pm 11$ & $7.2 \pm 1.2 $ & $-16.6 \pm 0.8 $ &\nodata & B3V & N/N & N/? \\
65 & HD 283875 & 72.795560 & 25.622469 &\nodata & $530 \pm 63$ & $2.1 \pm 1.7 $ & $-1.7 \pm 1.6 $ &\nodata & B8V & N & N/? \\
66 & HD 282537 & 72.898060 & 30.021865 &\nodata & $752 \pm 16$ & $-4.1 \pm 1.6 $ & $-9.5 \pm 1.6 $ &\nodata & B8 & N & N/? \\
67 & HD 283920 & 73.021685 & 26.917266 &\nodata & $497 \pm 3$ & $-0.8 \pm 1.7 $ & $-11.3 \pm 1.7 $ &\nodata & B7III & N & N/? \\
68 & HD 282548 & 73.297292 & 29.020944 &\nodata & $999 \pm 63$ & $-4.7 \pm 1.7 $ & $-3.4 \pm 1.7$ &\nodata & B9 & N & N/? \\
69 & HD 283952 & 73.318183 & 24.714376 &\nodata & $765 \pm 23$ & $2.2 \pm 1.3 $ & $-3.7 \pm 1.3$ &\nodata & B9V & N & N/? \\
70 & HD 284874 & 73.364583 & 22.251722 &\nodata & $793 \pm 37$ & $2.6 \pm 1.0 $ & $-1.6 \pm 1.0 $ &\nodata & B8 & N & N/? \\
71 & HD 31120 & 73.539369 & 23.066227 &\nodata & $374 \pm 28$ & $-4.8 \pm 1.1 $ & $-10.8 \pm 1.1 $ &\nodata & B9V & N & N/? \\
72 & HD 283932 & 73.677168 & 25.729056 &\nodata & $442 \pm 46$ & $0.7 \pm 1.5 $ & $-7.0 \pm 1.4 $ &\nodata & B8V & N & N/? \\
73 & HD 282653 & 73.945167 & 29.333056 &\nodata & $670 \pm 26$ & $-4.4 \pm 1.7 $ & $-9.3 \pm 1.7 $ &\nodata & B9 & N & N/? \\
74 & HD 31353 & 74.029726 & 24.004623 &\nodata & $294 \pm 12$ & $4.6 \pm 1.2 $ & $-8.8 \pm 1.3 $ &\nodata & B9 & N & N/? \\
75 & HD 282635 & 74.120904 & 29.994960 & $172_{-42}^{+81}$ & $490 \pm 21$ & $2.0 \pm 1.3 $ & $-5.4 \pm 1.1 $ &\nodata & B8 & Y/N & N/? \\
76 & HD 282617 & 74.150050 & 30.882917 &\nodata & $673 \pm 51$ & $-6.8 \pm 1.7 $ & $-6.7 \pm 1.7 $ &\nodata & B8 & N & N/? \\
77 & V722 Tau & 74.205683 & 27.718514 &\nodata & $2304 \pm 583$ & $1.0 \pm 1.8 $ & $-2.6 \pm 1.8$ &\nodata & B3e & N & N/? \\
78 & HD 283968 & 74.205773 & 24.043778 &\nodata & $643 \pm 103$ & $2.3 \pm 1.5 $ & $-1.9 \pm 1.5$ &\nodata & B9V & N & N/? \\
79 & HD 284993 & 74.302100 & 22.159576 &\nodata & $790 \pm 44$ & $2.4 \pm 1.2 $ & $-3.9 \pm 1.2 $ &\nodata & B9 & N & N/? \\
80 & HD 282633 & 74.362179 & 30.174772 &\nodata & $581 \pm 48$ & $-2.4 \pm 1.3 $ & $2.8 \pm 1.3 $ &\nodata & B8V & N & N/? \\
81 & HD 283971 & 74.382746 & 28.759850 &\nodata & $859 \pm 240$ & $-3.5 \pm 1.7 $ & $-5.4 \pm 1.5 $ &\nodata & B9V & N & N/? \\
82 & HD 284006 & 74.532658 & 26.298360 &\nodata & $358 \pm 30$ & $-3.1 \pm 1.5 $ & $-7.4 \pm 1.4 $ &\nodata & B9 & N & N/? \\
83 & HD 282754 & 74.593708 & 29.722028 &\nodata & $1033 \pm 128$ & $-1.0 \pm 1.7 $ & $-2.3 \pm 1.7$ &\nodata & B9 & N & N/? \\
84 & BD+30 748 & 74.639620 & 30.697656 &\nodata & $1202 \pm 115$ & $1.5 \pm 1.3 $ & $-2.5 \pm 1.3 $ & $7.2 \pm 1.4$ & B1.5V & N & N/N \\
85 & HD 284941 & 74.674310 & 23.585106 &\nodata & $1011 \pm 195$ & $-0.1 \pm 1.2 $ & $-3.5 \pm 1.2$ &\nodata & B9 & N & N/? \\
86 & HD 31679 & 74.719814 & 24.495712 & $379_{-137}^{+498}$ & $378 \pm 28$ & $2.1 \pm 0.9 $ & $-4.4 \pm 0.8 $ &\nodata & B5 & N/N & N/? \\
87 & HD 284012 & 74.898442 & 25.807481 & $258_{-60}^{+113}$ & $255 \pm 26$ & $0.7 \pm 1.1 $ & $-7.3 \pm 0.9 $ & $11 \pm 2.5$ & B8 & N/N & N/Y \\
88 & HD 31806 & 74.973961 & 27.325604 & $138_{-30}^{+52}$ & $232 \pm 6$ & $14.8 \pm 1.6 $ & $-24.6 \pm 1.7 $ &\nodata & B7V & Y/N & N/? \\
\hline
\\
\multicolumn{12}{c}{Additional B and early A stars proposed by Rebull et al. 2010 from infrared excess}\\
\\\hline
89 & HD 27659 & 65.727746 & 28.398614 &\nodata & $164 \pm 10$ & $-23.7 \pm 1.5 $ & $-17.8 \pm 1.4 $ &\nodata & A3V & Y & N/? \\
90 & HD 27923 & 66.329861 & 23.788020 & $194_{-49}^{+101}$ & $277 \pm 5$ & $5.9 \pm 1.1 $ & $-8.6 \pm 1.1 $ &\nodata & B9V & Y/N & N/? \\
91 & HD 283637 & 66.495250 & 27.617028 &\nodata & $855 \pm 279$ & $-3.6 \pm 1.7 $ & $-6.8 \pm 1.7 $ &\nodata & B9.5V & N & N/? \\
92 & 2MASS J04285940+2736254 & 67.247550 & 27.607081 &\nodata & $664 \pm 147$ & $1.7 \pm 2.3 $ & $-15.5 \pm 2.4$ &\nodata & A4III & N & Y/? \\
93 & 2MASS J04313313+2928565 & 67.888075 & 29.482378 &\nodata & $639 \pm 144$ & $-6.0 \pm 5.1 $ & $0.3 \pm 5.1$ &\nodata & A1V & N & N/? \\
94 & HD 284530 & 68.582858 & 23.447141 &\nodata & $347 \pm 9$ & $7.1 \pm 1.2 $ & $-15.5 \pm 1.2 $ &\nodata & B9.5V & N & Y/? \\
95 & HD 283751 & 69.353563 & 27.155458 &\nodata & $1010 \pm 85$ & $6.1 \pm 1.4 $ & $-1.9 \pm 1.4 $ &\nodata & B5Ve & N & N/? \\
96 & HD 283815 & 70.671592 & 24.688295 &\nodata & $268 \pm 40$ & $8.7 \pm 1.2 $ & $-20.4 \pm 1.2 $ &\nodata & A0 & N & Y/? \\
\hline
\\
\multicolumn{12}{c}{Additional early type candidates selected from 2MASS photometry}\\
\\\hline
97 & HD 25111 & 60.054439 & 23.149068 &\nodata &\nodata & $5.4 \pm 1.1$ & $-4.6 \pm 1.1$ &\nodata & A & ? & N/? \\
98 & HD 283286 & 60.959556 & 28.404501 &\nodata & $420 \pm 21$ & $-6.9 \pm 2.7$ & $-5.19 \pm 2.37$ &\nodata & A2 & N & N/? \\
99 & 04040178+2715454 & 61.007451 & 27.262593 &\nodata &\nodata & $-8.1 \pm 5.6$ & $-5.16 \pm 4.98$ &\nodata & ? & ? & N/? \\
100 & HD 25554 & 61.160276 & 30.884204 & $207_{-42}^{+70}$ & $280 \pm 9$ & $3 \pm 1$ & $1.97 \pm 0.69$ &\nodata & A0 & N/N & N/? \\
101 & HD 25620 & 61.208024 & 23.509276 &\nodata & $176 \pm 11$ & $2.7 \pm 1.1$ & $-0.92 \pm 1.01$ &\nodata & F0 & N & N/? \\
102 & HD 25626 & 61.301075 & 27.609715 & $202_{-38}^{+62}$ & $191 \pm 3$ & $16.7 \pm 1$ & $-26.23 \pm 0.89$ & $2 \pm 1.8$ & A2 & N/N & N/N \\
103 & HD 25694 & 61.413234 & 28.395374 &\nodata & $347 \pm 15$ & $1.6 \pm 1.4$ & $-5.01 \pm 1.32$ &\nodata & A0 & N & N/? \\
104 & 04055962+2956381 & 61.498439 & 29.943959 &\nodata &\nodata & $0.1 \pm 1.7$ & $-5.63 \pm 1.47$ &\nodata & ? & ? & N/? \\
105 & HD 283346 & 61.550881 & 25.362189 &\nodata & $373 \pm 32$ & $5.7 \pm 1.1$ & $-9.94 \pm 0.99$ &\nodata & A2 & N & N/? \\
106 & HD 26212 & 62.432002 & 24.072944 & $123_{-14}^{+17}$ & $115 \pm 4$ & $2.4 \pm 0.8$ & $-18.81 \pm 0.55$ & $20.3 \pm 3.9$ & A5V & Y/N & Y/Y \\
107 & HD 284189 & 62.614432 & 22.256375 &\nodata & $289 \pm 11$ & $5.4 \pm 1.1$ & $-12.86 \pm 1.02$ &\nodata & A3 & N & N/? \\
108 & HD 284191 & 62.819075 & 22.248622 &\nodata & $374 \pm 30$ & $3.6 \pm 1.2$ & $-10.64 \pm 1.11$ &\nodata & A0 & N & N/? \\
109 & HD 283467 & 62.988010 & 27.168309 &\nodata & $648 \pm 8$ & $2.1 \pm 1.5$ & $-2.22 \pm 1.33$ &\nodata & A0 & N & N/? \\
110 & 04115969+3046563 & 62.998756 & 30.782361 &\nodata &\nodata & $1.4 \pm 1.7$ & $-4.64 \pm 1.46$ &\nodata & ? & ? & N/? \\
111 & 04124695+2902138 & 63.195599 & 29.037144 &\nodata &\nodata & $0.7 \pm 1.7$ & $-1.05 \pm 1.49$ &\nodata & ? & ? & N/? \\
112 & HD 283457 & 63.356642 & 27.357025 &\nodata &\nodata & $-3.9 \pm 1.2$ & $-6.57 \pm 1.07$ &\nodata & \nodata & ? & N/? \\
113 & HD 283503 & 63.511797 & 24.813039 &\nodata & $404 \pm 68$ & $3.2 \pm 1.3$ & $-17.16 \pm 1.18$ &\nodata & A7 & N & Y/? \\
114 & FM Tau & 63.556610 & 28.213672 &\nodata & $808 \pm 710$ & $4.7 \pm 2.4$ & $-29.78 \pm 2.11$ &\nodata & K3 & Y & Y/? \\
115 & CW Tau & 63.570878 & 28.182714 &\nodata & $236 \pm 188$ & $18 \pm 5.1$ & $-24.94 \pm 4.5$ &\nodata & K5V:e... & Y & Y/? \\
116 & HD 281820 & 64.102354 & 29.149663 &\nodata &\nodata & $12.9 \pm 1.6$ & $-14.5 \pm 1.48$ &\nodata & A & ? & N/? \\
117 & 04171672+2518050 & 64.319686 & 25.301403 &\nodata &\nodata & $-1.1 \pm 2.3$ & $-4.25 \pm 2.08$ &\nodata & ? & ? & N/? \\
118 & DD Tau & 64.629677 & 28.274725 &\nodata & $396 \pm 326$ & $6.7 \pm 5.1$ & $-25.28 \pm 4.49$ &\nodata & K6V:e... & Y & Y/? \\
119 & HD 284308 & 64.646193 & 22.701229 &\nodata & $429 \pm 40$ & $1.6 \pm 1.2$ & $-5.07 \pm 1.11$ &\nodata & A3 & N & N/? \\
120 & FR Tau & 64.897754 & 28.456022 &\nodata &\nodata & $5.9 \pm 5.5$ & $-25.32 \pm 4.84$ &\nodata & \nodata & ? & Y/? \\
121 & HD 283568 & 65.094954 & 28.737826 &\nodata & $688 \pm 78$ & $-5.9 \pm 1.6$ & $-6.58 \pm 1.4$ &\nodata & A0 & N & N/? \\
122 & HD 284383 & 65.391471 & 22.119671 &\nodata & $684 \pm 28$ & $-2.9 \pm 1.3$ & $-2.22 \pm 1.2$ &\nodata & A0V & N & N/? \\
123 & HD 283571 & 65.489190 & 28.443195 & $134_{-30}^{+55}$ & $69 \pm 29$ & $9.1 \pm 1.6$ & $-25.94 \pm 1.41$ & $24.3 \pm 1.9$ & F8V:e... & Y/N & Y/N \\
124 & 04233478+2804292 & 65.894960 & 28.074773 &\nodata &\nodata & $-8.8 \pm 2.4$ & $2.56 \pm 2.12$ &\nodata & ? & ? & N/? \\
125 & FU Tau & 65.897490 & 25.050747 &\nodata & $118 \pm 103$ & $7.3 \pm 4.5$ & $-24.01 \pm 4.08$ &\nodata & M7.25 & Y & Y/? \\
126 & HD 27787 & 66.113276 & 30.124556 &\nodata & $380 \pm 33$ & $-3 \pm 1.5$ & $-7.35 \pm 1.3$ & $-24$ & A0V & N & N/N \\
127 & 2MASS J04244457+2610141 & 66.185733 & 26.170592 &\nodata & $254 \pm 191$ & $9.8 \pm 4.5$ & $-21.09 \pm 4.04$ &\nodata & M0 & Y & Y/? \\
128 & JH 15 & 66.209223 & 26.482108 &\nodata &\nodata & $-1 \pm 2.2$ & $-3.94 \pm 1.97$ &\nodata & \nodata & ? & N/? \\
129 & 04262631+2742225 & 66.609620 & 27.706273 &\nodata &\nodata & $3.8 \pm 1.8$ & $-3.9 \pm 1.59$ &\nodata & ? & ? & N/? \\
130 & HD 283625 & 66.715248 & 28.953061 &\nodata & $527 \pm 16$ & $1.1 \pm 1.4$ & $-5.86 \pm 1.23$ &\nodata & A1V & N & N/? \\
131 & DG Tau & 66.769586 & 26.104481 &\nodata &\nodata & $5.6 \pm 1.7$ & $-19.13 \pm 1.53$ &\nodata & GV:e... & ? & Y/? \\
132 & 04273688+2936338 & 66.903701 & 29.609408 &\nodata &\nodata & $-4.6 \pm 1.7$ & $-5.56 \pm 1.48$ &\nodata & ? & ? & N/? \\
133 & DH Tau & 67.423219 & 26.549413 &\nodata & $244 \pm 198$ & $9.3 \pm 4.5$ & $-21.56 \pm 4.03$ &\nodata & M0.5V:e & Y & Y/? \\
134 & IQ Tau & 67.464609 & 26.112596 &\nodata & $88 \pm 56$ & $-2.2 \pm 4.8$ & $-16.88 \pm 4.04$ &\nodata & M2 & Y & Y/? \\
135 & DK Tau & 67.684336 & 26.023512 &\nodata & $76 \pm 53$ & $5.5 \pm 2.4$ & $-14.56 \pm 2.16$ &\nodata & M0V:e & Y & Y/? \\
136 & 2MASS J04305028+2300088 & 67.709603 & 23.002469 &\nodata &\nodata & $3.9 \pm 4.5$ & $-40.32 \pm 4.14$ &\nodata & F1 & N & N/? \\
137 & HD 28697 & 68.090832 & 25.185490 & $109_{-11}^{+14}$ & $150 \pm 2$ & $-2 \pm 0.6$ & $-25.88 \pm 0.54$ &\nodata & A2 & N/Y & N/? \\
138 & HD 282267 & 68.099987 & 30.118652 &\nodata & $342 \pm 7$ & $0.4 \pm 1.5$ & $-12.46 \pm 1.38$ &\nodata & A2V & N & N/? \\
139 & HD 283688 & 68.108755 & 28.317406 &\nodata & $145 \pm 6$ & $-3 \pm 1.4$ & $-10.39 \pm 1.23$ &\nodata & F5 & Y & N/? \\
140 & FZ Tau & 68.132328 & 24.334158 &\nodata &\nodata & $1.6 \pm 4.5$ & $-27.61 \pm 4.1$ &\nodata & \nodata & ? & Y/? \\
141 & HD 284484 & 68.192963 & 22.103749 &\nodata & $287 \pm 36$ & $2.2 \pm 1.5$ & $-5.84 \pm 1.39$ &\nodata & A5 & N & N/? \\
142 & 2MASS J04325030+2942395 & 68.209596 & 29.710960 &\nodata &\nodata & $1.2 \pm 5.1$ & $-2.52 \pm 4.43$ &\nodata & \nodata & ? & N/? \\
143 & 2MASS J04325316+2948046 & 68.221551 & 29.801318 &\nodata &\nodata & $-0.1 \pm 6.7$ & $-5.81 \pm 5.81$ &\nodata & \nodata & ? & N/? \\
144 & HD 282270 & 68.275192 & 29.778140 &\nodata & $294 \pm 8$ & $8 \pm 1.6$ & $-8.77 \pm 1.39$ &\nodata & A1V & N & N/? \\
145 & HD 283718 & 68.366065 & 24.923085 &\nodata & $693 \pm 213$ & $-5.5 \pm 1.3$ & $-2.9 \pm 1.18$ &\nodata & A3V/A0 & N & N/? \\
146 & GK Tau & 68.394029 & 24.351647 &\nodata &\nodata & $10.7 \pm 13.2$ & $-1.55 \pm 12.3$ &\nodata & K7 & N & Y/? \\
147 & IS Tau & 68.403326 & 26.163645 &\nodata & $233 \pm 158$ & $11.9 \pm 4.5$ & $-22.53 \pm 4.04$ &\nodata & K7 & Y & Y/? \\
148 & DL Tau & 68.412898 & 25.344065 &\nodata &\nodata & $2.4 \pm 4.5$ & $-12.74 \pm 4.07$ &\nodata & GV:e... & ? & Y/? \\
149 & HD 283684 & 68.570530 & 28.436188 &\nodata & $392 \pm 29$ & $-5.7 \pm 2.5$ & $-6.68 \pm 2.29$ &\nodata & A7 & N & N/? \\
150 & HQ Tau & 68.947255 & 22.839317 &\nodata &\nodata & $6.5 \pm 4.5$ & $-19.81 \pm 4.15$ &\nodata & \nodata & ? & Y/? \\
151 & HP Tau & 68.969924 & 22.906426 &\nodata & $101 \pm 79$ & $-0.9 \pm 10.3$ & $-22.02 \pm 9.49$ &\nodata & K3 & Y & Y/? \\
152 & HD 282334 & 69.033170 & 30.394761 &\nodata & $686 \pm 66$ & $-5.5 \pm 1.4$ & $-1.9 \pm 1.21$ &\nodata & A0 & N & N/? \\
153 & HD 29333 & 69.608312 & 29.387357 & $152_{-25}^{+37}$ & $132 \pm 7$ & $3.7 \pm 0.8$ & $0.78 \pm 0.7$ &\nodata & A2 & Y/Y & N/? \\
154 & DO Tau & 69.619111 & 26.180440 &\nodata &\nodata & $-7.1 \pm 4.6$ & $-29.61 \pm 4.13$ &\nodata & GV:e... & ? & Y/? \\
155 & HD 283739 & 69.682933 & 28.075829 &\nodata & $251 \pm 17$ & $-4 \pm 1.4$ & $2.29 \pm 1.24$ &\nodata & A7 & N & N/? \\
156 & HD 282387 & 69.695646 & 29.475297 &\nodata & $282 \pm 59$ & $5.7 \pm 1.6$ & $-11.32 \pm 1.39$ &\nodata & A0 & N & N/? \\
157 & GN Tau & 69.837159 & 25.750568 &\nodata & $138 \pm 99$ & $3.6 \pm 4.6$ & $-22.16 \pm 4.14$ &\nodata & M2.5 & Y & Y/? \\
158 & HD 283746 & 69.842850 & 27.765421 &\nodata & $292 \pm 8$ & $-1.7 \pm 1.4$ & $-7.52 \pm 1.33$ &\nodata & A3 & N & N/? \\
159 & HD 29459 & 69.846455 & 25.218274 & $111_{-9}^{+10}$ & $68 \pm 3$ & $18 \pm 0.6$ & $-10.68 \pm 0.45$ & $17.8 \pm 3.1$ & A5Vn & N/N & N/Y \\
160 & HD 29631 & 70.245274 & 23.939795 &\nodata &\nodata & $1.5 \pm 1.1$ & $-9.23 \pm 1.1$ &\nodata & F & ? & N/? \\
161 & HD 29646 & 70.332333 & 28.614989 & $103_{-8}^{+10}$ & $82 \pm 1$ & $35.5 \pm 0.3$ & $-27.56 \pm 0.44$ & $25.3$ & A2V & N/N & N/N \\
162 & 04415107+2914109 & 70.462775 & 29.236408 &\nodata &\nodata & $-5.2 \pm 1.7$ & $-8.38 \pm 1.48$ &\nodata & ? & ? & N/? \\
163 & HD 284648 & 70.572097 & 23.268516 &\nodata & $428 \pm 5$ & $5.7 \pm 0.9$ & $-8.45 \pm 0.83$ &\nodata & A0 & N & N/? \\
164 & DP Tau & 70.657043 & 25.260368 &\nodata & $205 \pm 133$ & $-3.7 \pm 5.1$ & $-19.63 \pm 4.61$ &\nodata & M0V:e & Y & Y/? \\
165 & 04433905+2353578 & 70.912683 & 23.899431 &\nodata &\nodata & $-0.1 \pm 1.5$ & $-14.45 \pm 1.37$ &\nodata & ? & ? & Y/? \\
166 & HD 282424 & 71.037023 & 30.866020 &\nodata &\nodata & $-1.4 \pm 1.6$ & $-9.18 \pm 1.37$ &\nodata & \nodata & ? & N/? \\
167 & 04454979+2442422 & 71.457514 & 24.711720 &\nodata &\nodata & $4.2 \pm 2.3$ & $-11.08 \pm 2.09$ &\nodata & ? & ? & N/? \\
168 & HD 30168 & 71.550277 & 26.035498 & $266_{-58}^{+104}$ & $166 \pm 4$ & $8.2 \pm 0.7$ & $-34.14 \pm 0.63$ &\nodata & A0 & N/Y & N/? \\
169 & HD 283868 & 71.778034 & 26.179319 &\nodata & $16 \pm 2$ & $3.6 \pm 1.4$ & $-2.33 \pm 1.26$ & $30$ & K3pv/G6e & N & N/N \\
170 & HD 283823 & 71.796194 & 28.964173 &\nodata & $406 \pm 51$ & $3.8 \pm 1.6$ & $-12.6 \pm 1.4$ &\nodata & A2 & N & N/? \\
171 & HD 30309 & 71.865612 & 24.354856 &\nodata & $398 \pm 26$ & $3 \pm 1.4$ & $-2.64 \pm 1.28$ &\nodata & A0/F5 & N & N/? \\
172 & HD 284763 & 71.940583 & 22.685857 &\nodata & $458 \pm 46$ & $-2 \pm 1.5$ & $-1.48 \pm 1.38$ &\nodata & F0 & N & N/? \\
173 & DS Tau & 71.952482 & 29.419766 &\nodata & $98 \pm 63$ & $7.8 \pm 2.7$ & $-29.62 \pm 2.35$ & $0$ & K4V:e & Y & Y/N \\
174 & HD 283861 & 71.975276 & 26.560652 &\nodata & $463 \pm 12$ & $1.1 \pm 1.5$ & $-4.47 \pm 1.43$ &\nodata & A0 & N & N/? \\
175 & HD 30466 & 72.316676 & 29.571363 & $163_{-22}^{+30}$ & $177 \pm 5$ & $8 \pm 0.8$ & $-28.18 \pm 0.61$ & $17$ & A0p & Y/N & N/Y \\
176 & 04492661+2730388 & 72.360927 & 27.510755 &\nodata &\nodata & $1.3 \pm 1.4$ & $-5.41 \pm 1.24$ &\nodata & ? & ? & N/? \\
177 & HD 283842 & 72.548615 & 27.677257 &\nodata &\nodata & $-2 \pm 1.7$ & $-3.54 \pm 1.51$ &\nodata & A & ? & N/? \\
178 & HD 283830 & 72.585658 & 28.377142 &\nodata & $675 \pm 154$ & $-3.1 \pm 1.7$ & $-12.58 \pm 1.5$ &\nodata & A2 & N & N/? \\
179 & HD 283885 & 72.743801 & 24.278521 &\nodata & $343 \pm 35$ & $3.1 \pm 1.5$ & $-5.65 \pm 1.37$ &\nodata & A5 & N & N/? \\
180 & UY Aur & 72.947410 & 30.787076 &\nodata & $115 \pm 44$ & $4.9 \pm 2.3$ & $-19.24 \pm 2.06$ & $18 \pm 3$ & G5V:e... & Y & Y/Y \\
181 & HD 283945 & 73.028640 & 25.437055 &\nodata & $515 \pm 66$ & $-4.4 \pm 1.7$ & $-8.31 \pm 1.54$ &\nodata & A2 & N & N/? \\
182 & HD 283889 & 73.110373 & 28.729730 &\nodata & $592 \pm 65$ & $-3.2 \pm 1.6$ & $-6.75 \pm 1.4$ &\nodata & A0 & N & N/? \\
183 & BD+26 758 & 73.135736 & 27.027236 &\nodata & $158 \pm 12$ & $-12.3 \pm 1.5$ & $-22.63 \pm 1.34$ &\nodata & A5 & Y & N/? \\
184 & HD 283890 & 73.185411 & 28.619612 &\nodata & $301 \pm 48$ & $-1.4 \pm 1.7$ & $-4.83 \pm 1.49$ &\nodata & A3 & N & N/? \\
185 & HD 283941 & 73.289352 & 25.491105 &\nodata & $487 \pm 69$ & $4.4 \pm 1.5$ & $-7.67 \pm 1.44$ &\nodata & A0 & N & N/? \\
186 & HD 283893 & 73.392339 & 28.456354 &\nodata & $616 \pm 88$ & $-2.5 \pm 1.4$ & $-11.43 \pm 1.23$ &\nodata & A0 & N & N/? \\
187 & HD 284873 & 73.431141 & 22.178329 &\nodata &\nodata & $6.4 \pm 1$ & $-6.11 \pm 0.93$ &\nodata & A & ? & N/? \\
188 & HD 283911 & 73.731551 & 27.630496 &\nodata & $411 \pm 57$ & $1.1 \pm 1.5$ & $-8.24 \pm 1.33$ &\nodata & A2 & N & N/? \\
189 & HD 31329 & 73.914289 & 22.187528 &\nodata & $281 \pm 17$ & $-1.2 \pm 0.9$ & $-8.15 \pm 0.93$ &\nodata & A2 & N & N/? \\
190 & HD 282624 & 73.997451 & 30.567085 & $152_{-34}^{+63}$ & $451 \pm 1$ & $5.4 \pm 1.4$ & $-20.32 \pm 0.95$ & $23.2 \pm 2.8$ & G2III & Y/N & Y/N \\
191 & HD 284989 & 74.128224 & 22.591051 &\nodata & $775 \pm 86$ & $1.5 \pm 1.4$ & $-5.63 \pm 1.38$ &\nodata & A0 & N & N/? \\
192 & HD 31581 & 74.581425 & 29.847393 &\nodata & $194 \pm 9$ & $-1.7 \pm 1.5$ & $-3.9 \pm 1.21$ &\nodata & A2 & N & N/? \\
193 & HD 31648 & 74.692775 & 29.843609 & $131_{-18}^{+24}$ & $124 \pm 32$ & $5.6 \pm 0.9$ & $-21.16 \pm 0.69$ &\nodata & A3Ve/A2 & Y/Y & Y/? \\
194 & HD 284946 & 74.893341 & 23.502550 &\nodata & $479 \pm 21$ & $-1.8 \pm 1.2$ & $-5.04 \pm 1.1$ &\nodata & A0 & N & N/? \\
195 & HD 284035 & 74.941567 & 24.314204 &\nodata & $523 \pm 61$ & $-7 \pm 5.9$ & $-7.11 \pm 5.65$ &\nodata & A0 & N & N/? \\
\hline
\\
\multicolumn{12}{c}{O,B,A stars from Knapp et al. {\it SDSS} data}\\
\\\hline
196 & 04095167+2520112 & 62.465271 & 25.336500 &\nodata & $4389 \pm 605$ & $-1.6 \pm 3.7 $ & $-4.8 \pm 3.7$ & $-0.1 \pm 1.6$ & A0 & N & N/N \\
197 & 04111342+2447170 & 62.805889 & 24.788080 &\nodata & $8576 \pm 2737$ & $0.3 \pm 3.8 $ & $-3.3 \pm 3.8$ & $37.5 \pm 3.2$ & A0 & N & N/N \\
198 & 04113476+2524136 & 62.894859 & 25.403830 &\nodata & $12774 \pm 2784$ & $-3.7 \pm 4.3 $ & $-2.5 \pm 4.3$ & $-2.5 \pm 2.5$ & A0 & N & N/N \\
199 & 04122067+2430477 & 63.086201 & 24.513310 &\nodata & $6397 \pm 895$ & $-2.0 \pm 3.8 $ & $-3.9 \pm 3.8$ & $-7.4 \pm 1.5$ & A0 & N & N/N \\
200 & 04132281+2620282 & 63.345039 & 26.341200 &\nodata & $11426 \pm 1981$ & $1.3 \pm 4.2 $ & $-6.2 \pm 4.2$ & $-1.1 \pm 1.7$ & A0 & N & N/N \\
201 & 04135173+2631257 & 63.465488 & 26.523861 &\nodata & $11234 \pm 838$ & $1.1 \pm 4.4 $ & $-4.2 \pm 4.4$ & $69.9 \pm 2.6$ & A0 & N & N/N \\
202 & 04135440+2609097 & 63.476631 & 26.152700 &\nodata &\nodata & $3.2 \pm 3.9 $ & $-9.9 \pm 3.9$ & $64.2 \pm 2.7$ &\nodata & ? & N/N \\
203 & 04142194+2500478 & 63.591438 & 25.013350 &\nodata & $8722 \pm 1026$ & $0.1 \pm 4.3 $ & $-3.2 \pm 4.3$ & $-14.5 \pm 3$ & A0 & N & N/N \\
204 & 04142864+2608046 & 63.619362 & 26.134621 &\nodata & $7392 \pm 1367$ & $3.0 \pm 3.8 $ & $-8.2 \pm 3.8$ & $2.3 \pm 1.8$ & A0 & N & N/N \\
205 & 04152014+2629584 & 63.833901 & 26.499580 &\nodata & $8897 \pm 748$ & $3.8 \pm 4.1 $ & $-7.5 \pm 4.1$ & $25.6 \pm 2.2$ & A0 & N & N/N \\
206 & 04152167+2522393 & 63.840290 & 25.377581 &\nodata & $18112 \pm 1920$ & $10.3 \pm 4.4 $ & $-5.3 \pm 4.4$ & $101.5 \pm 5.1$ & B6 & N & N/N \\
207 & 04153610+2538570 & 63.900219 & 25.649200 &\nodata &\nodata & $7.4 \pm 4.5 $ & $-0.3 \pm 4.5$ & $-66.4 \pm 7.3$ & O & ? & N/N \\
208 & 04154968+3035156 & 63.957008 & 30.587580 &\nodata & $20128 \pm 6470$ & $4.7 \pm 5.5 $ & $-0.5 \pm 5.5$ & $62.6 \pm 3.7$ & A0p & N & N/N \\
209 & 04155589+2941428 & 63.982899 & 29.695271 &\nodata & $14164 \pm 7605$ & $2.0 \pm 4.4 $ & $2.7 \pm 4.4$ & $2.8 \pm 2.6$ & A0 & N & N/N \\
210 & 04160013+2730051 & 64.000603 & 27.501440 &\nodata & $5608 \pm 403$ & $-0.9 \pm 4.3 $ & $0.9 \pm 4.3$ & $110.4 \pm 2.6$ & A0 & N & N/N \\
211 & 04161037+3053487 & 64.043190 & 30.896900 &\nodata & $14170 \pm 5336$ & $2.3 \pm 4.6 $ & $-3.0 \pm 4.6$ & $-3 \pm 2.3$ & A0 & N & N/N \\
212 & 04162411+2434450 & 64.100487 & 24.579250 &\nodata & $7766 \pm 575$ & $6.0 \pm 3.9 $ & $-3.8 \pm 3.9$ & $21.9 \pm 1.9$ & A0 & N & N/N \\
213 & 04162961+2643589 & 64.123383 & 26.733021 &\nodata & $10475 \pm 1941$ & $5.8 \pm 4.0 $ & $-6.5 \pm 4$ & $31.8 \pm 3.2$ & A0 & N & N/N \\
214 & 04163672+2650392 & 64.152969 & 26.844250 &\nodata & $11636 \pm 1775$ & $1.4 \pm 4.4 $ & $-3.7 \pm 4.4$ & $0 \pm 3$ & A0 & N & N/N \\
215 & 04165196+2601449 & 64.216522 & 26.029141 &\nodata & $12571 \pm 2322$ & $6.9 \pm 4.3 $ & $-8.3 \pm 4.3$ & $26.8 \pm 2.7$ & A0 & N & N/N \\
216 & 04170011+2522475 & 64.250412 & 25.379850 &\nodata & $5729 \pm 756$ & $4.1 \pm 3.9 $ & $1.6 \pm 3.9$ & $-10 \pm 2.8$ & A0 & N & N/N \\
217 & 04170272+2644290 & 64.261322 & 26.741400 &\nodata & $16618 \pm 6674$ & $-0.9 \pm 4.2 $ & $-1.6 \pm 4.2$ & $-42 \pm 1.9$ & A0 & N & N/N \\
218 & 04174593+2618579 & 64.441330 & 26.316080 &\nodata & $12914 \pm 2916$ & $11.6 \pm 4.2 $ & $-6.2 \pm 4.1$ & $-23.7 \pm 2.6$ & A0 & N & N/N \\
219 & 04180495+2952363 & 64.520653 & 29.876770 &\nodata & $5337 \pm 739$ & $0.4 \pm 4.3 $ & $-1.0 \pm 4.3$ & $56.2 \pm 2.9$ & A0 & N & N/N \\
220 & 04180763+2846201 & 64.531853 & 28.772261 &\nodata &\nodata & $-3.0 \pm 4.3 $ & $2.4 \pm 4.3$ & $-8.2 \pm 4$ &\nodata & ? & N/N \\
221 & 04182140+2552023 & 64.589111 & 25.867371 &\nodata & $13651 \pm 2176$ & $0.7 \pm 4.5 $ & $-2.1 \pm 4.5$ & $19.9 \pm 2.3$ & A0 & N & N/Y \\
222 & 04182615+2454459 & 64.608917 & 24.912741 &\nodata & $6099 \pm 481$ & $4.1 \pm 3.9 $ & $-0.7 \pm 3.9$ & $23.2 \pm 1.9$ & A0 & N & N/N \\
223 & 04183386+2547250 & 64.641022 & 25.790310 &\nodata & $7144 \pm 1040$ & $1.2 \pm 3.9 $ & $-1.1 \pm 3.9$ & $40.5 \pm 2$ & A0 & N & N/N \\
224 & 04183552+3006115 & 64.647987 & 30.103270 &\nodata & $11840 \pm 6382$ & $5.1 \pm 4.2 $ & $-3.7 \pm 4.2$ & $-28.9 \pm 1.6$ & A0 & N & N/N \\
225 & 04183818+2735261 & 64.658989 & 27.590740 &\nodata & $5840 \pm 429$ & $0.1 \pm 4.3 $ & $-2.7 \pm 4.3$ & $93.5 \pm 4.2$ & A0 & N & N/N \\
226 & 04184109+2449442 & 64.671127 & 24.828951 &\nodata & $6709 \pm 1285$ & $5.4 \pm 3.9 $ & $-6.5 \pm 3.9$ & $-28.6 \pm 3.2$ & A0 & N & N/N \\
227 & 04184261+2550172 & 64.677528 & 25.838110 &\nodata & $4135 \pm 2695$ & $4.5 \pm 4.4 $ & $-1.5 \pm 4.4$ & $60.3 \pm 2.2$ & A0 & N & N/N \\
228 & 04185780+3004129 & 64.740830 & 30.070339 &\nodata & $7548 \pm 2854$ & $-2.8 \pm 4.2 $ & $-1.0 \pm 4.2$ & $16.4 \pm 1.5$ & A0 & N & N/Y \\
229 & 04185955+3050026 & 64.748200 & 30.834129 &\nodata & $13418 \pm 4181$ & $-5.9 \pm 4.7 $ & $-2.4 \pm 4.7$ & $17.3 \pm 3.3$ & A0 & N & N/Y \\
230 & 04190128+2918288 & 64.755364 & 29.308020 &\nodata & $17037 \pm 6908$ & $0.3 \pm 4.7 $ & $-0.9 \pm 4.7$ & $-42 \pm 2.8$ & A0 & N & N/N \\
231 & 04190952+2907266 & 64.789658 & 29.124069 &\nodata & $10467 \pm 2703$ & $0.5 \pm 4.4 $ & $-1.4 \pm 4.4$ & $-24.5 \pm 2$ & A0 & N & N/N \\
232 & 04191436+2552469 & 64.809807 & 25.879721 &\nodata & $7069 \pm 461$ & $3.8 \pm 3.9 $ & $-0.2 \pm 3.9$ & $22.5 \pm 2.3$ & A0 & N & N/N \\
233 & 04191985+2749395 & 64.832733 & 27.827740 &\nodata &\nodata & $-2.6 \pm 4.3 $ & $-4.5 \pm 4.3$ & $37 \pm 2.7$ &\nodata & ? & N/N \\
234 & 04192106+2931069 & 64.837784 & 29.518650 &\nodata & $5810 \pm 1609$ & $0.1 \pm 4.3 $ & $-4.8 \pm 4.3$ & $-67 \pm 1.5$ & A0 & N & N/N \\
235 & 04193853+2949396 & 64.910599 & 29.827740 &\nodata &\nodata & $-12.3 \pm 5.8 $ & $-6.3 \pm 5.8$ & $-5.3 \pm 2.2$ & 0 & ? & N/N \\
236 & 04195512+2801576 & 64.979652 & 28.032820 &\nodata & $20143 \pm 8223$ & $-4.1 \pm 5.0 $ & $-13.7 \pm 5$ & $49.8 \pm 5.6$ & B9 & N & N/N \\
237 & 04200363+2950586 & 65.015122 & 29.849670 &\nodata & $22508 \pm 13749$ & $-2.8 \pm 4.4 $ & $1.1 \pm 4.4$ & $25.8 \pm 3.4$ & A0 & N & N/N \\
238 & 04200803+2843222 & 65.033478 & 28.722879 &\nodata & $5852 \pm 1409$ & $-3.4 \pm 4.3 $ & $-2.6 \pm 4.3$ & $13.5 \pm 1.5$ & A0 & N & N/Y \\
239 & 04203271+3015272 & 65.136299 & 30.257601 &\nodata & $11597 \pm 1905$ & $2.5 \pm 4.8 $ & $-5.1 \pm 4.8$ & $-15.5 \pm 3.2$ & A0 & N & N/N \\
240 & 04203551+2945073 & 65.147903 & 29.752081 &\nodata & $6242 \pm 572$ & $-4.8 \pm 4.4 $ & $-3.4 \pm 4.4$ & $-21.6 \pm 1.3$ & A0 & N & N/N \\
241 & 04204960+3010153 & 65.206642 & 30.170919 &\nodata & $5912 \pm 1380$ & $4.0 \pm 4.3 $ & $-3.2 \pm 4.3$ & $-1 \pm 1.8$ & A0 & N & N/N \\
242 & 04205096+2840119 & 65.212318 & 28.670050 &\nodata & $5659 \pm 1353$ & $1.2 \pm 4.3 $ & $-1.7 \pm 4.3$ & $1.5 \pm 1.6$ & A0 & N & N/N \\
243 & 04210068+2711172 & 65.252724 & 27.188320 &\nodata & $15034 \pm 10003$ & $1.8 \pm 3.8 $ & $-8.5 \pm 3.8$ & $-168.4 \pm 2.2$ & A0 & N & N/? \\
244 & 04210658+2546557 & 65.277359 & 25.782129 &\nodata & $16067 \pm 2065$ & $-9.8 \pm 4.1 $ & $-15.4 \pm 4.1$ & $64 \pm 2.8$ & B6 & N & N/N \\
245 & 04211397+2850064 & 65.308197 & 28.835079 &\nodata & $12704 \pm 3762$ & $1.0 \pm 4.7 $ & $3.1 \pm 4.7$ & $-45.6 \pm 4$ & A0 & N & N/N \\
246 & 04212796+3002107 & 65.366501 & 30.036320 &\nodata & $6888 \pm 1080$ & $-4.7 \pm 4.2 $ & $-2.8 \pm 4.2$ & $-33.4 \pm 1.4$ & A0 & N & N/N \\
247 & 04213150+2520399 & 65.381271 & 25.344431 &\nodata & $4475 \pm 673$ & $2.7 \pm 3.8 $ & $-3.2 \pm 3.8$ & $-11.3 \pm 1.7$ & A0 & N & N/N \\
248 & 04213151+2917334 & 65.381310 & 29.292620 &\nodata & $9353 \pm 1500$ & $0.6 \pm 4.5 $ & $-0.1 \pm 4.5$ & $-1.6 \pm 1.9$ & A0 & N & N/N \\
249 & 04213578+2936482 & 65.399117 & 29.613489 &\nodata & $10898 \pm 2482$ & $-1.7 \pm 4.5 $ & $-4.6 \pm 4.5$ & $-22.9 \pm 2$ & A0 & N & N/N \\
250 & 04222582+2616149 & 65.607529 & 26.271099 &\nodata & $13098 \pm 3069$ & $-1.3 \pm 4.5 $ & $-3.3 \pm 4.5$ & $-11.5 \pm 2.7$ & A0 & N & N/N \\
251 & 04224008+2957006 & 65.667023 & 29.950230 &\nodata & $8232 \pm 1932$ & $-1.0 \pm 4.4 $ & $1.4 \pm 4.4$ & $4.8 \pm 1.7$ & A0 & N & N/N \\
252 & 04224882+2904442 & 65.703377 & 29.079029 &\nodata & $14458 \pm 3048$ & $0.1 \pm 5.0 $ & $-4.6 \pm 5$ & $-1.4 \pm 2.1$ & A0 & N & N/N \\
253 & 04225306+2613188 & 65.721130 & 26.222340 &\nodata &\nodata & $-2.9 \pm 4.4 $ & $-8.6 \pm 4.4$ & $-28.4 \pm 4.3$ & 0 & ? & N/N \\
254 & 04225686+2939043 & 65.736900 & 29.651279 &\nodata & $9610 \pm 1149$ & $-6.2 \pm 4.6 $ & $2.3 \pm 4.6$ & $20 \pm 2.1$ & A0 & N & N/Y \\
255 & 04230562+2538494 & 65.773392 & 25.647060 &\nodata & $5550 \pm 424$ & $2.4 \pm 3.8 $ & $-0.2 \pm 3.8$ & $1.9 \pm 2.4$ & A0 & N & N/N \\
256 & 04231596+2941221 & 65.816551 & 29.689520 &\nodata & $9546 \pm 1035$ & $8.9 \pm 4.7 $ & $-8.6 \pm 4.7$ & $57.1 \pm 2.4$ & A0 & N & Y/N \\
257 & 04231716+2757432 & 65.821480 & 27.962071 &\nodata & $1417 \pm 1307$ & $-6.5 \pm 4.3 $ & $-9.1 \pm 4.3$ & $34.9 \pm 1.9$ & B9 & Y: & N/N \\
258 & 04233427+2947149 & 65.892807 & 29.787510 &\nodata & $6340 \pm 1232$ & $-1.5 \pm 4.3 $ & $-5.6 \pm 4.3$ & $12.7 \pm 1.4$ & A0 & N & N/Y \\
259 & 04233438+3056585 & 65.893272 & 30.949650 &\nodata & $3320 \pm 420$ & $4.8 \pm 4.3 $ & $1.4 \pm 4.3$ & $-20.8 \pm 1.8$ & A0 & N & N/N \\
260 & 04234983+2532157 & 65.957558 & 25.537720 &\nodata & $11883 \pm 2984$ & $2.2 \pm 4.2 $ & $-1.7 \pm 4.2$ & $49.3 \pm 3.2$ & A0 & N & N/N \\
261 & 04240540+2744507 & 66.022369 & 27.747620 &\nodata &\nodata & $-0.3 \pm 4.3 $ & $-5.5 \pm 4.3$ & $53.3 \pm 2.8$ &\nodata & ? & N/N \\
262 & 04240902+2611351 & 66.037521 & 26.193510 &\nodata &\nodata & $4.2 \pm 3.9 $ & $-4.8 \pm 3.9$ & $-17.2 \pm 5$ & 0 & ? & N/N \\
263 & 04243937+2946563 & 66.164047 & 29.782440 &\nodata & $12882 \pm 4098$ & $3.2 \pm 4.8 $ & $-5.9 \pm 4.8$ & $12.9 \pm 2.4$ & A0 & N & N/Y \\
264 & 04244689+2826203 & 66.195351 & 28.439030 &\nodata & $16870 \pm 6801$ & $0.7 \pm 4.9 $ & $-4.7 \pm 4.9$ & $12.1 \pm 2.3$ & A0 & N & N/Y \\
266 & 04245454+3014488 & 66.227280 & 30.246929 &\nodata & $9024 \pm 2583$ & $-1.4 \pm 4.3 $ & $-1.6 \pm 4.3$ & $-20.4 \pm 2$ & A0 & N & N/N \\
267 & 04254650+2805257 & 66.443710 & 28.090530 &\nodata & $10098 \pm 1561$ & $-2.5 \pm 4.4 $ & $-4.9 \pm 4.4$ & $29.9 \pm 4.2$ & B9 & N & N/N \\
268 & 04254749+3039409 & 66.447868 & 30.661350 &\nodata & $10176 \pm 3362$ & $-3.3 \pm 4.4 $ & $-6.1 \pm 4.4$ & $2.8 \pm 2.6$ & A0 & N & N/N \\
269 & 04255111+3041231 & 66.462982 & 30.689720 &\nodata & $5837 \pm 1217$ & $-6.7 \pm 4.3 $ & $-1.8 \pm 4.3$ & $41.9 \pm 2.5$ & A0 & N & N/N \\
270 & 04262320+2629082 & 66.596626 & 26.486160 &\nodata &\nodata & $-1.4 \pm 3.9 $ & $-3.8 \pm 3.9$ & $15.9 \pm 12.6$ & 0 & ? & N/Y \\
271 & 04263605+2904190 & 66.650147 & 29.072081 &\nodata & $5375 \pm 1263$ & $-2.2 \pm 4.3 $ & $-2.2 \pm 4.3$ & $-16.1 \pm 1.4$ & A0 & N & N/N \\
272 & 04263695+3035247 & 66.654030 & 30.590200 &\nodata & $9300 \pm 1949$ & $-4.4 \pm 4.2 $ & $-3.0 \pm 4.2$ & $7.2 \pm 2.8$ & A0 & N & N/N \\
273 & 04264332+3054283 & 66.680603 & 30.907900 &\nodata & $6815 \pm 1694$ & $-2.3 \pm 4.1 $ & $1.2 \pm 4.1$ & $18.7 \pm 1.9$ & A0 & N & N/Y \\
274 & 04265712+2920054 & 66.737999 & 29.334990 &\nodata & $4125 \pm 496$ & $0.0 \pm 4.3 $ & $-8.3 \pm 4.3$ & $3.2 \pm 1.4$ & A0 & N & N/N \\
275 & 04300090+2511506 & 67.503799 & 25.197390 &\nodata & $11767 \pm 4813$ & $-0.1 \pm 3.9 $ & $-2.1 \pm 3.9$ & $11.1 \pm 3.1$ & A0 & N & N/Y \\
276 & 04304916+2527382 & 67.704887 & 25.460671 &\nodata & $11299 \pm 2639$ & $0.4 \pm 4.3 $ & $-3.1 \pm 4.3$ & $30.3 \pm 3.2$ & A0 & N & N/N \\
277 & 04324026+2537105 & 68.167717 & 25.619631 &\nodata & $9909 \pm 1761$ & $2.9 \pm 4.2 $ & $-2.5 \pm 4.2$ & $92.8 \pm 3.8$ & A0 & N & N/N \\
278 & 04325041+2613573 & 68.210083 & 26.232540 &\nodata &\nodata & $1.8 \pm 3.8 $ & $-8.8 \pm 3.8$ & $-33.1 \pm 7$ &\nodata & ? & N/N \\
279 & 04343357+2538478 & 68.639870 & 25.646650 &\nodata &\nodata & $0.7 \pm 3.8 $ & $-11.1 \pm 3.8$ & $41.1 \pm 4.9$ &\nodata & ? & Y/N \\
280 & 04344093+2526178 & 68.670540 & 25.438280 &\nodata & $9457 \pm 2601$ & $-0.2 \pm 3.9 $ & $-0.9 \pm 3.9$ & $13 \pm 3.3$ & A0 & N & N/Y \\
281 & 04352460+2511584 & 68.852524 & 25.199591 &\nodata &\nodata & $0.6 \pm 3.9 $ & $-0.4 \pm 3.9$ & $1.3 \pm 3.4$ &\nodata & ? & N/N \\
282 & 04352506+2505371 & 68.854431 & 25.093679 &\nodata & $8742 \pm 1788$ & $1.9 \pm 4.0 $ & $-1.8 \pm 4$ & $9.5 \pm 3.8$ & A0 & N & N/Y \\
283 & 04355804+2429239 & 68.991859 & 24.490030 &\nodata & $4305 \pm 611$ & $-2.0 \pm 3.7 $ & $-2.1 \pm 3.7$ & $39.6 \pm 2.3$ & A0 & N & N/N \\
284 & 04360432+2512256 & 69.018021 & 25.207150 &\nodata & $7098 \pm 1703$ & $3.5 \pm 3.7 $ & $-0.5 \pm 3.7$ & $83.4 \pm 4$ & A0 & N & N/N \\
285 & 04364715+2508462 & 69.196533 & 25.146160 &\nodata & $7047 \pm 1224$ & $-1.4 \pm 3.9 $ & $-0.6 \pm 3.9$ & $21.8 \pm 3.1$ & A0 & N & N/N \\
286 & 04374665+2415466 & 69.444412 & 24.262991 &\nodata & $8970 \pm 2523$ & $-3.6 \pm 3.9 $ & $-3.3 \pm 3.9$ & $-24.1 \pm 2.9$ & A0 & N & N/N \\
287 & 04384454+2438388 & 69.685593 & 24.644190 &\nodata &\nodata & $-0.3 \pm 3.8 $ & $-5.4 \pm 3.8$ & $16 \pm 3.6$ &\nodata & ? & N/Y \\
288 & 04384756+2524115 & 69.698174 & 25.403200 &\nodata &\nodata & $1.4 \pm 3.8 $ & $-3.1 \pm 3.8$ & $34.2 \pm 6.9$ &\nodata & ? & N/N \\
289 & 04391699+2520299 & 69.820900 & 25.341660 &\nodata &\nodata & $-6.7 \pm 3.8 $ & $2.0 \pm 3.8$ & $30.2 \pm 7.7$ &\nodata & ? & N/N \\
290 & 04393029+2457401 & 69.876183 & 24.961180 &\nodata &\nodata & $17.8 \pm 3.8 $ & $-2.4 \pm 3.8$ & $11.9 \pm 2.5$ &\nodata & ? & N/Y \\
291 & 04400871+2451438 & 70.036377 & 24.862181 &\nodata &\nodata & $0.9 \pm 3.8 $ & $2.5 \pm 3.8$ & $28.1 \pm 5$ &\nodata & ? & N/N \\
292 & 04402154+2629461 & 70.089851 & 26.496161 &\nodata &\nodata & $-3.5 \pm 3.8 $ & $-6.1 \pm 3.8$ & $-21.3 \pm 4.2$ &\nodata & ? & N/N \\
293 & 04402572+2635205 & 70.107269 & 26.589060 &\nodata &\nodata & $-6.6 \pm 3.8 $ & $5.4 \pm 3.8$ & $-26 \pm 5.4$ &\nodata & ? & N/N \\
294 & 04415750+2451568 & 70.489609 & 24.865770 &\nodata &\nodata & $-0.1 \pm 4.0 $ & $1.5 \pm 4$ & $66.5 \pm 3.5$ &\nodata & ? & N/N \\
295 & 04422622+2459000 & 70.609337 & 24.983379 &\nodata &\nodata & $9.5 \pm 4.0 $ & $-7.6 \pm 4$ & $22.7 \pm 4.2$ &\nodata & ? & N/N \\
296 & 04422916+2422346 & 70.621521 & 24.376360 &\nodata &\nodata & $2.6 \pm 4.2 $ & $-4.8 \pm 4.2$ & $3.9 \pm 1.9$ &\nodata & ? & N/N \\
297 & 04423176+2601455 & 70.632378 & 26.029329 &\nodata &\nodata & $2.1 \pm 4.0 $ & $-5.7 \pm 4$ & $2.5 \pm 2.6$ &\nodata & ? & N/N \\
298 & 04435975+2529297 & 70.998970 & 25.491650 &\nodata &\nodata & $-1.3 \pm 4.3 $ & $-5.1 \pm 4.3$ & $1.5 \pm 3.2$ &\nodata & ? & N/N \\
299 & 04441407+2415542 & 71.058632 & 24.265110 &\nodata & $12108 \pm 5721$ & $3.2 \pm 4.2 $ & $-3.8 \pm 4.2$ & $31.7 \pm 3$ & B9 & N & N/N \\
300 & 04442409+2609464 & 71.100418 & 26.162939 &\nodata & $14445 \pm 5743$ & $-3.0 \pm 4.7 $ & $-7.9 \pm 4.7$ & $6.8 \pm 2.7$ & A0 & N & Y/N \\
301 & 04443402+2537401 & 71.141808 & 25.627850 &\nodata &\nodata & $-2.0 \pm 4.3 $ & $-9.5 \pm 4.3$ & $54.6 \pm 3.9$ &\nodata & ? & Y/N \\
302 & 04445791+2601458 & 71.241364 & 26.029421 &\nodata & $4464 \pm 486$ & $-2.1 \pm 4.3 $ & $-5.3 \pm 4.3$ & $51.3 \pm 1.8$ & A0 & N & N/N \\
303 & 04450067+2619327 & 71.252762 & 26.325800 &\nodata & $9289 \pm 1437$ & $-0.1 \pm 4.7 $ & $-7.3 \pm 4.7$ & $-17.3 \pm 4$ & A0 & N & Y/N \\
304 & 04452569+2417505 & 71.357018 & 24.297430 &\nodata & $11112 \pm 3705$ & $1.4 \pm 4.4 $ & $-7.1 \pm 4.4$ & $24.1 \pm 3.2$ & A0 & N & Y/N \\
305 & 04453199+2548133 & 71.383301 & 25.803770 &\nodata &\nodata & $-2.2 \pm 4.3 $ & $-7.7 \pm 4.3$ & $89.7 \pm 2.8$ &\nodata & ? & N/N \\
306 & 04461962+2434148 & 71.581787 & 24.570881 &\nodata &\nodata & $1.3 \pm 4.3 $ & $-4.1 \pm 4.3$ & $-23.5 \pm 5$ &\nodata & ? & N/N \\
307 & 04463405+2413551 & 71.641983 & 24.232010 &\nodata & $11533 \pm 1056$ & $9.7 \pm 4.9 $ & $1.3 \pm 4.9$ & $22.3 \pm 3.4$ & B9 & N & N/N \\
308 & 04464411+2620004 & 71.683830 & 26.333450 &\nodata & $3561 \pm 2188$ & $-6.0 \pm 4.4 $ & $-2.3 \pm 4.4$ & $-7.9 \pm 2.6$ & A0 & N & N/N \\
309 & 04464556+2550507 & 71.689850 & 25.847441 &\nodata &\nodata & $-4.6 \pm 4.3 $ & $-9.0 \pm 4.3$ & $8.5 \pm 2.5$ &\nodata & ? & N/Y \\
310 & 04464746+2419597 & 71.697853 & 24.333281 &\nodata & $6966 \pm 1113$ & $-2.5 \pm 4.4 $ & $-2.8 \pm 4.4$ & $20.5 \pm 2.6$ & A0 & N & N/Y \\
311 & 04464757+2448496 & 71.698288 & 24.813789 &\nodata &\nodata & $-3.2 \pm 4.3 $ & $-2.6 \pm 4.3$ & $50.7 \pm 2.5$ &\nodata & ? & N/N \\
312 & 04471316+2452301 & 71.804802 & 24.875071 &\nodata &\nodata & $9.5 \pm 4.4 $ & $-13.3 \pm 4.4$ & $-25.5 \pm 4.6$ &\nodata & ? & Y/N \\
313 & 04475295+2407358 & 71.970642 & 24.126699 &\nodata & $14372 \pm 3532$ & $3.1 \pm 4.8 $ & $-2.0 \pm 4.8$ & $14.1 \pm 2.8$ & B9 & N & N/Y \\
314 & 04481479+2412522 & 72.061668 & 24.214569 &\nodata & $12093 \pm 1796$ & $0.2 \pm 5.1 $ & $-6.6 \pm 5.1$ & $-24.5 \pm 2.6$ & A0 & N & Y/N \\
315 & 04484223+2405315 & 72.175957 & 24.092159 &\nodata & $5295 \pm 1494$ & $3.4 \pm 4.3 $ & $-8.7 \pm 4.3$ & $25.2 \pm 2.3$ & B9 & N & Y/N \\
316 & 04485467+2603349 & 72.227799 & 26.059731 &\nodata &\nodata & $3.8 \pm 4.3 $ & $-8.2 \pm 4.3$ & $-15.3 \pm 2.3$ &\nodata & ? & Y/N \\
317 & 04490164+2603474 & 72.256897 & 26.063191 &\nodata &\nodata & $-2.1 \pm 4.3 $ & $-4.7 \pm 4.3$ & $-8.1 \pm 3.1$ &\nodata & ? & N/N \\
318 & 04495186+2546445 & 72.466042 & 25.779091 &\nodata & $4478 \pm 1063$ & $-1.4 \pm 4.3 $ & $-4.7 \pm 4.3$ & $18 \pm 2.2$ & B9 & N & N/Y \\
319 & 04501677+2455582 & 72.569870 & 24.932921 &\nodata &\nodata & $7.7 \pm 4.3 $ & $-14.6 \pm 4.3$ & $205.7 \pm 3.1$ &\nodata & ? & Y/N \\
320 & 04502462+2534113 & 72.602654 & 25.569889 &\nodata &\nodata & $-0.1 \pm 4.3 $ & $-12.8 \pm 4.3$ & $39.3 \pm 4.2$ &\nodata & ? & Y/N \\
321 & 04502592+2609475 & 72.607941 & 26.163250 &\nodata &\nodata & $-0.5 \pm 4.3 $ & $-9.7 \pm 4.3$ & $-8.8 \pm 1.7$ &\nodata & ? & Y/N \\
322 & 04503511+2516324 & 72.646263 & 25.275740 &\nodata &\nodata & $5.1 \pm 4.4 $ & $-5.4 \pm 4.4$ & $2.4 \pm 4.3$ & \nodata& ? & N/N \\
323 & 04505387+2426576 & 72.724533 & 24.449440 & \nodata& $7078 \pm 1947$ & $-1.5 \pm 4.3 $ & $-3.9 \pm 4.3$ & $28.7 \pm 2.2$ & A0 & N & N/N \\
324 & 04511381+2522599 & 72.807442 & 25.383381 & \nodata& $6538 \pm 1250$ & $5.6 \pm 4.3 $ & $-4.0 \pm 4.3$ & $12.4 \pm 2.5$ & B9 & N & N/Y \\
325 & 04513070+2438129 & 72.877831 & 24.636990 & \nodata& $3468 \pm 296$ & $10.7 \pm 4.3 $ & $-7.0 \pm 4.3$ & $32 \pm 2.5$ & A0 & N & N/N \\
326 & 04521892+2520501 & 73.078812 & 25.347281 & \nodata& $5194 \pm 884$ & $4.8 \pm 4.3 $ & $-3.0 \pm 4.3$ & $63.8 \pm 2$ & A0 & N & N/N \\
327 & 04522135+2506498 & 73.088966 & 25.113840 & \nodata& $9897 \pm 2710$ & $4.1 \pm 4.4 $ & $-4.8 \pm 4.4$ & $26.4 \pm 3.6$ & A0 & N & N/N \\
\hline
\\
\multicolumn{12}{c}{Other stars }\\
\\\hline
328 & HD 31305 & 73.950958 & 30.337911 &\nodata & $174 \pm 11$ & $6.4 \pm 1.7 $ & $-21.9 \pm 1.6$ &\nodata & A1V & N & Y/?\\
329 &  HD 31293 &  73.941022 &  30.551191 & $139_{-16}^{+22}$  &  $121 \pm 50$  &  $1.9 \pm 0.9 $ &  $-24.4 \pm 0.7$  &  $8.9 \pm 0.9$  &  $\sim$A0V &  Y & Y/Y
\enddata
\\
\hline
\\
\multicolumn{12}{p{7in}}{
(1) Error on $d_{_{\rm SPEC}}$ is the standard deviation among distances calculated using B,V,R,J,H, and K magnitudes, as available, and underestimates true error values; see text. 
(2) Errors on other columns are taken from original references; see text.
(3) Spectral types are from SIMBAD and Kharchenko et al. 2009 in the top three sections, 
with a slash ($/$) denoting any discrepancy between these two compilations, and from {\it SDSS} in the fourth section.
In cases where we have derived new spectral types ourselves in this paper (see Table~\ref{tab.spec}), our types supersede
those from the references.
(4) Last two columns state whether or not (Y/N) the star is a probable Taurus member based on the two distance 
estimates ($d$), and the two kinematic assessments: proper motion ($\mu$) and radial velocity (RV); see text.
}

\end{deluxetable*}

\clearpage
\begin{table*}
\tabletypesize{\footnotesize}
\caption{Spectral types derived through new spectroscopy. }
\scriptsize{
\begin{tabular}{p{0.1cm}p{1.6cm} p{0.3cm} lp{0.55cm}p{0.45cm}l
lp{0.55cm}p{0.5cm}ll
p{0.45cm}lllp{0.55cm}
llll}
\hline \hline
B\# & Star & \multirow{3}{*}{$\frac{\rm SNR}{100}$} & \multicolumn{16}{c}{Equivalent Widths}   & Derived\\ 
\cline{4-19}
 &  &  & Ca II K & N II & He I & He I & H$_{\delta}$ & He I & Si II & He I & H$_{\gamma}$ & He I & He I & MgII & H$_{\beta}$ & He I & H$_{\alpha}$ & He I &  SpT\\
 &  &  & 3933.7 & 3995.0 & 4009.3 & 4026.2 & 4101.7 & 4120.8 & 4128.1, 4130.9 & 4143.7 & 4340.4 & 4387.9 & 4471.5 & 4481.2 & 4862.3 & 4921.9 & 6562.8 & 6678.2 & \\
\hline
\multicolumn{20}{c}{Spectral Type Standards}\\
\hline
 &  HD 36960  & 3.0 & 0.12 & 0.09 & 0.42 & 1.2 & 4.0 & 0.48 & 0.02 & 0.55 & 4.2 & 0.58 & 0.98 & 0.15 & 3.6 & 0.78 & 3.4 & 0.68 &   B0.5V\\
 &  HD 19374  & 2.6 & 0.15 & 0.08 & 0.56 & 1.5 & 5.7 & 0.35 & 0.08 & 0.82 & 5.2 & 0.87 & 1.6 & 0.20 & 5.4 & 0.97 & 4.2 & 0.66 &   B1.5V\\
 &  HD 35912  & 3.2 & 0.14 & 0.04 & 0.65 & 1.5 & 5.9 & 0.26 & 0.12 & 0.81 & 5.9 & 0.91 & 1.6 & 0.25 & 5.9 & 0.88 & 4.4 & 0.56 &   B2V\\
 &  HD 28375  & 2.6 & 0.13 & 0.01 & 0.30 & 1.1 & 7.6 & 0.20 & 0.21 & 0.45 & 7.6 & 0.61 & 1.1 & 0.28 & 7.4 & 0.56 & 5.5 & 0.31 &   B3V\\
 &  HD 16219  & 2.7 & 0.13 & 0.01 & 0.19 & 0.82 & 8.1 & 0.13 & 0.26 & 0.33 & 8.0 & 0.48 & 0.85 & 0.32 & 8.0 & 0.43 & 5.7 & 0.21  &  B5V\\
 &  HD 21071  & 2.7 & 0.14 & $<$0.01 & 0.15 & 0.84 & 8.4 & 0.11 & 0.24 & 0.29 & 8.4 & 0.44 & 0.80 & 0.30 & 8.7 & 0.41 & 5.8 & 0.20 &   B7V\\
 &  HD 14272  & 1.7 & 0.19 & \nodata & $<$0.02 & 0.40 & 10.2 & $^a$ & 0.24 & 0.11 & 9.3 & 0.11 & 0.25 & 0.31 & 10.1 & 0.28 & 6.4 & 0.12 &   B8V\\
 &  HD 16350  & 1.9 & 0.57 & \nodata & \nodata & $^a$ & 12.3 & $^a$ & 0.26 & 0.08 & 17.8 & $^a$ & 0.10 & 0.38 & 11.9 & $<$0.23 & 8.7 & 0.04 &   B9.5V\\
 &  HD 14171  & 1.7 & 0.29 & \nodata & \nodata & \nodata & 14.6 & $^a$ & 0.30 & $^a$ & 13.0 & \nodata & 0.023 & 0.33 & 13.6 & $^a$ & 9.4 & $^a$ &   B9.5V\\
\hline
\multicolumn{20}{c}{O,B, and A0 stars from SIMBAD}\\
\hline
50   &  HD 283809  & 1.8 & 0.24 & 0.09 & 0.70 & 1.4 & 4.9 & 0.35 & 0.10 & 0.86 & 4.9 & 0.90 & 1.4 & 0.27 & 5.0 & 0.93 & 3.8 & 0.64 &   B1.5-2V\\
51  &  $\tau$ Tau  & 2.4 & 0.29 & 0.01 & 0.43 & 1.2 & 8.2 & 0.21 & 0.15 & 0.56 & 7.9 & 0.72 & 1.0 & 0.27 & 7.8 & 0.68 & 5.6 & 0.38 &   B3V\\
25  &  72 Tau  & 3.1 & 0.12 & 0.02 & 0.16 & 0.82 & 9.8 & $>$0.1 & 0.24 & 0.27 & 9.2 & 0.49 & 0.75 & 0.31 & 8.6 & 0.41 & 6.0 & 0.18 &   B7V\\
88  &  HD 31806  & 3.4 & 0.12 & 0.01 & 0.21 & 0.77 & 9.7 & $^a$ & 0.22 & 0.27 & 8.7 & 0.45 & 0.72 & 0.34 & 8.8 & 0.39 & 6.4 & 0.19 &   B7V\\
6   &  HD 283304  & 1.2 & 0.42 & \nodata & 0.03 & $^a$ & 8.9 & $^a$ & 0.79 & $^c$ & 8.2 & 0.15 & 0.07 & 0.16 & 8.2 & 0.20 & 6.6 & $<$0.05 &   B8V Si\\
33  &  HD 282276  & 1.4 & 0.28 & 0.02 & 0.05 & 0.31 & 8.9 & $^c$ & 0.41 & 0.09 & 8.2 & $^a$ & 0.29 & 0.26 & 8.0 & 0.24 & 6.5 & 0.12 &   B8V\\
36  &  HR 1445  & 2.3 & 0.19 & $^c$ & 0.04 & 0.41 & 9.6 & $^a$ & 0.30 & 0.13 & 9.1 & $<$0.20 & 0.32 & 0.32 & 8.7 & 0.26 & 6.7 & 0.05 &   B8V \\
13  &  V1137 Tau  & 2.9 & 0.18 & 0.01 & 0.11 & 0.29 & 6.4 & $>$0.05 & 0.37 & 0.11 & 5.9 & $<$0.22 & 0.30 & 0.22 & 6.5 & 0.22 & 5.9 & 0.09 &   B8III\\
48  &  HD 29647  & 1.6 & 0.31 & 0.04 & $<$0.1 & 0.24 & 8.7 & $^a$ & 0.34 & 0.09 & 7.8 & $<$0.19 & 0.26 & 0.39 & 7.7 & 0.22 & 6.1 & 0.06 &   B9III \\
81  &  HD 283971  & 1.6 & 0.92 & 0.01 & 0.03 & 0.25 & 11.1 & $^a$ & 0.25 & 0.07 & 9.5 & $^a$ & 0.19 & 0.38 & 10.6 & $<$0.22 & 7.9 & $<$0.05 &   B9V\\
63  &  HD 30378  & 3.0 & 0.28 & $<$0.01 & 0.01 & 0.24 & 13.6 & $^a$ & 0.24 & 0.06 & 13.0 & $<$0.1 & 0.27 & 0.44 & 12.7 & 0.18 & 9.1 & $<$0.05 &   B9.5V\\
4   &  BD+23607  & 2.3 & 1.1 & $<$0.01 & $^c$ & $<$0.01 & 16.6 & $^a$ & $^a$ & $^a$ & 14.0 & $^a$ & 0.08 & 0.38 & 13.9 & $<$0.19 & 9.7 & $<$0.02 &   A0V\\
12  &  HD 283449  & 1.2 & 0.75 & $^c$ & $^c$ & 0.23 & 14.3 & $^a$ & 0.26 & $<$0.09 & 13.7 & $^c$ & 0.11 & 0.46 & 14.3 & 0.14 & 10.7 & $<$0.05 &   A0V\\
53  &  V1081 Tau  & 3.1 & 0.59 & $^{a,c}$ & $^c$ & 0.11 & 16.8 & $^a$ & $^a$ & 0.01 & 16.8 & $^a$ & 0.13 & 0.37 & 15.8 & $<$0.16 & 10.7 & $<$0.03 &   A0V\\
19  &  V892 Tau  & 0.5 & $^a$ & \nodata & $^c$ & 0.2 & 14 & $^c$ & $^c$ & $^c$ & $^b$ & $^c$ & $^c$ & 0.2 & $^b$ & $^c$ & $^b$ & $^c$ &   $\sim$A0Ve\\
\hline
\multicolumn{20}{c}{Early-type stars from \cite{rebull2010}}\\
\hline
95  &  HD 283751  & 1.5 & 0.39 & 0.49 & 0.26 & 0.82 & 8.0 & 0.13 & 0.23 & 0.36 & $^b$ & 0.50 & 0.79 & 0.32 & $^b$ & 0.31 & $^b$ & 0.21 &   B5Ve\\
90  &  HD 27923  & 2.3 & 0.24 & $^c$ & 0.04 & 0.33 & 12.9 & $^a$ & 0.18 & 0.07 & 11.8 & $<$0.21 & 0.27 & 0.35 & 11.5 & 0.18 & 8.2 & $<$0.05 &   B9V\\
91  &  HD 283637  & 1.6 & 0.91 & $^c$ & 0.1 & $<$0.32 & 15.9 & $^a$ & $>$0.04 & $^{a,c}$ & 13.7 & $^a$ & 0.1 & 0.37 & 13.6 & 0.22 & $^b$ & 0.1 &   B9.5eV\\
94  &  HD 284530  & 2.1 & 0.34 & $<$0.01 & $^c$ & 0.30 & 13.6 & $^a$ & 0.30 & 0.09 & 11.7 & $<$0.35 & 0.25 & 0.38 & 11.5 & 0.24 & 8.5 & $<$0.07 &   B9.5V\\
93  &  2MASS0431+29 & 1.9 & 0.48 &\nodata  & $<$0.03 & $^a$ & 16.6 & $^a$ & $^a$ & $^c$ & 17.2 & $^{a/c}$ & $<$0.06 & 0.27 & 15.6 & $<$0.24 & 10.9 & $<$0.01 &   A1V\\
89  &  HD 27659  & 1.0 & 2.6 &\nodata  & $<$0.19 & $\leqslant$0.16 & 10.4 & $^a$ & 0.4 & $\leqslant$0.19 & 10.4 & $<$0.39 & $<$0.23 & 0.47 & 11.6 & $<$0.48 & 10.1 & 0.1 &   A3V\\
92  &  2MASS0428+27 & 0.3 & 3.5 & $^a$ & $<$0.04 & $^c$ & 12 & $^a$ & 0.5 & $<$0.2 & 12 & $^a$ & $<$0.6 & 0.87 & 14 & $<$0.11 & 9.6 & 0.14 &   A4III\\
\hline
\multicolumn{20}{c}{Early-type stars selected using 2MASS colors}\\
\hline
106 &  HD 26212  & 1.2 & 3.9 & $^a$ & $^a$ & $^a$ & 11.8 & $^a$ & 0.4 & $\leqslant$0.16 & 8.9 & $<$0.33 & $<$0.21 & 0.41 & 9.0 & $\leqslant$0.42 & 9.6 & 0.1 &   A5V\\
\hline
\multicolumn{20}{c}{Other}\\
\hline
328 &  HD 31305  & 1.8 & 1.7 &\nodata  & $^c$ & $<$0.27 & 17.6 & $^a$ & 0.3 & $<$0.09 & 16.3 & $<$0.28 & $<$0.16 & 0.46 & 15.0 & $<$0.27 & 9.6 & $<$0.02 &   A1V\\
329 &  AB Aur  & 2.4 & $^b$ &\nodata  &\nodata  &\nodata  &\nodata  &\nodata  & 0.2 &\nodata  &\nodata  &\nodata  &\nodata  & 0.40 &\nodata  &\nodata  & $^b$ & $^b$ &   A0Ve\\
\hline
\multicolumn{20}{p{7in}}{
B\# is repeated from Table ~\ref{tab.allBstars}. SNR is the signal-to-noise ratio of the obtained spectrum.
Notes- 
(a) embedded in an adjacent absorption line, (b) emission line present within the absorption line, (c) hard to distinguish from noise.}
\end{tabular}
}
\label{tab.spec}
\end{table*}

\begin{table}[htp]
\centering
\tabletypesize{\footnotesize}
\caption{Physical parameters derived through model atmosphere fitting.}
\begin{tabular}{crrrl}
\hline
\hline
Star & $T_{\rm eff}$ & $v \cdotp$ sin $i$  & log g  & SpT\\
    & (K)& (km s$^{-1}$)  & (cgs)  & \\
\hline
HD 283809               & 21000         & $<50$                 & 4.0       & B2V\\
HD 29763 = $\tau$ Tau   & 18000         & 150                   & 4.0       & B3V\\  
HD 283751               & 15000         & 50                    & 4.0       & B5V\\
HD 28149 = 72 Tau       & 14000         & 75                    & 4.0       & B7V\\  
HD 31806                & 14000         & 100                   & 4.0       & B7V\\
HD 28929 = HR 1445      & 13000         & $<50$                 & 4.0       & B7V\\  
HD 26571 = V1137 Tau    & 13000         & $<50$                 & 3.5       & B7III\\  
HD 284530               & 12000         & $<50$                 & 4.0       & B7.5V\\
HD 283971               & 12000         & 50                    & 4.0       & B7.5V\\
HD 283304               & 12000         & $<50$                 & 3.5       & B8III\\
HD 282276               & 12000         & $<50$                 & 3.5       & B8III\\
HD 29647                & 11500         & $<50$                 & 3.5       & B8III\\
HD 27923                & 11000         & $<50$                 & 4.0       & B8.5V\\
AB Aur                  & 11000         & 200                   & 4.5       & B8.5V\\
2MASS J04313313+2928565 & 11000         & 250                   & 4.5       & B8.5V\\
HD 31305                & 11000         & 150                   & 4.5       & B8.5V\\
HD 30378                & 11000         & $<50$                 & 4.0       & B8.5V\\
V892 Tau                & 11000         & 100                   & 4.5       & B8.5V\\
HD 29935 = V1081 Tau    & 11000         & 200                   & 4.5       & B8.5V\\
HD 283637               & 11000         & 50                    & 4.0       & B8.5V\\
BD+23 607               & 10000         & $<50$                 & 4.0       & A0V\\
HD 283449               & 10000         & $<50$                 & 4.0       & A0V\\
HD 27659                &  9000         & $<50$                 & 3.5       & A1III\\
2MASS J04285940+2736254 &  9000         & $<50$                 & 3.5       & A4III\\
HD 26212                &  8500         & $<50$                 & 3.5       & A5III\\
\hline
\end{tabular}
\label{tab.kurucz_param}
\end{table}

\begin{table*}
\centering
\scriptsize{
\caption{Final list of early-type stars showing indications of membership with Taurus.}
\begin{tabular}{p{0.1cm}cccccllrcp{4.5cm}}
\hline \hline
B\#  & HD Number  & Alt. Identifier & $\alpha_{\rm J2000}$& $\delta_{\rm J2000}$   & SpT    & $d_{_{HIP}}$      & $d_{_{SPEC}}$& $P(\chi^2)$ & RV              & Comments \\
     &             &                  & (h,m,s)             & (\degr,\arcmin,\arcsec)&           & (pc)              & (pc)         & (\%)        & (km s$^{-1}$)   & \\
\hline
\multicolumn{11}{c}{Probable members based on our analysis of distance and/or kinematics}\\
\hline
51   & HD 29763      & $\tau$ Tau      & 04 42 14.70         & 22 57 24.9             & B3V       & $123^{+13}_{-11}$ & $137\pm9$    & 5.1         & $12.3\pm4.1$    & \\
25   & HD 28149      & 72 Tau          & 04 27 17.45         & 22 59 46.8             & B7V       & $127^{+13}_{-11}$ & $161\pm3$    & 2.4         & \,\,$7.3\pm2.6$ & IR nebula; cool dust SED\\
36   & HD 28929      & HR 1445         & 04 34 37.99         & 28 57 40.1             & B8V       & $143^{+20}_{-16}$ & $157\pm3$    & 11.5        & $12.6\pm2.2$    & Weak nebula\\
19   &  ----         & V892 Tau        & 04 18 40.62         & 28 19 15.5             & $\sim$B8.5--A0Ve &  \nodata   & \dag         & 25.7   &        & IR nebula; Class I SED\\
329  &  HD 31293     & AB Aur          & 04 55 45.85         & 30 33 04.3             & A0Ve      & $139_{-16}^{+22}$ & $120\pm50$   & 36.1        & $8.9\pm0.9$     & Class II SED\\
328  & HD 31305      & IRAS 04526+3015 & 04 55 48.23         & 30 20 16.5             & A1V       &    \nodata        & $174\pm11$   & 21.5        &                 & Cool dust SED\\
193  &  HD 31648     & MWC 480         & 04 58 46.27         & 29 50 37.0             & A3Ve      & $137^{+31}_{-21}$ & $186\pm48$   & 18.8        &                 & Class II SED\\
106  & HD 26212      &   ----          & 04 09 43.68         & 24 04 22.6             & A5V       & $123^{+17}_{-14}$ &              & 3.3         & $20.3\pm3.9$ &              \\
\hline
\multicolumn{11}{c}{Candidates with several membership indicators but that are not secure distance and kinematic members}\\
\hline
46   &  ----         & IC 2087-IR         & 04 39 55.75         & 25 45 02.0             & B5-B8     &  \nodata          & $ $          & 46.7        &                 & IR and visible nebula; Class I SED; member\\
96   &  HD 283815    & ----            & 04 42 41.18         & 24 41 17.9             & A0        &      \nodata      & $268\pm40$   & 12.8        &                 & Meets proper motion but not distance criteria \\ 
89   &  HD 27659     & ----            & 04 22 54.66         & 28 23 55.0             & A3V       &      \nodata      & $164\pm10$   &  $<0.1$     & $ $             & Meets distance but not proper motion criteria; cool dust SED \\
\hline
\multicolumn{11}{c}{Stars illuminating infrared nebulae but that can not be associated with Taurus based on distance and kinematic criteria}\\
\hline
33   & HD 282276     &   ----          & 04 33 04.23         & 29 21 49.9             & B8V       &    \nodata        & $422\pm52$    & $<0.1$      &               & IR nebula; cool dust SED \\
48   & HD 29647      & IRAS 04380+2553 & 04 41 08.05         & 25 59 34.0             & B9III     & $177^{+43}_{-29}$ & $160\pm1$     & $<0.1$      &               & IR nebula; cool dust SED\\
\hline
\multicolumn{11}{p{7in}}{
(1) B\# is repeated from Table ~\ref{tab.allBstars}.
(2) SpT is the spectral type as revised in this work or from literature.
(3) As noted elsewhere, error on d$_{SPEC}$ corresponds to variance among calculations and underestimates the true error.
(4) The probability of proper motion membership, $P(\chi^2)$, is as discussed in Section~\ref{subsubsec.pm_rv}.
(5) The membership criteria used in this work are: $P(\chi^2) > 1\%$; $128 < d < 162$ pc within $1\sigma$ error; and $9.8 \leq RV \leq 17.5$ km s$^{-1}$ wherever radial velocity is available.
(6) For the $^\dag$ source V892 Tau, $d_{_{SPEC}} = 1697\pm1548$ pc, but $d_B=4720$ pc while $d_K=135$ pc.
}
\end{tabular}
\label{tab.members}
}
\end{table*}

\end{document}